\pdfoutput=1
\documentclass[PAPER, atlasdraft=false, cernpreprint, UKenglish, texmf, orcidlogo]{atlasdoc}
\usepackage{atlaspackage}
\usepackage{multirow}
\usepackage{atlasbiblatex}

\usepackage{atlasphysics}
\graphicspath{{logos/}{figures/}}

\usepackage{ANA-HIGP-2024-09-PAPER-defs}
\AtlasTitle{Probing the Higgs boson CP properties in vector-boson fusion production in the $H\rightarrow\tau^+\tau^-$ channel with the ATLAS detector}

\AtlasAbstract{%
The CP properties of the Higgs boson are studied in the vector-boson fusion production mode. The analysis exploits the decay mode of the Higgs boson into two $\tau$-leptons using 140~fb$^{-1}$ of proton--proton collision data at $\sqrt{s}=13$~TeV collected by the ATLAS experiment at the Large Hadron Collider.  Results are obtained using the Optimal Observable method. CP-violating interactions between the Higgs boson and electroweak gauge bosons are considered in the effective field theory framework, with the interaction strength described in the HISZ basis by $\tilde{d}$, and in the Warsaw basis by $c_{H\tilde{W}}$, $c_{H\tilde{B}}$, and $c_{H\tilde{W}B}$.  No deviations relative to the Standard Model are observed, and limits are obtained on the strength parameters. The $\tilde{d}$ parameter is constrained to the interval [$-0.012, 0.044$] at the 95\% confidence level while $c_{H\tilde{W}}$ is constrained to [$-0.24, 0.83$], when considering both linear and quadratic effects of physics beyond the Standard Model.
}
\author{The ATLAS Collaboration}

\AtlasRefCode{HIGP-2024-09}

\PreprintIdNumber{CERN-EP-2025-127}

\AtlasDate{\today}

\AtlasJournal{JHEP}
\AtlasJournalRef{\JHEP 10 (2025) 92}
\AtlasDOI{10.1007/JHEP10(2025)092}

\hypersetup{pdftitle={ATLAS document},pdfauthor={The ATLAS Collaboration}}

\begin{document}

\maketitle

\tableofcontents

\section{Introduction}

The origin of the observed baryon asymmetry in the universe is one of the fundamental open questions in our current understanding of particle physics.  In 1967 Sakharov proposed three necessary conditions~\cite{Sakharov:1967dj} to explain the  generation of the baryon asymmetry. One of these conditions includes the violation in fundamental interactions of the invariance under combined charge (C) and parity (P) conjugation.  In the Standard Model (SM) of particle physics, CP-violating contributions arise from a complex phase in the quark mixing matrix. However, the magnitude of these effects is not enough to explain the observed value of the baryon asymmetry in the universe~\cite{Huet:1994jb,Gavela:1993ts,Cohen:1993nk,Riotto:1999yt,Hou:2008xd}.

The discovery of the Higgs boson at the Large Hadron Collider (LHC)~\cite{Evans:2008zzb} by the ATLAS and CMS collaborations~\cite{HIGG-2012-27,CMS-HIG-12-028} has opened a new sector in which possible CP-violating interactions can be investigated. %
In the SM, the Higgs boson is CP-even (scalar) in the interaction with other SM particles, both in production and decay.  Any deviation from the SM predictions would be a sign of physics beyond the SM (BSM). In particular, a promising avenue for testing CP invariance is in the interaction between the Higgs boson ($H$) and electroweak gauge bosons, namely the $HVV$ vertex  (where $V=W^{\pm},Z$ vector bosons)~\cite{Brehmer:2017lrt}.
The current results from the ATLAS and CMS Collaborations strongly support the scalar nature of the Higgs boson, consistent with SM expectation~\cite{HIGG-2018-30-fix,HIGG-2020-08,HIGP-2024-07,CMS-HIG-20-007}. %
In order to search for a clear sign of physics beyond the SM, the results are interpreted in the effective field theory (EFT) framework, and limits are set on the Wilson coefficients describing the strength of the CP-violating interactions terms with the Higgs boson.
The BSM contributions to the SM tensor structure are parameterised with additional higher dimensional operators $O_i^{(d)}$ of dimension $d$ and combine the SM Lagrangian $\mathcal{L}_{\textrm{SM}} $ to form an EFT Lagrangian:
\begin{equation*}
\mathcal{L}_{\textrm{EFT}}=\mathcal{L}_{\textrm{SM}}+\sum_i\frac{c_i^{(d)}}{\Lambda^{(d-4)}}O_i^{(d)}
\end{equation*}
with dimension $d>4$, $\Lambda$ the scale of the new physics, and $c_i^{(d)}$  the Wilson coefficients describing the strength of the new interactions. In this paper only operators of dimension six are considered.  Operators with odd dimension imply lepton and baryon number violation, and are therefore ignored. Higher dimension operators such as dimension-8 are also not considered, since they are suppressed by an additional factor $\Lambda^2 $. %
From this formulation of the effective Lagrangian it is possible to obtain a complete set of independent operators.  There is no unique way to form such a set of operators,  and one of the most used is the HISZ basis~\cite{PhysRevD.48.2182,Hankele:2006ma}.  In this formulation, after electroweak symmetry breaking, the Lagrangian, corresponding to the  CP-violating terms in the $HVV$ vertex, can be written in the mass basis as described in Ref.  \cite{BUCHMULLER1986621,Hankele:2006ma} as:

\begin{equation*}
\mathcal{L}_{\textrm{EFT}}=
\mathcal{L}_{\textrm{SM}}
+\tilde{c}_{H\gamma\gamma}H \tilde{A}_{\mu\nu }A^{\mu\nu }
+\tilde{c}_{H\gamma Z}H \tilde{A}_{\mu\nu }Z^{\mu\nu }
+\tilde{c}_{HZZ}H \tilde{Z}_{\mu\nu }Z^{\mu\nu }
+\tilde{c}_{HWW}H \tilde{W}^+_{\mu\nu }W^{-\mu\nu }
\end{equation*}

where H is the Higgs boson,  $V^{\mu\nu}$ (with $V=W^{\pm},Z,A$) denotes the field strength tensor for the gauge fields $W,Z,\gamma$, and $\tilde{V}^{\mu\nu}=\epsilon^{\mu\nu\rho\sigma} V_{\rho\sigma}$ is the corresponding dual field strength tensor. Only two of these four couplings are independents due to constraints imposed by U(1)$_Y$ and SU(2)$_{I_W,L} $ invariance,  and can be described by two dimensionless parameters $\tilde{d} $ and $\tilde{d}_B $,

\begin{linenomath}
\begin{eqnarray*}
\tilde{c}_{H\gamma\gamma}  =   \frac{g}{2 m_W} (\tilde{d} \sin^2 \theta_W + \tilde{d}_B \cos^2 \theta_W) & &
\tilde{c}_{H\gamma Z}  =   \frac{g}{2 m_W} \sin 2\theta_W (\tilde{d} -\tilde{d}_B)\\
\tilde{c}_{HZZ}  =  \frac{g}{2 m_W} (\tilde{d} \cos^2 \theta_W + \tilde{d}_B \sin^2 \theta_W) & &
\tilde{c}_{HWW}  =   \frac{g}{m_W} \tilde{d} \,,
\end{eqnarray*}
\end{linenomath}

where $\theta_W$ is the Weinberg angle, and $g$ is the SU(2) coupling constant. Assuming the arbitrary condition $\dtil=\dtil_B$\footnote{The assumption $\dtil=\dtil_B$ is a common one in literature, and allows comparison and combination with other analyses such as Ref.~\cite{HIGG-2018-14,HIGG-2018-30-fix,HIGG-2020-08}.}, the CP-violating effect can then be parameterised by a single parameter \dtil.
Another widely used EFT parameterisation is the Warsaw basis~\cite{Grzadkowski:2010es}, adopted by both ATLAS~\cite{HIGG-2020-16,HIGG-2020-08}  and CMS~\cite{CMS-HIG-23-016} in the SMEFT~\cite{Brivio:2020onw} formalism. In this basis, three CP-odd terms contribute to the $HVV$ vertex.
These CP-odd operators with their corresponding Wilson coefficients are:
\begin{equation*}
\frac{\chwtil}{\Lambda^2} H^{\dag}H\tilde{W}_{\mu\nu}^IW^{\mu\nu I}, \frac{\chbtil}{\Lambda^2} H^{\dag}H\tilde{B}_{\mu\nu}B^{\mu\nu }, \frac{\chwbtil}{\Lambda^2} H^{\dag}H\sigma^I\tilde{W}_{\mu\nu}^IB^{\mu\nu},
\end{equation*}
where the Higgs doublet and its complex conjugate are defined by $H$ and $H^{\dag}$,  $W^I_{\mu\nu}$ and  $B_{\mu\nu}$ are the field strength tensor for the $W^I$ ($I=1,2,3$) and $B$ gauge fields, and $ \tilde{W}^I_{\mu\nu}=\epsilon^{\mu\nu\rho\sigma}W^I_{\rho\sigma} (\tilde{B}_{\mu\nu} = \epsilon^{\mu\nu\rho\sigma}B_{\rho\sigma})$ are their dual, and $\sigma^I $ are the Pauli matrices.
The two representations are not independent, and it is possible to transform one into the other. If it is assumed that $\chwtil=\chbtil$  and $ \chwbtil=0 $, it is possible to show that $\dtil=\frac{v^2}{\Lambda^2}\chwtil=\frac{v^2}{\Lambda^2}\chbtil $, where $v$ is the Higgs vacuum expectation value.
The changes in the cross-section due to BSM effects can be parameterised with an interference term that depends linearly on the Wilson coefficient, and is responsible for the CP-odd effects, and a quadratic term that is purely from BSM physics, but CP-even. In this analysis,  the interpretations use both the linear and quadratic terms, but the case where only the linear term is included in the BSM parameterisation is also considered.
The Wilson coefficients in both parameterisations have been constrained by different analyses from LEP experiments~\cite{DELPHI:2008uqu,LEPDtil2001,20057},   and ATLAS and CMS, in both Run 1 and Run 2, and in different channels. These include measurements of CP-odd observables in $H\rightarrow \gamma\gamma$ in the vector-boson fusion (VBF) production channel~\cite{HIGG-2020-08}.
Another ATLAS analysis based on CP-odd observables investigated the channel $H\rightarrow ZZ^*\rightarrow 4\ell$ ~\cite{HIGG-2018-30-fix}, where the $HVV$ coupling enters in both the production and decay for VBF production, and only in the decay for gluon--gluon fusion (ggF) production.  The same final state was investigated by CMS,  although using a different basis for the BSM interpretation~\cite{CMS-HIG-19-009}. A similar analysis was also performed in the channel $H\rightarrow WW^* \rightarrow \ell\nu\ell\nu$~\cite{HIGP-2024-07}. %
The $H\rightarrow\tau^+\tau^- $ channel\footnote{For simplicity a pair of $\tau^+\tau^-$ are referred to as $\tau\tau$ in the rest of the paper.}  has also been used by ATLAS to constrain the Wilson coefficients in the HISZ basis in both Run~1~\cite{HIGG-2015-06} and in a 36~\ifb partial Run~2 dataset~\cite{HIGG-2018-14}.  A similar analysis in the $H\rightarrow\tau\tau $  final state was performed by CMS~\cite{CMS-HIG-20-007} in Run 2,  following the same BSM formalism as Ref.~\cite{CMS-HIG-19-009}, described in Ref. ~\cite{CMS-HIG-14-018}\footnote{In the CMS analysis in Ref.~\cite{CMS-HIG-14-018} the CP-violating strength parameter is $a_3^{VV}$, but the effective cross section ratio $f_{a3}$~\cite{CMS-HIG-19-009} is instead measured.  The \dtil parameter is related to the cross section ratio by: $\dtil^2=\frac{f_{a3}}{(\frac{\sigma_{a_3}}{\sigma_{SM}})\cdot(1-f_{a3})} $, where $\sigma_i$ are the cross sections when only the related terms are present in the Lagrangian. }.
A differential measurement of the  $H\rightarrow\tau\tau$ channel was done by ATLAS using the full Run~2 dataset~\cite{HIGG-2022-07}, where the CP-sensitive azimuthal angle between the two VBF jets, $\Delta\phi_{jj}^{\textrm{sign}}$ (defined in Section~\ref{sec:theoryandmethodology}), was unfolded and provided constraints on different EFT parameters.
To date, no hints of CP violations have been observed and the best constraints on \dtil have been set in Ref.~\cite{HIGG-2018-30-fix}, with limits $[-0.017, 0.014] $ at 68\% confidence level (CL) including both linear and quadratic terms in the parameterisation of the BSM effects.
For \chwtil, \chbtil, and \chwbtil, limits at the 68\% CL are set at $[-0.07, 1.09]$, $[-0.42, 0.31]$,  and $[-0.56, 0.53]$, respectively, assuming a scale of new physics of $ \Lambda =1~\TeV$ and considering only the linear term in the BSM effects. For both bases, the limits obtained considering the linear only and both linear and quadratic terms in the parameterisation of the BSM effects are found to be in agreement.

In this analysis, a search for CP violation in the VBF production vertex of the Higgs boson is performed, using the $H\rightarrow\tau\tau $ decay mode.
With a comparatively high branching ratio of 6\%~\cite{Zyla:2020zbs-fix}, this channel provides a phase space with a high signal-to-background ratio and a high efficiency for Higgs boson candidate reconstruction.  The analysis is performed using a dataset of 140~\ifb of proton--proton (\(pp\)) collisions recorded by the ATLAS  experiment during the Run 2 of the LHC between 2015 and 2018. The analysis is based on the Optimal Observable method~\cite{Atwood:1991ka,Davier:1992nw,Diehl:1993br}. This observable combines the full information of the multidimensional production phase space, independent of the decay mode, providing a full description of the production process. Other CP-odd observables are also used in the analysis for comparison.

The analysis presented in this paper follows closely the publication in Ref.~\cite{HIGG-2018-14}, but with several improvements: new background estimates were performed and novel machine learning techniques were implemented to increase the signal-to-background ratio.  The results are interpreted in the context of the HISZ and Warsaw bases.

The paper is structured as follows: Section~\ref{sec:theoryandmethodology} describes the CP-odd observables used in the analysis. Section~\ref{sec:det} describes the ATLAS detector and  Section~\ref{sec:samples} the Monte Carlo (MC) simulations and data samples used in the analysis.
Section~\ref{sec:selection} describes the event selection and the implementation of the new machine learning techniques, while Sections~\ref{sec:bkg} and~\ref{sec:systematics} describe the background estimation and the main sources of systematic uncertainties.   Section~\ref{sec:fitmodel} describes the fit setup and in Section~\ref{sec:results} the results are shown.


\section{CP-odd observables and methodology}
\label{sec:theoryandmethodology}

Tests of CP-invariance can be performed in a model independent way by measuring the mean value of a CP-odd observable $<\mathcal{O}_{\textrm{CP}} >$. If CP-invariance holds, the mean value vanishes: $<\mathcal{O}_{\textrm{CP}} >\; = 0$. An observation of a non-vanishing mean value is then a clear sign of CP-violation.
The production kinematics consisting of the reconstructed Higgs boson four-momentum and the two jets from the VBF production (tagging jets) can, under the assumption of a fixed Higgs mass and massless jets, be characterised by a seven-dimensional phase space.
The Optimal Observable method combines the information of these seven phase space variables into a single observable, which is shown to have the highest sensitivity to small values of the CP-violating strength parameter.
It is defined by the ratio of the interference term in the matrix element (ME) to the SM contribution:
\begin{equation*}
\OO =\frac{2\Re(\mathcal{M}_{\textrm{SM}}^{*}\mathcal{M}_{\textrm{CP-odd}})}{|\mathcal{M}_{\textrm{SM}}|^{2}}.
\end{equation*}

The mean value of the Optimal Observable can be experimentally measured, and if no CP violation is present $<\OO>\; = 0$, as the observable is a CP-odd (and $\hat{T}$-odd\footnote{ $\hat{T}$ denotes the time reversal according to Ref.~\cite{Brehmer:2017lrt},which inverts the directions of momenta and spins. }) variable.  This assumes that any effects from rescattering (i.e., new particles being on mass shell in the loop corrections to the $HVV$ vertex) are negligible~\cite{Brehmer:2017lrt}, since the initial-state of VBF production of the Higgs boson is not a CP-symmetric process.
Figure~\ref{fig:oodem} shows the distribution of the \OO for SM VBF Higgs production and for two example distributions  where $\dtil\neq 0$, on simulated samples after the selections described in Section~\ref{sec:selection}.
This Figure shows how the mean of the \OO is shifted to higher or lower values, depending on the sign of the \dtil parameter.

\begin{figure}[!t]
\centering
\subfloat[]{  \includegraphics[width=.33\textwidth]{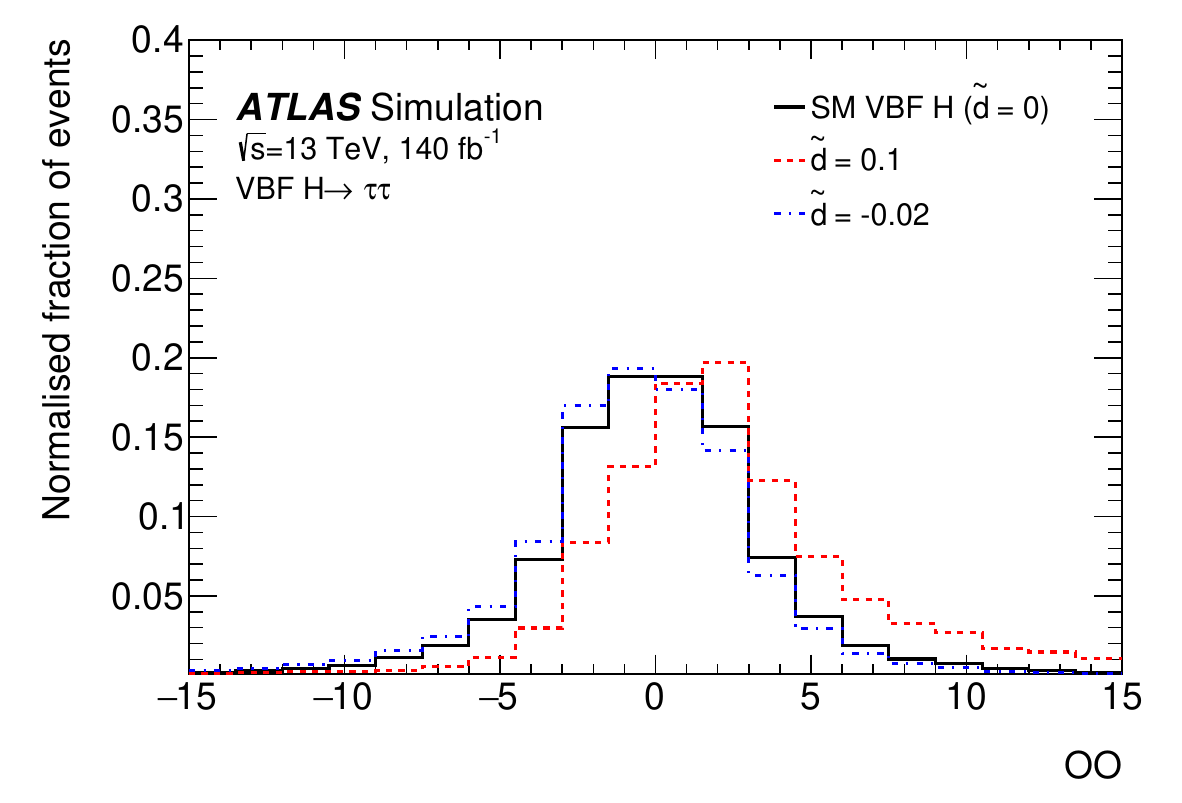}\label{fig:oodem}}
\subfloat[]{  \includegraphics[width=.33\textwidth]{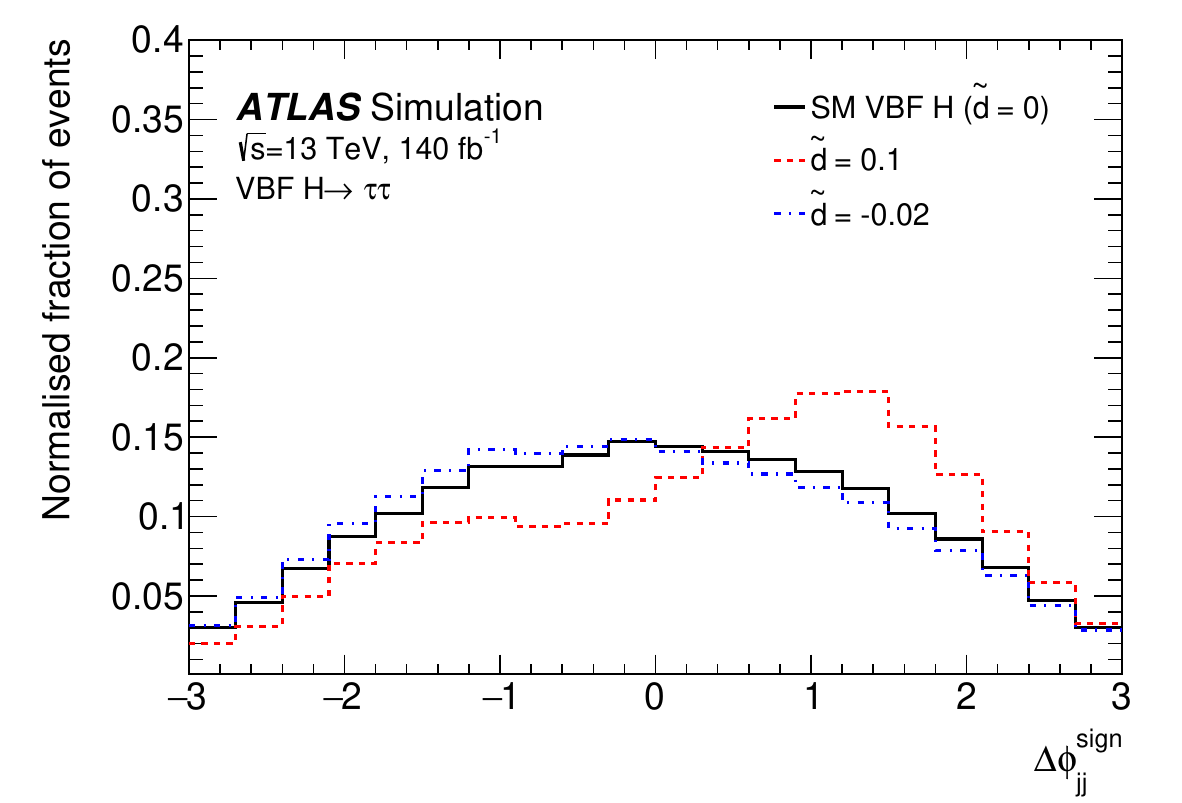}\label{fig:dphijj_dem}}
\subfloat[]{  \includegraphics[width=.33\textwidth]{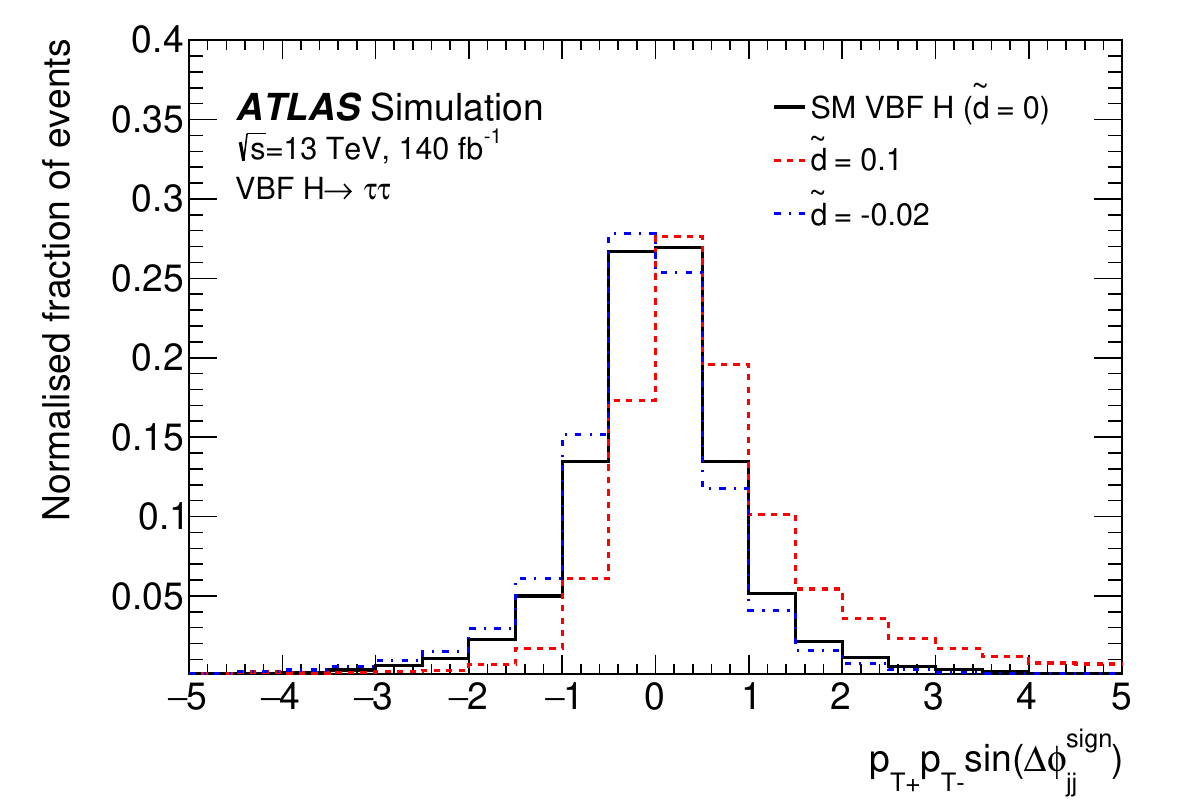}\label{fig:ooreg_dem}}

\caption{Distributions for SM VBF Higgs boson production and two VBF Higgs boson samples with $\dtil\neq 0$, for (a) the Optimal Observable, \OO, (b) $\Delta\phi_{jj}^{\textrm{sign}}$, and (c) $\OOReg$, shown for final states where one $\tau$-lepton decays hadronically, and the other one, leptonically.
The MC simulations used are those described in Section~\ref{sec:samples}.}
\end{figure}

The values of the leading-order matrix elements needed for the calculation of the Optimal Observable are extracted from HAWK~\cite{Ciccolini:2007jr,Ciccolini:2007ec,Denner:2014cla}, using the HISZ basis with condition $\tilde{d}=\tilde{d}_B$. The evaluation requires the four vectors of the Higgs boson and the two tagging jets. The Bjorken $x$ values of the incoming partons in positive ($x_1$) and negative ($x_2$) $z$-direction can be derived exploiting energy-momentum conservation from the Higgs boson and tagging jet  four-momenta as
\begin{equation*}
x_{1/2}^{\textrm{reco}} = \frac{m_{\textrm{Hjj}}}{\sqrt{s}}e^{\pm y_{\textrm{Hjj}}} \,,
\end{equation*}
where $m_{\textrm{Hjj}}$ ($y_{\textrm{Hjj}}$) is the mass (rapidity) of the vectorial sum of the tagging jets and the Higgs boson four-momentum, the latter estimated by using the missing mass calculator (MMC)~\cite{Elagin:2010aw}.
As the flavour of the incoming and outgoing partons cannot be determined experimentally, the sum over all possible flavour configurations $ij \rightarrow kl H$ weighted by the parton distribution functions
is calculated separately for the matrix elements in numerator and denominator:
\begin{eqnarray*}
2\Re(\mathcal{M}_{\textrm{SM}}^{*}\mathcal{M}_{\textrm{CP-odd}}) &=& \sum_{i,j,k,l}
f_i(x_1) f_j(x_2) 2\Re(\mathcal{M}_{\textrm{SM}}^{*}\mathcal{M}_{\textrm{CP-odd}})
(ij \rightarrow kl H) \\
|\mathcal{M}_{\textrm{ SM}}|^2 &=&  \sum_{i,j,k,l} f_i(x_1) f_j(x_2) |\mathcal{M}_{\textrm{ SM}}|^2
(ij \rightarrow kl H) \,.
\end{eqnarray*}

Another simple CP-odd observable suggested in Ref.~\cite{PhysRevD.74.095001} is defined based on the ``signed'' azimuthal angular difference between the two tagging VBF jets:
\begin{equation*}
\epsilon_{\mu\nu\rho\sigma}b^{\mu}_+p^{\nu}_+b^{\rho}_-p^{\sigma}_-=2p_{\textrm{T}+}p_{\textrm{T}-}\sin(\phi_+-\phi_-)=2p_{\textrm{T}+}p_{\textrm{T}-}\sin(\Delta\phi_{jj}^{\textrm{sign}}) \,,
\end{equation*}
where $b^{\mu}_{+/-}$ denotes the normalised four-momenta of the two incoming protons, $p^{\mu}_{+/-} (\phi_{+/-})$ denote the four-momenta (azimuthal angles) of the two outgoing VBF tagging jets, and where $p_{+/-}$ points into the same detector hemisphere as $b^{\mu}_{+/-}$.
In the literature, $\Delta\phi_{jj}^{\textrm{sign}}$ is also a commonly proposed CP-odd observable; however, the combined term $p_{\textrm{T}+}p_{\textrm{T}-}\sin(\Delta\phi_{jj}^{\textrm{sign}})$ is usually more sensitive than $\Delta\phi_{jj}^{\textrm{sign}}$, due to the additional information of the momentum of the tagging jets.  In this analysis \OOReg is also considered for comparison to the Optimal Observable.
Figures~\ref{fig:dphijj_dem} and~\ref{fig:ooreg_dem} show the distributions of $\Delta\phi_{jj}^{\textrm{sign}}$ and $p_{\textrm{T}+}p_{\textrm{T}-}\sin(\Delta\phi_{jj}^{\textrm{sign}})$, respectively.


\section{ATLAS detector}
\label{sec:det}

%
\newcommand{\AtlasCoordFootnote}{%
ATLAS uses a right-handed coordinate system with its origin at the nominal interaction point (IP)
in the centre of the detector and the \(z\)-axis along the beam pipe.
The \(x\)-axis points from the IP to the centre of the LHC ring,
and the \(y\)-axis points upwards.
Polar coordinates \((r,\phi)\) are used in the transverse plane,
\(\phi\) being the azimuthal angle around the \(z\)-axis.
The pseudorapidity is defined in terms of the polar angle \(\theta\) as \(\eta = -\ln \tan(\theta/2)\) and is equal to the rapidity
$ y = \frac{1}{2} \ln \left( \frac{E + p_z}{E - p_z} \right) $ in the relativistic limit.
Angular distance is measured in units of \(\Delta R \equiv \sqrt{(\Delta y)^{2} + (\Delta\phi)^{2}}\).}

The ATLAS detector~\cite{PERF-2007-01} at the LHC covers nearly the entire solid angle around the collision point.\footnote{\AtlasCoordFootnote}
It consists of an inner tracking detector surrounded by a thin superconducting solenoid, electromagnetic and hadronic calorimeters,
and a muon spectrometer incorporating three large superconducting air-core toroidal magnets.

The inner-detector system (ID) is immersed in a \qty{2}{\tesla} axial magnetic field
and provides charged-particle tracking in the range \(|\eta| < 2.5\).
The high-granularity silicon pixel detector covers the vertex region and typically provides four measurements per track,
the first hit generally being in the insertable B-layer (IBL) installed before Run~2~\cite{ATLAS-TDR-19,PIX-2018-001}.
It is followed by the SemiConductor Tracker (SCT), which usually provides eight measurements per track.
These silicon detectors are complemented by the transition radiation tracker (TRT),
which enables radially extended track reconstruction up to \(|\eta| = 2.0\).
The TRT also provides electron identification information
based on the fraction of hits (typically 30 in total) above a higher energy-deposit threshold corresponding to transition radiation.

The calorimeter system covers the pseudorapidity range \(|\eta| < 4.9\).
Within the region \(|\eta|< 3.2\), electromagnetic calorimetry is provided by barrel and
endcap high-granularity lead/liquid-argon (LAr) calorimeters,
with an additional thin LAr presampler covering \(|\eta| < 1.8\)
to correct for energy loss in material upstream of the calorimeters.
Hadronic calorimetry is provided by the steel/scintillator-tile calorimeter,
segmented into three barrel structures within \(|\eta| < 1.7\), and two copper/LAr hadronic endcap calorimeters.
The solid angle coverage is completed with forward copper/LAr and tungsten/LAr calorimeter modules
optimised for electromagnetic and hadronic energy measurements, respectively.

The muon spectrometer (MS) comprises separate trigger and
high-precision tracking chambers measuring the deflection of muons in a magnetic field generated by the superconducting air-core toroidal magnets.
The field integral of the toroids ranges between \num{2.0} and \qty{6.0}{\tesla\metre}
across most of the detector.
Three layers of precision chambers, each consisting of layers of monitored drift tubes, cover the region \(|\eta| < 2.7\),
complemented by cathode-strip chambers in the forward region, where the background is highest.
The muon trigger system covers the range \(|\eta| < 2.4\) with resistive-plate chambers in the barrel, and thin-gap chambers in the endcap regions.

The luminosity is measured mainly by the LUCID--2~\cite{LUCID2} detector that records Cherenkov light produced in the quartz windows of photomultipliers located close to the beampipe.

Events were selected by the first-level trigger system implemented in custom hardware,
followed by selections made by algorithms implemented in software in the high-level trigger~\cite{TRIG-2016-01}.
The first-level trigger accepted events from the \qty{40}{\MHz} bunch crossings at a rate close to \qty{100}{\kHz},
which the high-level trigger further reduced in order to record complete events to disk at about \qty{1.25}{\kHz}.

A software suite~\cite{SOFT-2022-02} is used in data simulation, in the reconstruction
and analysis of real and simulated data, in detector operations, and in the trigger and data acquisition
systems of the experiment.


\section{Data and simulated event samples}
\label{sec:samples}

This analysis uses the full dataset collected by the ATLAS detector during Run 2 (2015--2018) of the LHC,
consisting of events produced from \(pp\) collisions at a centre-of-mass energy of $\sqrt{s} =13~\TeV$.
Events are selected only if they are recorded under stable beam conditions where all relevant detector subsystems are in good operating condition~\cite{DAPR-2018-01}, and satisfy data-quality requirements.
The dataset corresponds to a total integrated luminosity of 140~\ifb~\cite{DAPR-2021-01}.

Monte Carlo simulated event samples are used to model the $H \to \tau\tau$ signal and most of the backgrounds from SM processes, including \Zttjets, top-quark processes and multi-boson production. A summary of the generators used for the simulation is shown in Table~\ref{tab:SamplesUsed}. Backgrounds arising from misidentified objects are estimated with a data-driven method described in Section~\ref{sec:bkg}. The same event generators as in Ref.~\cite{HIGG-2018-14} are used, with the exception of the generator used for the parton shower (PS) of the \ttbar and single top simulations.
The normalisation of all Higgs boson samples accounts for the decay branching ratio calculated with HDECAY~\cite{Djouadi:1997yw,Spira:1997dg,Djouadi:2006bz-fix} and \PROPHECY~\cite{Bredenstein:2006ha,Bredenstein:2006rh,Bredenstein:2006nk-fix}.
For all samples, excluding those simulated using \SHERPA, the decays of bottom and charm hadrons were performed by \EVTGEN~\cite{Lange:2001uf}.

\begin{table}[tp]
\centering
\caption{
Summary of nominal samples used including the information about the matrix element (ME) generators, the parton shower (PS), parton distribution function (PDF) sets used to compute the ME and the PS, the set of tuned parameters (Tune), and the order of the cross-section calculation used for normalisation of the samples (Order). Further details are given in the text.
}
\begin{scriptsize}
\begin{tabular}{ lcccccc }
\toprule
Process &  \multicolumn{2}{c}{\qquad Generator}&  \multicolumn{2}{c}{PDF set }& Tune &  Order \\
&  ME & PS & ME & PS & & (QCD / EWK) \\
\midrule
\midrule
Higgs \\
\midrule
ggF &\POWHEGBOX[v2] & \PYTHIA[8.2] & \PDFforLHC[15nnlo] & \CTEQ[6L1] &  \AZNLO & N$^3$LO / NLO \\
VBF &\POWHEGBOX[v2] & \PYTHIA[8.2] & \PDFforLHC[15nlo] & \CTEQ[6L1] & \AZNLO & NNLO / NLO \\
$VH$ &\POWHEGBOX[v2] & \PYTHIA[8.2] & \PDFforLHC[15nlo] & \CTEQ[6L1] & \AZNLO & (N)NLO / NLO \\
\(t\bar{t}H\)  &\POWHEGBOX[v2] & \PYTHIA[8.2] &  \NNPDF[3.0nlo] & \NNPDF[2.3lo] & A14 & NNLO / NLO \\
\tHq & \textsc{MG5\_aMC@NLO 2.6.0} &  \PYTHIA[8.2] &  \NNPDF[3.0nlo] & \NNPDF[2.3lo] & A14 & NLO / NLO \\
\midrule
Non-Higgs\\
\midrule
\ttbar / single top & \POWHEGBOX[v2] & \PYTHIA[8.2] &  \NNPDF[3.0nlo] &  \NNPDF[2.3lo] &  A14 & NNLL / NNLO \\
\(V+\)jets &    \SHERPA[2.2.1]  &    \SHERPA[2.2.1]   & \NNPDF[3.0nnlo]  & \NNPDF[3.0nnlo]   & \SHERPA & NNLO \\
EWK \(V+\)jets &    \SHERPA[2.2.1] &  \SHERPA[2.2.1]  &  \NNPDF[3.0nnlo] &  \NNPDF[3.0nnlo]   & \SHERPA & LO \\
Diboson & \SHERPA[2.2.2]/2.2.1 & \SHERPA[2.2.2]/2.2.1 &  \NNPDF[3.0nnlo] &  \NNPDF[3.0nnlo]  & \SHERPA & NLO \\
\bottomrule
\end{tabular}
\end{scriptsize}
\label{tab:SamplesUsed}
\end{table}

Higgs boson production via gluon--gluon fusion was simulated at next-to-next-to-leading-order (NNLO) accuracy in QCD using \POWHEGBOX[v2]~\cite{Nason:2009ai,Alioli:2010xd,Nason:2004rx,Frixione:2007vw}, with the \PDFforLHC[15nnlo] parton distribution function (PDF) set~\cite{Butterworth:2015oua}.
The simulation achieves NNLO accuracy for arbitrary inclusive \(gg\to H\) observables by reweighting the Higgs boson rapidity spectrum in \textsc{Hj}-\MINLO~\cite{Hamilton:2012np,Campbell:2012am,Hamilton:2012rf} to that of HNNLO~\cite{Catani:2007vq}.
For PS, the \CTEQ[6L1]~\cite{Pumplin:2002vw} PDF and the  set of tuned parameters (tune) \AZNLO~\cite{STDM-2012-23} of \PYTHIA[8.2]~\cite{Sjostrand:2014zea} were used.

The gluon--gluon fusion prediction from the MC samples was normalised to the next-to-next-to-next-to-leading-order (N$^3$LO) cross-section in QCD plus electroweak corrections at next-to-leading-order (NLO)~\cite{deFlorian:2016spz,Anastasiou:2016cez,Anastasiou:2015ema,Dulat:2018rbf,Harlander:2009mq,Harlander:2009bw,Harlander:2009my,Pak:2009dg,Actis:2008ug,Actis:2008ts,Bonetti:2018ukf,Bonetti:2018ukf}.

Higgs boson production via vector-boson fusion was simulated with \POWHEGBOX[v2] and interfaced with \PYTHIA[8.2] for PS and non-perturbative effects, with parameters set according to the \AZNLO tune.
The \POWHEGBOX prediction is accurate at NLO and uses the \PDFforLHC[15nlo] PDF set.
It was normalised to an approximate-NNLO QCD cross-section with NLO electroweak corrections~\cite{Ciccolini:2007jr,Ciccolini:2007ec,Bolzoni:2010xr}.

Higgs boson production in association with a vector boson was simulated using \POWHEGBOX[v2] and interfaced with \PYTHIA[8.2] for PS and non-perturbative effects.
The \POWHEGBOX prediction is accurate at NLO for \(VH\) boson plus one-jet production.
The loop-induced \(gg\to ZH\) process was generated separately at leading order. The \PDFforLHC[15nlo] PDF set and the \AZNLO tune of \PYTHIA[8.2] were used.
The MC prediction was normalised to cross-sections calculated at NNLO in QCD with NLO electroweak corrections for \(q\bar{q}/qg \to VH\) and at NLO and next-to-leading-logarithm accuracy in QCD for \(gg \to ZH\)~\cite{Ciccolini:2003jy,Brein:2003wg,Brein:2011vx,Altenkamp:2012sx,Denner:2014cla,Brein:2012ne,Harlander:2014wda}.
The production of \ttH events was modelled using the \POWHEGBOX[v2] generator at NLO with the \NNPDF[3.0nlo] PDF set.
The events were interfaced to \PYTHIA[8.2] using the A14 tune~\cite{ATL-PHYS-PUB-2014-021} and the \NNPDF[2.3lo]~\cite{Ball:2012cx-fix} PDF set.
The production of \tHq events was modelled using the \MGNLO[2.6.0]~\cite{Alwall:2014hca} generator at NLO with the \NNPDF[3.0nlo] PDF\@.
The events were interfaced with \PYTHIA[8.2] using the A14 tune and the \NNPDF[2.3lo] PDF set.
The production of \(V+\)jets was simulated with the \SHERPA[2.2.1]~\cite{Bothmann:2019yzt-fix} generator using NLO ME for up to two partons, and leading-order (LO) ME for up to four partons calculated with the Comix~\cite{Gleisberg:2008fv} and \OPENLOOPS~\cite{Buccioni:2019sur,Cascioli:2011va,Denner:2016kdg} libraries.
They were matched with the \SHERPA PS~\cite{Schumann:2007mg} using the \MEPSatNLO prescription~\cite{Hoeche:2011fd,Hoeche:2012yf,Catani:2001cc,Hoeche:2009rj} using the set of tuned parameters developed by the \SHERPA authors.
The \NNPDF[3.0nnlo] set of PDFs was used and the samples were normalised to an NNLO prediction~\cite{Anastasiou:2003ds}.

Electroweak production of \(\ell\ell jj\), \(\ell\nu jj\) and \(\nu\nu jj\) final states (EW $V+$jets) was simulated with \SHERPA[2.2.1] using leading-order ME with up to one additional parton emission.
The ME were merged with the \SHERPA PS following the prescription described above.
The \NNPDF[3.0nnlo] set of PDFs was employed.
The samples were produced using the VBF approximation, which avoids overlap  with semileptonic diboson topologies by requiring a \(t\)-channel colour-singlet exchange.
The production of \ttbar events was modelled using the \POWHEGBOX[v2] generator at NLO with the \NNPDF[3.0nlo] PDF set and the \hdamp parameter\footnote{The \hdamp parameter is a resummation damping factor and one of the parameters that controls the matching of \POWHEG ME to the PS and thus effectively regulates the high-transverse momentum (\pT) radiation against which the \ttbar system recoils.} set to 1.5\,\mtop~\cite{ATL-PHYS-PUB-2016-020}.
The events were interfaced to \PYTHIA[8.2] to model the PS, hadronisation, and underlying event, with parameters set according to the A14 tune and using the \NNPDF[2.3lo] set of PDFs.
The \ttbar sample was normalised to the cross-section prediction at NNLO
in QCD including the resummation of next-to-next-to-leading logarithmic (NNLL) soft-gluon terms calculated using
\TOPpp[2.0]~\cite{Beneke:2011mq,Cacciari:2011hy,Baernreuther:2012ws,Czakon:2012zr,Czakon:2012pz,Czakon:2013goa,Czakon:2011xx}.

Single-top \(t\)-channel production was modelled using the \POWHEGBOX[v2] generator at NLO in QCD using the four-flavour scheme and the corresponding \NNPDF[3.0nlo] set of PDFs.
The events were interfaced with \PYTHIA[8.2] using the A14 tune and the \NNPDF[2.3lo] set of PDFs.
The inclusive cross-section was corrected to the theory prediction calculated at NLO in QCD with NNLL soft-gluon corrections~\cite{Kidonakis:2010ux,Kidonakis:2013zqa}.

The associated production of top quarks with \(W\) bosons (\(tW\)) was modelled by the \POWHEGBOX[v2] generator at NLO in QCD using the five-flavour scheme and the \NNPDF[3.0nlo] set of PDFs.
The diagram removal scheme~\cite{Frixione:2008yi} was used to remove interference and overlap with \ttbar production.
The related uncertainty was estimated by comparison with an alternative sample generated using the diagram subtraction scheme~\cite{Frixione:2008yi,ATL-PHYS-PUB-2016-020}.
The events were interfaced to \PYTHIA[8.2] using the A14 tune and the \NNPDF[2.3lo] set of PDFs.

Samples of diboson final states (\(VV\)) were simulated with the \SHERPA[2.2.1] or 2.2.2 generator depending on the process,
including off-shell effects and Higgs boson contributions, where appropriate.
Fully leptonic final states and semileptonic final states, where one boson decays leptonically and the other hadronically, were generated using ME at NLO accuracy in QCD for up to one additional parton and at LO accuracy for up to three additional parton emissions.
Samples for the loop-induced processes \(gg \to VV\) were generated using LO-accurate ME for up to one additional parton emission for both the cases of fully leptonic and semileptonic final states.
The ME calculations were matched and merged with the \SHERPA PS based on Catani--Seymour dipole factorisation~\cite{Gleisberg:2008fv,Schumann:2007mg} using the MEPS@NLO prescription~\cite{Hoeche:2011fd,Hoeche:2012yf,Catani:2001cc,Hoeche:2009rj}.
The virtual QCD corrections were provided by the \OPENLOOPS library~\cite{Buccioni:2019sur,Cascioli:2011va,Denner:2016kdg}.
The \NNPDF[3.0nnlo] set of PDFs was used, along with the dedicated set of tuned PS parameters developed by the \SHERPA authors.

The effect of multiple interactions in the same and neighbouring bunch crossings (\pileup) was modelled by overlaying the simulated hard-scattering event with inelastic proton--proton events generated with \PYTHIA[8.1]~\cite{Sjostrand:2007gs} using the \NNPDF[2.3lo] set of PDFs and the A3 set of tuned parameters~\cite{ATL-PHYS-PUB-2016-017}.
All samples were processed through an ATLAS detector simulation~\cite{SOFT-2010-01} based on \GEANT~\cite{Agostinelli:2002hh}.

In this analysis, only Higgs boson production via VBF is considered as signal. This includes the decay into $H\rightarrow\tau\tau $ and the decay $H\rightarrow WW^*\rightarrow\ell\nu\ell\nu$. The analysis is not sensitive to the CP violating effects in the $H\rightarrow\tau\tau $ and $H\rightarrow WW^*$ decay vertex, since the Optimal Observable contains only production kinematics information, and it is the same for all the decay channels. All other Higgs boson production mechanisms are considered as backgrounds, with all other couplings set to the SM values. Non-VBF Higgs boson production, along with $Z\to\ell\ell$ and $W\to\tau\nu + \mathrm{jets}$, are for the remainder of the text grouped together and denoted ``other backgrounds''.

The production of the signal samples with non-vanishing values of CP violation in the $HVV$ vertex is obtained with a reweighting procedure, based on the EFT interpretation considered. For the HISZ basis, an event-based weight is obtained from the ratio of the squared ME value of the VBF process associated with a specific amount of CP mixing (given in terms of \dtil) to that obtained from the SM\@. These weights are obtained from HAWK, considering the $2\rightarrow 2 +H  $ and $2\rightarrow 3 +H  $ processes separately.  The MEs are evaluated using the parton-level information of the four momentum and flavour of the initial and final state partons and the Higgs boson.

For the Warsaw basis, a procedure similar to the one described in Ref.~\cite{HIGG-2020-08} is used, where
an event-weight is obtained based on the \OO distribution at different values of the Warsaw basis operator considered. The weights are obtained as
\begin{equation*}
\frac{d\sigma}{d\OO}=\left.\left( \frac{d\sigma}{d\OO}\right)^{\textrm{NLO}} \times\left( \frac{d\sigma}{d\OO}\right)^{\textrm{MG5}}_{c_i}\middle/\left( \frac{d\sigma}{d\OO}\right)^{\textrm{MG5}}_{c_i=0} \,,\right.
\end{equation*}
where the term with ``NLO'' is the SM VBF signal prediction, while those with ``MG5'' describe the predictions obtained with \MADGRAPH~\cite{Alwall:2014hca,Frederix:2012ps} for a given value of the CP-violating coupling constant considered ($c_i$), and fixing all the other Wilson coefficient to zero.  The \MADGRAPH samples for $c_i\neq 0$ are obtained by setting the scale of the new physics to $\Lambda=1~\TeV $ and using the $m_W$ input scheme.
The \MADGRAPH samples are generated at LO and effects of higher-order QCD and electroweak corrections are assumed to factor out in the ratio. The procedure is done separately for each Wilson coefficient considered.

\FloatBarrier


\section{Event selection}
\label{sec:selection}

This analysis targets final states with at least two jets and a $H\to\tau\tau$ decay candidate.
Three combinations of $\tau$-lepton decays are considered: fully hadronic ($\thadhad$), where both $\tau$-leptons decay to hadrons (with $\tauhadvis$ being the visible part) and a $\nu$; semileptonic ($\tlhad$), where one decays hadronically and the other to $\ell\nu\bar{\nu}$, $\ell = e,\,\mu$; and leptonic ($\tll$), where one decays to $e\nu\bar{\nu}$ and the other to $\mu\nu\bar{\nu}$.  Same-flavour leptonic decays are rejected due to the large contribution from \Zee and \Zmm backgrounds.

The definition of the reconstructed objects follows what is described in more detail in Ref.~\cite{HIGG-2022-07}. The full description of the requirements in each channel is given in Table~\ref{tab:eventselection}.


\begin{table}[tbhp]
\caption{
Summary of the event selection for all sub-channels of the analysis. For electrons and muons, the lowest \pT requirement depends on the trigger used, which changes with the year of data taking. The \pT requirements per year are listed in Ref.~\cite{HIGG-2022-07}.
}
\begin{center}
\resizebox{\textwidth}{!}{\begin{tabular}{@{}r|cccc@{}}
\toprule
Criteria
& \temu
& \multicolumn{2}{c}{\tlhad}
& \thadhad
\\
\cmidrule(lr){3-4}
&
& \tehad
& \tmhad
&
\\
\textbf{Preselection} & & & & \\ \midrule
$N(e)$
& 1
& 1
& 0
& 0
\\
\midrule
$N(\mu)$
& 1
& 0
& 1
& 0
\\
\midrule
$N(\tauhadvis)$
& 0
& 1
& 1
& 2
\\

\midrule
$N(b\textrm{-jets})$
& 0 (85$\%$ WP)
& 0 (85$\%$ WP)
& 0 (85$\%$ WP)
& 0 (70$\%$ WP)
\\
&
&
\\
\midrule
\multirow{2}{*}{$\pT\,[\GeV]$}
& \multirow{2}{*}{$e/\mu$: $\pt >10$ to $27.3$}
& \multicolumn{2}{c}{$e/\mu$: $\pt >10$ to $27.3$,}
& \multirow{2}{*}{$\tauhadvis$: $\pt>40,30$}
\\
&
&
\multicolumn{2}{c}{$\tauhadvis$: $\pt > 30 $}
&
\\[3pt]
\midrule
Identification
& $e$/$\mu$: Medium
& \multicolumn{2}{c}{$e$/$\mu$: Medium, $\tauhadvis$: RNN Medium}
& \tauhadvis: RNN Medium
\\
&
& \multicolumn{2}{c}{ }
& \\
\midrule
Isolation
& $e$: Loose, $\mu$: Tight
& $e$: Loose
& $\mu$: Tight
&
\\
&
& \multicolumn{2}{c}{ }
&
\\
\midrule
Electric charge
& \multicolumn{4}{c}{Opposite charge}
\\
\midrule
\MET\,[\GeV]
& \multicolumn{4}{c}{\textgreater\,20}
\\
&
& \multicolumn{2}{c}{ }
&
\\
\midrule
\multirow{2}{*}{Kinematics}
& $m_{\tau\tau}^{\mathrm{coll}} > (m_Z - 25)\,\GeV$
& \multicolumn{2}{c}{$\mT < \SI{70}{\GeV}$}
&
\\
& $\SI{30}{\GeV} < m_{e\mu} < \SI{100}{\GeV}$
& \multicolumn{2}{c}{ }
&
\\
&
& \multicolumn{2}{c}{ }
&
\\
\midrule
Leading jet
& \multicolumn{3}{c}{$\pT > \SI{40}{\GeV}$}
& $\pT > \SI{70}{\GeV}$, $|\eta| < 3.2$
\\
&
& \multicolumn{2}{c}{ }
&
\\
\midrule
\multirow{2}{*}{Angular}
& $\Delta R_{e\mu} < 2.5$
& \multicolumn{2}{c}{$\Delta R_{\ell\tauhadvis} < 2.5$}
& $0.6 < \Delta R_{\tauhadvis\tauhadvis} < 2.5$
\\
& $|\Delta\eta_{e\mu}| < 1.5$
& \multicolumn{2}{c}{  $|\Delta\eta_{\ell\tauhadvis}| < 1.5$ }
& $|\Delta\eta_{\tauhadvis\tauhadvis}| < 1.5$
\\
&
& \multicolumn{2}{c}{ }
&
\\
\midrule
\multirow{2}{*}{Coll.\ app.\ $x_{1}/x_{2}$} & $0.1 < x_{1} < 1.0$
& \multicolumn{2}{c}{$0.1 < x_{1} < 1.4$}
& $0.1 < x_{1} < 1.4$
\\
& $0.1 < x_{2} < 1.0$
& \multicolumn{2}{c}{ $0.1 < x_{2} < 1.2$}
& $0.1 < x_{2} < 1.4$
\\
&
& \multicolumn{2}{c}{ }
&
\\
\midrule
\textbf{VBF inclusive} & \multicolumn{4}{c}{sub-leading jet $\pT > \SI{30}{\GeV}$} \\

& \multicolumn{4}{c}{$\mjj > \SI{350}{\GeV}$, $|\detajj| > 3$} \\
& \multicolumn{4}{c}{$\eta(j_{1}) \times \eta(j_{2}) < 0$} \\

& \multicolumn{4}{c}{lepton centrality: all visible decay products of the two $\tau-$leptons lie between the two jets in $\eta$}\\

& \multicolumn{4}{c}{ } \\
\bottomrule
\end{tabular}}
\end{center}
\label{tab:eventselection}
\end{table}


Events were collected with the lowest unprescaled single-lepton, dilepton, or di-hadronic $\tau$-lepton triggers~
\cite{TRIG-2018-05,TRIG-2018-01,TRIG-2019-03,TRIG-2019-02}.
Each trigger chain has a different online requirement on the \pT of the reconstructed leptons, depending on the year.  The offline requirements are set tighter to ensure that the analysis is on the trigger efficiency plateau, and follow the selection used in Ref.~\cite{HIGG-2022-07}.
All channels require exactly two identified and isolated $\tau$-lepton decay candidates, and events with an additional $\tau$-lepton candidate are rejected.
The two $\tau$-candidates are required to have opposite sign electric charge.
All channels are required to have a missing transverse energy of $\MET>20~\GeV$, to improve the resolution of the MMC\@. In the \LL channel an additional lower limit on the collinear mass~\cite{Ellis:1987xu}  of the two-$\tau$-lepton system is required,  $\mttcol>(m_Z-25)~\GeV $ (where $m_Z$ is the $Z$ boson mass), to ensure orthogonality with the $H\rightarrow WW^* $ analysis~\cite{HIGG-2021-20}.  The effect on signal acceptance is negligible. Moreover,  it is also required that the dilepton invariant mass lies in the range of $30~\GeV<m_{e\mu}<100~\GeV$ to enhance the fraction of Higgs events and reduce the number of background events. In the \LH channel a requirement on the transverse mass between the lepton and the missing transverse energy of $\mT< 70~\GeV $ is instead imposed to reduce the contribution from leptonic $W$ decays.

In all channels, the contribution from events containing a top-quark is reduced by vetoing events with $b$-tagged jets; the \LL and \LH channels use an 85\% working point (WP) of the DL1r $b$-tagging algorithm, while in the \HH channel a 70\% WP~\cite{FTAG-2019-07} is used, to ensure similar performance in all the channels.
The \LL and \LH channels require the leading jet to have a transverse momentum of $\pT^{j_1}>40~\GeV$,  to enhance the VBF contribution. In the \HH channel, the requirement is raised to $\pT^{j_1}>70~\GeV$, and it is also required that the leading jet be within the pseudorapidity range of $|\eta|<3.2$. This is done to ensure candidate events are on the di-hadronic $\tau$-lepton triggers efficiency plateau.  Requirements on the angular distance between the reconstructed decays of the two $\tau$-candidates are applied in all channels,  to reject non-resonant background events. Additional requirements are applied on the momentum fraction, $x_1/x_2$,  of each $\tau$-lepton carried by its visible decay products in the collinear approximation to enhance the sensitivity to the Higgs signal and reduce background contributions.

A set of requirements to define the VBF topology is applied. Each event is required to have at least two jets, where in addition to the leading jet requirements above, the sub-leading jet must have a transverse momentum $\pT^{j_2}$ greater than $30~\GeV$, the invariant mass of the two leading jets $m_{jj}$ must be greater than $350~\GeV$, and the pseudorapidity separation $|\Delta\eta_{jj}|$ between the two jets must be greater than $3.0$, with each jet in a separate hemisphere of the detector. Additionally, all visible decay products of the $\tau-$leptons must lie in $\eta$ between the two leading jets.

To construct an enriched signal region (SR), a multivariate classifier based on a neural network (NN) is used. The network is implemented in \textsc{Keras}~\cite{chollet2015keras} using \textsc{TensorFlow}~\cite{tensorflow} as backend.
For each channel, a separate NN is optimised to best exploit the different compositions and features of the channels. All three channels have similar features in architecture and training procedure: the ADAM optimizer algorithm~\cite{adam} is used, with ReLU activation functions~\cite{relu}, and L2 regularisation~\cite{l2} and $k$-fold (with $k=5$) to prevent overfitting.  A binary classifier is built for the \LH and \HH channel, while in the  \LL channel a multi-class classifier is optimised, with the categories being: signal, \Ztt, misidentified $\tau$-lepton decays, and other backgrounds.  This is done to better describe the background coming from misidentified $\tau$-lepton decays. The training is performed assuming the SM predictions, therefore no CP-odd component is added, only the VBF Higgs SM sample is used. The NN is then tested for different CP-violating hypotheses, and is found to be consistent with the SM case, indicating the NN is not inducing any biases in the choice of events based on the CP hypothesis considered. Each channel uses as input a different set of observables, optimised to achieve the best separation between signal and background. The list of input variables used in the training is shown in Table~\ref{tab:features}. The different input variables can be categorised as follows:
\begin{itemize}
\item[•] Four-vectors of the two $\tau$-decay products, and of the two leading VBF tagging jets.
\item[•] Missing transverse energy vector and its ratio to the transverse momentum of other objects. Additionally, the \MET centrality\footnote{
The $\MET-\phi-$centrality quantifies the angular direction of the missing transverse momentum relative to the visible $\tau$-decay products in the transverse plane, and is constructed as: $\MET-\phi\textrm{-centrality}/\sqrt{2}= \frac{r+s}{\sqrt{r^2+s^2}}  $, where $r=\sin(\phi(\MET)-\phi(\tau_1))/\sin(\phi(\tau_2)-\phi(\tau_1))$ and  $s=\sin(\phi(\tau_2)-\phi(\MET))/\sin(\phi(\tau_2)-\phi(\tau_1))$.
} and missing transverse energy constructed using only the hard-scatter particles, \METhpto, are considered in the training. The transverse masses using the \MET and both $\tau$-decay products are also considered.
\item[•] Properties of the resonant di-$\tau$ system are used,  such as the invariant mass of the $\tau$-decay products (using both the missing mass calculator, $\mmmc$~\cite{Elagin:2010aw}, and using only the visible part of the $\tau$-decay products \mttvis ),  and the angular distances, $\Delta R_{\tau_1\tau_2}$, $\Delta\eta_{\tau_1\tau_2}$,  and $\Delta\phi_{\tau_1\tau_2}$. In addition,  the scalar sum, $\pT(\tau_1)+\pT(\tau_2)$, and the ratio, $\pT(\tau_1)/\pT(\tau_2)$, of the transverse momenta of the two  $\tau$-decay products are considered. The ratio of the difference and the sum of transverse momenta of the di-$\tau$ decay products, $[\pT(\tau_1)-\pT(\tau_2)]/[\pT(\tau_1)+\pT(\tau_2)]$, is also used.

\item[•] Properties of the VBF topology: invariant mass of the dijet system, \mjj,  the product and the difference of the pseudorapidity of the two leading jets and the transverse momentum of their vectorial sum, $\eta-$centrality of the $\tau$-decay products, $C_{jj}(\tau_1)$, $C_{jj}(\tau_2)$\footnote{
$C_{jj}(\tau) = \exp \left[ \frac{-4}{(\eta_{j_1} - \eta_{j_2})^2}\left(\eta_{\tau} - \frac{\eta_{j_1} + \eta_{j_2}}{2}\right)^2\right]$, where $\eta_{\tau}$, $\eta_{j_{1/2}}$ are the pseudorapidities of the $\tau$-decay products and the two leading jets, respectively. This variable has a value of unity when the object is halfway in $\eta $ between the two jets, $1/e$ when the object is aligned with one of the jets, and $<1/e$ when the object is not between the jets in $\eta$.}, with respect to the pseudorapidity of the two tagging jets. Furthermore, the training also uses momenta of different systems composed of the objects considered in the VBF topology definition (di-$\tau$-decay products, two leading jets, and \MET). The azimuthal separation and the invariant mass between the di-$\tau$-system and the leading jet is also considered.

\end{itemize}
The most important input features in the training are the reconstructed masses \mmmc, \mttvis, \mjj, and the transverse momentum of the system composed of the $\tau$-decay products, jets and \MET, $\pT(\tau_1,\tau_2,j_1,j_2,\MET)$.

Figure~\ref{fig:nnscore} shows the distribution of the NN signal score for the three channels considered, after the fit described in Section~\ref{sec:fitmodel}.  All systematic uncertainties described in Section~\ref{sec:systematics} are included in the hatched area.  This figure shows the ability of the NN to select a region with a high signal-to-background ratio ($s/b$).  The MC predictions are in good agreement with the observed data in all channels. Figure~\ref{fig:nnscore} also shows the NN score distributions for signal samples with different \dtil hypotheses. These BSM samples are all in agreement, showing how the NN is insensitive to the value of the CP-violating strength parameter.
Two orthogonal signal regions are defined for each channel, based on the values of the NN signal score.  The high NN score SR (\SRHighNN) is defined by all the events with NN signal score greater than a threshold value. The optimal threshold value, along with the optimal \OO binning,  is found by choosing the value that gives the most stringent expected 68\% CL  \dtil interval.
A second SR, called low NN score SR (\SRLowNN), is then defined by selecting the events with an NN signal score between the \SRHighNN optimal cut and the NN signal score below which the ratio $s/b$ is less than 5\%.  Despite having a lower signal-to-background ratio than \SRHighNN,  \SRLowNN still contributes to the sensitivity, and also helps to better constrain the impact of the systematic uncertainties. The region with an NN signal score lower than the \SRLowNN threshold is used as a validation region, \VRLowNN\@. Table~\ref{tab:NNValues} shows the definition of the SRs and VRs used, and the corresponding ratio $s/\sqrt{s+b}$.

\begin{table}[htbp!]
\centering
\caption{Summary of the set of input variables used in the different channels. The order does not reflect the importance of each variable.
}
\begin{tabular}{ccccc}
\toprule
& Variable &  \tll & \tlhad & \thadhad \\
\midrule
\multirow{4}{*}{$\tau$-lepton properties}
& $\eta(\tau_1),\eta(\tau_2)$         &                      &  $\bullet$   & \\
& $\pT(\tau_1),\pT(\tau_2)$          &        		     &  $\bullet$   & \\
& $\phi(\tau_1),\phi(\tau_2)$         &         	     &  $\bullet$   & \\
& $E(\tau_1),E(\tau_2)$         &         	     &  $\bullet$   & \\
& $C_{jj}(\tau_1)$                &  $\bullet$        & $\bullet$           & $\bullet$ \\
& $C_{jj}(\tau_2)$                &  $\bullet$        & $\bullet$            & $\bullet$ \\
\midrule
\multirow{6}{*}{Mass}
& $\mmmc$                       &     $\bullet$        &  $\bullet$   & $\bullet$ \\
& $\mttvis$                     &  	$\bullet$    &  $\bullet$   & $\bullet$ \\
& $\mjj$                          &   $\bullet$        &  $\bullet$   & $\bullet$ \\
& $\mT(\tau_1,\MET)$              &  	$\bullet$    &  $\bullet$   & $\bullet$ \\
& $\mT(\tau_2,\MET)$              &   $\bullet$        &  $\bullet$   & $\bullet$ \\
& $m\left(\tau_1, \tau_2, j_1\right)$	&	$\bullet$    &	$\bullet$			& $\bullet$ \\
\midrule
\multirow{6}{*}{Angular distances}
& $\eta^{j_1} \times \eta^{j_2}$	&	$\bullet$		 &	$\bullet$			& $\bullet$ \\
& $\Delta R_{\tau_1\tau_2}$           &    $\bullet$       & $\bullet$    & $\bullet$ \\
& $\Delta\eta_{\tau_1\tau_2}$         &    $\bullet$       & $\bullet$         	& $\bullet$ \\
& $\Delta\phi_{\tau_1\tau_2}$         &    $\bullet$       & $\bullet$          	& $\bullet$ \\
& \detajj                         &    $\bullet$       & $\bullet$          	& $\bullet$ \\
& $\Delta\phi_{H,j_1}$				&    $\bullet$       & $\bullet$            & $\bullet$ \\
\midrule
\multirow{16}{*}{Jet properties}
& $\pT(\tau_1,\tau_2,j_1,j_2)$       &                   &  $\bullet$               &           \\
& $\pt\left(j_1, j_2\right)$      &    $\bullet$	    &  $\bullet$                       & $\bullet$ \\
& $\pT(\tau_1, \tau_2,j_1,j_2,\MET)$   &   $\bullet$       &  $\bullet$ & $\bullet$ \\

& $\pT(j_2)$, $\pT(j_3)$                       &  $\bullet$        &   &   $\bullet$        \\
& $\eta (j_1)$, $\eta (j_2)$                    &           		&  $\bullet$ &           \\
& $\phi (j_1)$, $\phi (j_2)$                     &           		&  $\bullet$ &           \\
& $E (j_1)$, $E (j_2)$                       &           		&  $\bullet$ &           \\

& $\pT(\tau_1)+\pT(\tau_2)$		&	  $\bullet$ 	&  		 & $\bullet$ \\
& $\pT(\tau_1,\tau_2)$				&	$\bullet$		& $\bullet$			 & $\bullet$ \\
& $\pT(\tau_1, \tau_2,\MET)$			&	$\bullet$		& $\bullet$			 & $\bullet$ \\
& $\pT(\tau_1, \tau_2,j_1,\MET)$		&    $\bullet$      &  $\bullet$          & $\bullet$ \\
& $\pT(\tau_1)/\pT(\tau_2)$		&   $\bullet$       &            & $\bullet$ \\
& $\Delta\pT(\tau_1,\tau_2)/(\pT(\tau_1)+\pT(\tau_2))$ &    $\bullet$  &            & $\bullet$ \\
\midrule
\multirow{8}{*}{Missing transverse energy}
& $\MET $              			&       &  $\bullet$ &    \\
& $\MET \phi $              		&           		&  $\bullet$ &    \\
& $\MET/\pT(\tau_1)$              &  $\bullet$        &  $\bullet$ & $\bullet$ \\
& $\MET/\pT(\tau_2)$              &  $\bullet$        &  $\bullet$ & $\bullet$ \\
& $\MET-\phi-$centrality$/\sqrt{2}$      &  $\bullet$        & $\bullet$           & $\bullet$ \\
& $\METhpto$						&   $\bullet$         		&            & $\bullet$ \\

\bottomrule
\end{tabular}
\label{tab:features}
\end{table}

\begin{figure}[!tb]
\centering

\subfloat[]{    \includegraphics[width=.33\textwidth]{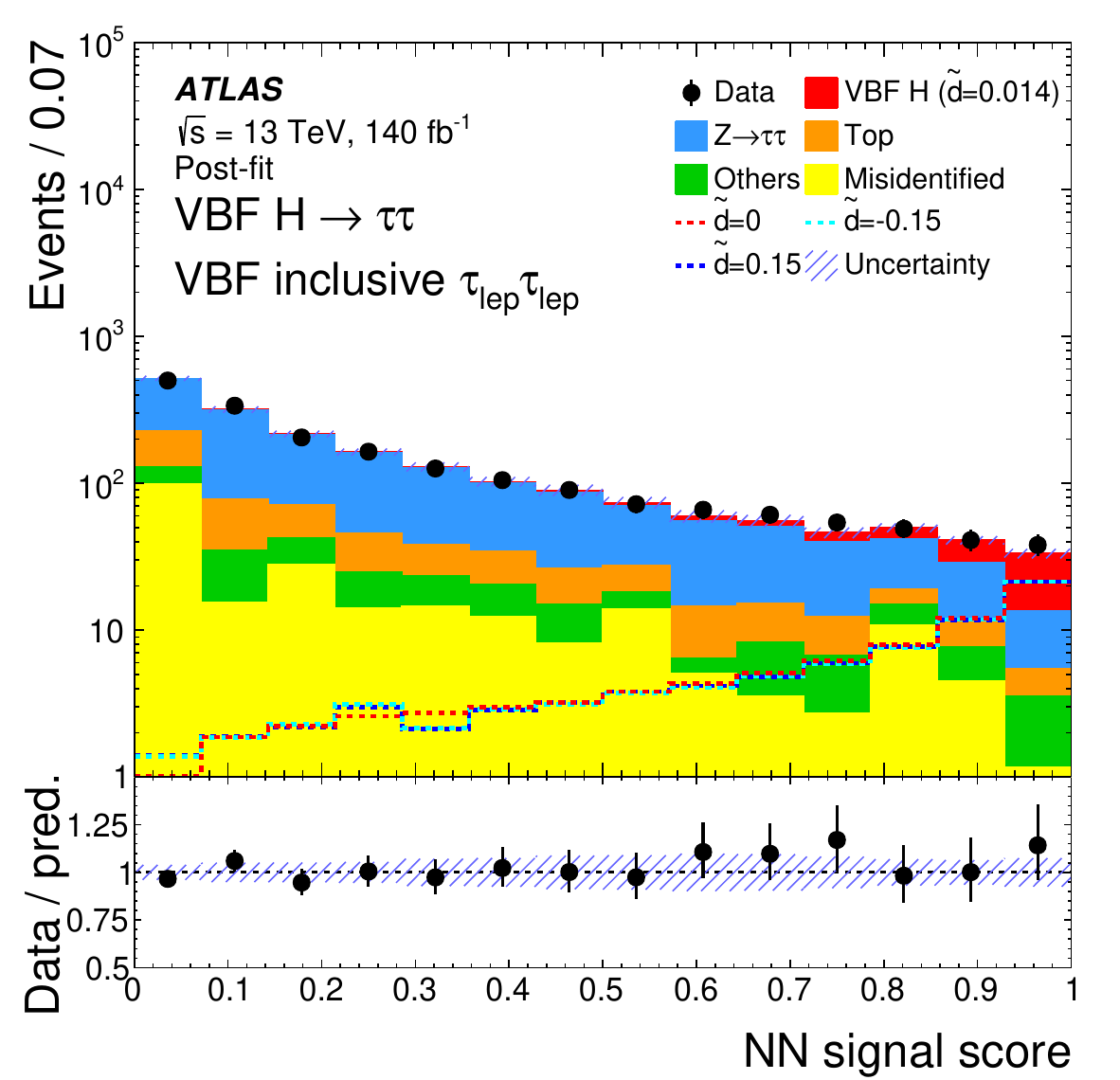}}
\subfloat[]{    \includegraphics[width=.33\textwidth]{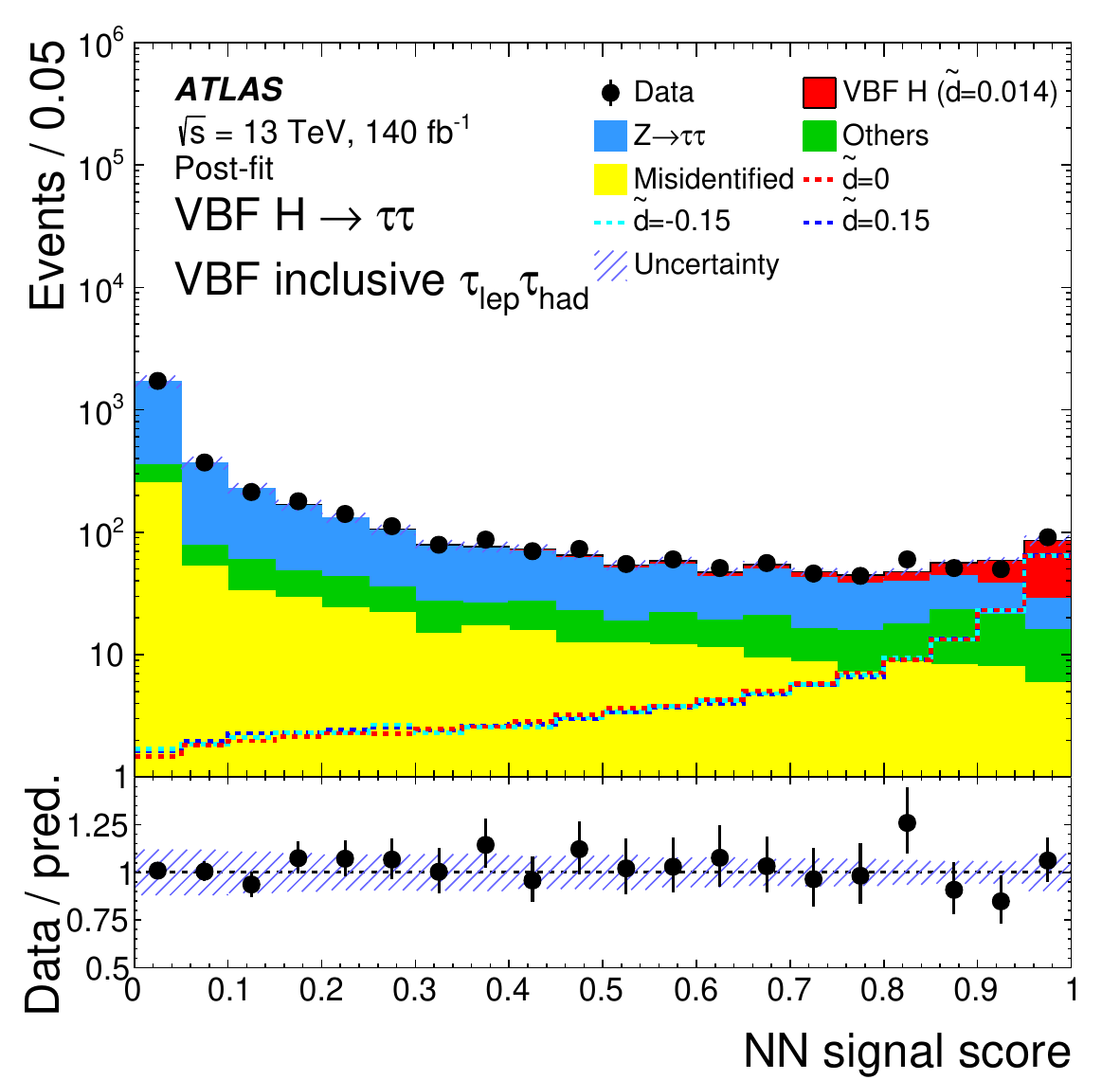}}
\subfloat[]{    \includegraphics[width=.33\textwidth]{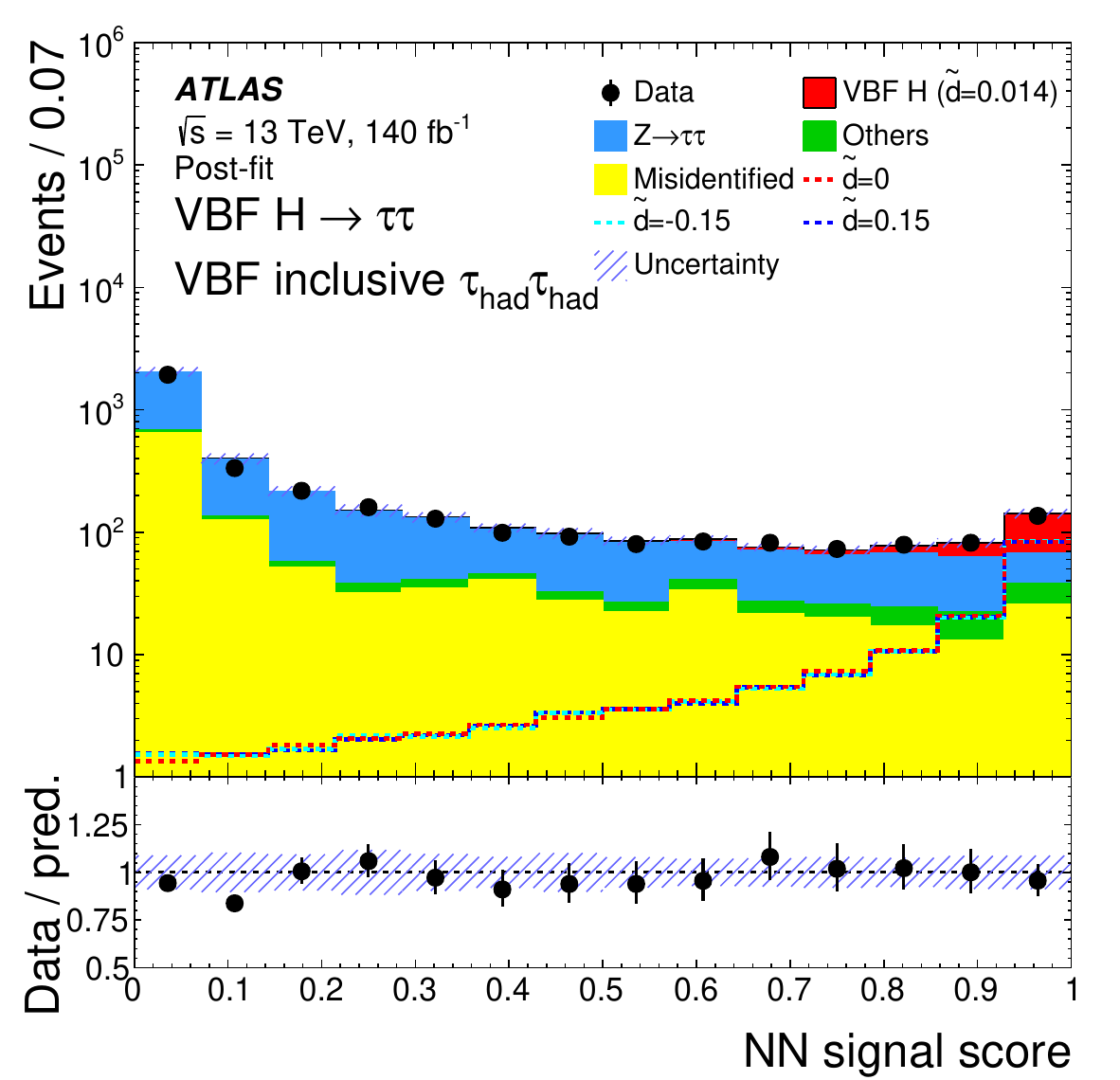}}

\caption{Post-fit distribution of the neural network signal score for the three channels,  (a) \LL, (b) \LH, and (c) \HH.  The ratio of the SM predictions to the observed data is shown in the bottom panel. The ``Other bkg'' includes diboson, $Z\rightarrow\ell\ell$, and $W\rightarrow\tau\nu+$jets, and other non-VBF-Higgs processes. The hatched area represents the impact of the experimental, theoretical and statistical uncertainties in the Standard Model predictions.  The dashed lines represent examples of signal samples, for different values of \dtil.
}
\label{fig:nnscore}
\end{figure}

\begin{table}[b!]
\centering
\caption{Summary of NN signal score values used for defining the SRs and the VR for each channel, and their corresponding ratio $s/\sqrt{s+b}$.
}
\begin{tabular}{ccccc}
\toprule
Channel & Region &  NN signal score & $s/\sqrt{s+b}$ \\

\midrule
\multirow{ 3}{*}{\LL} & \SRHighNN &  $>0.91$ & 3.50\\
& \SRLowNN & 0.67 -- 0.91 &1.98 \\
& \VRLowNN &  $<  0.67$ & 0.51\\
\midrule
\multirow{ 3}{*}{\LH} & \SRHighNN &  $>0.91$ & 6.93\\
&  \SRLowNN & 0.63 -- 0.91 &  2.69 \\
&\VRLowNN &  $<0.63$ & 0.30\\
\midrule
\multirow{ 3}{*}{\HH} 		& \SRHighNN &  $>0.93$ & 6.86\\
& \SRLowNN & 0.60 -- 0.93 &  2.42 \\
& \VRLowNN &  $<0.60$ & 0.34\\
\bottomrule

\end{tabular}

\label{tab:NNValues}
\end{table}


\section{Background estimation}
\label{sec:bkg}

The major backgrounds that contribute to the total yields in the SRs of this analysis are \Zttjets, misidentified $\tau$ candidates (\textit{fakes}), top-quark processes, and multi-boson production. These backgrounds are estimated with different techniques based on MC simulations and data-driven methods.

The main background for each channel comes from \Zttjets. These events have the same signature as the Higgs signal,  and similar kinematics. For these reasons, designing a signal-free control region (CR) where to constrain \Zttjets at high NN value is extremely challenging.  Instead this background is estimated following a procedure called \textit{embedding}~\cite{HIGG-2014-09},  where the \Zttjets yields in the SR predicted by the MC simulations are normalised by a scale factor that is obtained in a dedicated control region,  following closely the procedure in Ref.~\cite{HIGG-2022-07}, called \textit{object-level embedding}.

The embedding method relies on the assumption that, at Born level, \Zll (where $\ell$ is only $e$ or $\mu$) and \Ztt are kinematically identical processes.  This means that by substituting the leptons ($e$ or $\mu$) in  \Zll data events with $\tau$-lepton decays from simulation, and then recomputing all the necessary variables in the event, it is possible to get a template of the \Ztt background directly from data.  These templates are obtained starting from data and MC events in an enriched \Zlljets region, where the momentum of the reconstructed leptons are scaled using a parameterisation that accounts for the differences in efficiency of triggers, the reconstruction procedure and kinematics between the prompt leptons and the $\tau$-lepton decays. After this procedure, all the kinematic variables are recomputed to account for the rescaled lepton momentum, including the \etmiss from the additional neutrinos of the $\tau$-leptons decays. For each SR a dedicated \Zttjets CR is defined with the same selections but using the template samples, \ZttCRHighNN for the high NN score SR and \ZttCRLowNN for the low NN score SR\@. These CRs are then included in the fit to constrain the \Zttjets in the SR.

Dedicated systematic uncertainties are added to account for the modelling differences between the embedded \Zttjets MC and the nominal \Ztt in the SR\@.  The \VRLowNN is then used to check the modelling of the \Zttjets after the normalisation with the embedding procedure. Figure~\ref{fig:VRLowNN} shows the \OO distribution in each VR, after a background-only fit.

\begin{figure}[!tb]
\centering
\subfloat[]{  \includegraphics[width=.33\textwidth]{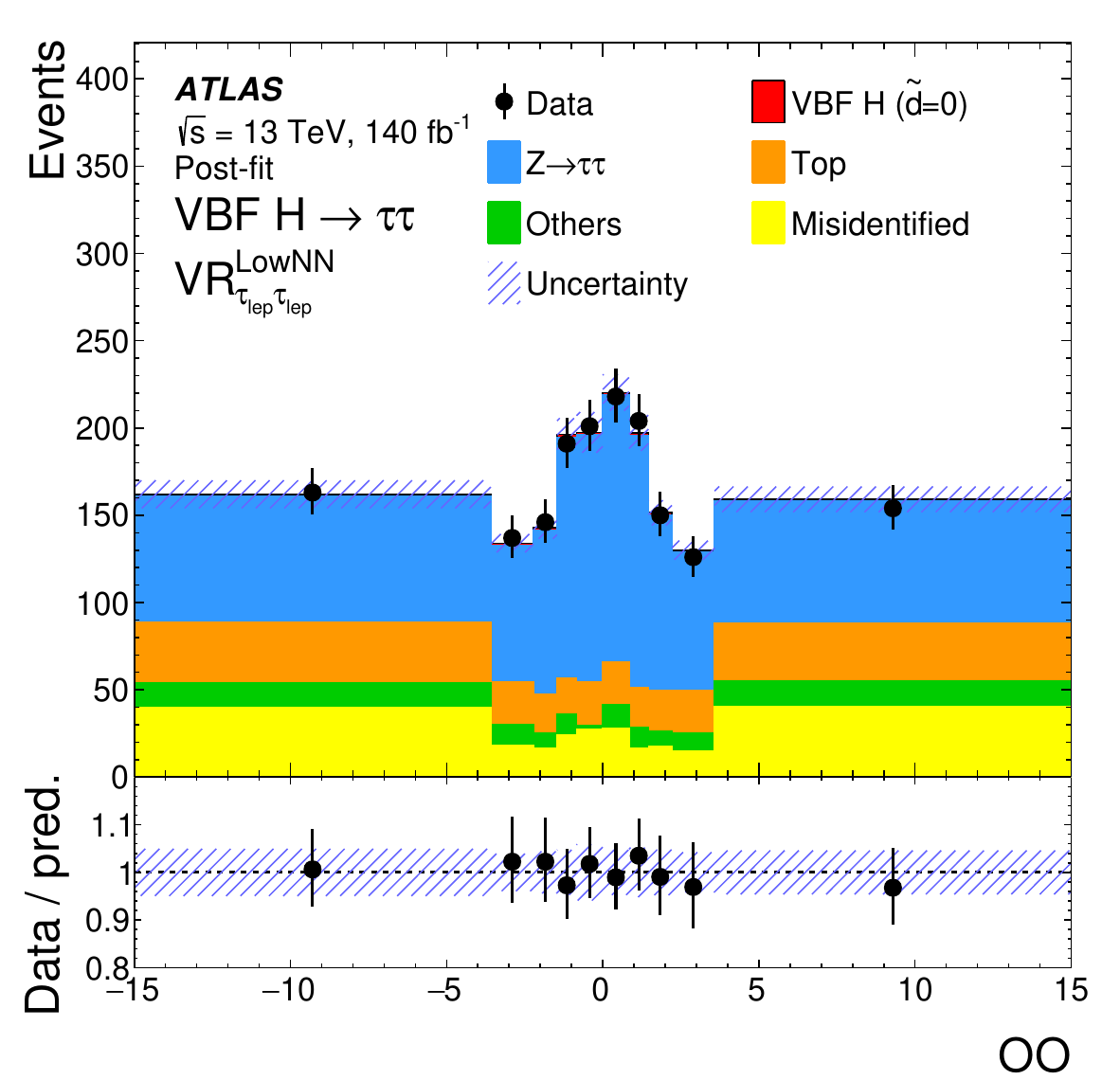}}
\subfloat[]{  \includegraphics[width=.33\textwidth]{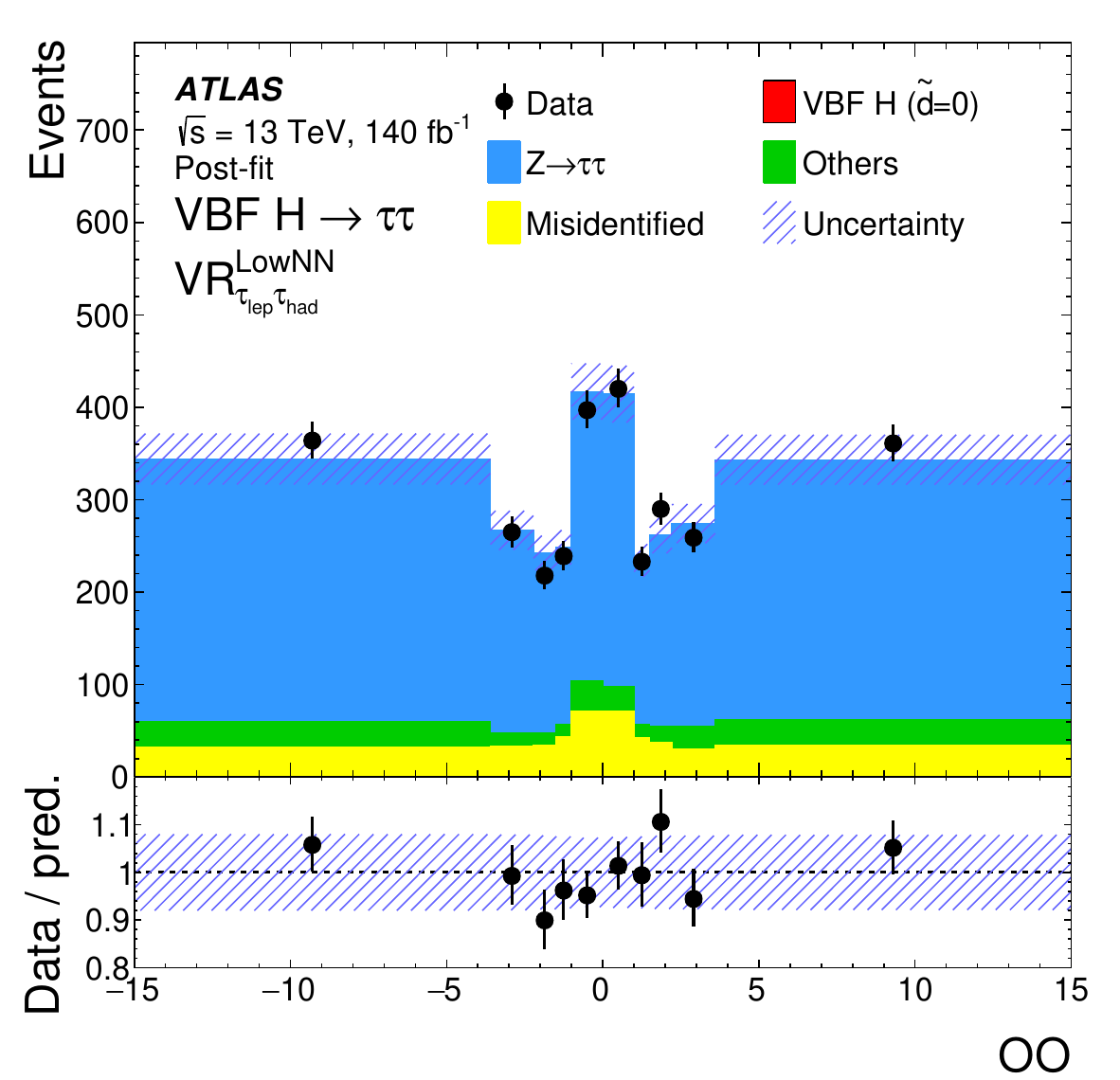}}
\subfloat[]{\includegraphics[width=.33\textwidth]{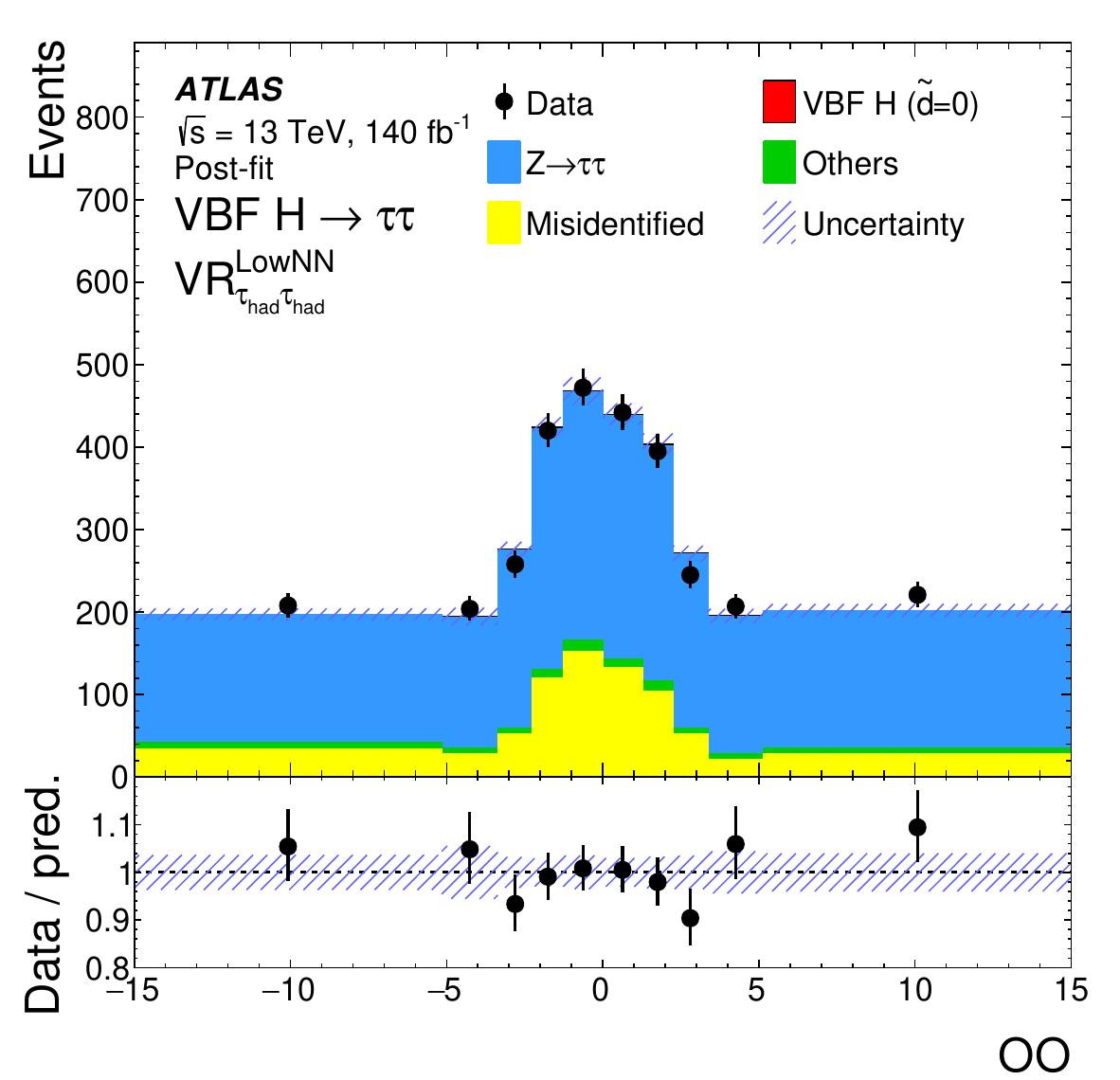}}

\caption{
Distributions of the \OO in the VR for the three channels: (a) \LL, (b) \LH, (c) \HH, after a background-only fit. The value of the parameter of interest, the signal strength, the normalisation of backgrounds, and the systematic uncertainties are set to the value from the best fit result. In the bottom panel, the hatched area represent the total post-fit systematic uncertainty in the final SM prediction.}
\label{fig:VRLowNN}
\end{figure}

The second largest background component comes from events where one or two $\tau$-leptons are incorrectly categorised. In these events, jets are misidentified as electrons,  muons, or as \tauh. These backgrounds account for 4\% to 30\% of the total SM background, depending on the channel and region.  Background events with misidentified $\tau$ candidates are estimated by using a data driven method, with different methodologies depending on the channel.

In the \LL channel,  the matrix method~\cite{EGAM-2019-01} is used to estimate the background from misidentified $e$ or $\mu$.  In this channel, events are selected from two control regions (\CRRealll and \CRFakell) similar to the SR, but without the requirements on the isolation and loosened identification criteria of the leptons (Loose ID WP for $e$ and Medium ID WP for $\mu$).  The number of misidentified leptons in the SR is then related to the number of leptons passing the loose and tight isolation and identification criteria, based on the efficiencies for real and fake leptons.  The real efficiencies are obtained from MC simulations, while the fake-lepton efficiencies are measured from data in a region with two leptons with the same sign.  The method is validated in a region with the same selection criteria as the SR but with same-sign leptons. Any difference in this region is added as a systematic uncertainty. Additional uncertainties are assigned to this background taking into account the statistical uncertainties in the fake and real efficiencies,  the uncertainty in the background subtraction when computing the efficiencies,  and the different flavour composition between the SR and the loosened \CRFakell.

In the \LH and \HH channels, misidentified $\tau$ candidates are estimated by using the fake-factor method~\cite{EGAM-2019-01}.  Events are selected from a control region (\antiTauCR) with the same selection as the SR, but with a loosened requirement on the \tauh identification criterion, and failing to meet the nominal identification requirement from Table~\ref{tab:eventselection}.  Contributions from real \tauh in this region are estimated with MC simulations and subtracted. The final number of misidentified $\tau$ candidates in the SR is then obtained by multiplying the background-subtracted events in the \antiTauCR by a \textit{fake-factor}, $\mathcal{F}$.  This factor is the ratio of the number of $\tau$ candidates passing the identification criterion in Table~\ref{tab:eventselection} to the number of those failing the same criterion.  These factors are obtained in separate dedicated CRs.  Misidentified $\tau$ candidates originate mainly from $W+$jets and dijet events, and the ratio of passing to failing identification criteria can be different for the two processes, in different kinematic regions. For this reason, one fake-factor is computed for each process, in a different CR\@. The final fake-factor is then a weighted mean of the two based on the relative composition in the SR\@. Additionally, the fake-factors are estimated separately for 1-prong  and 3-prong $\tau$ candidates.  The final estimate is validated in a region with the same selection as the SR, but with same-sign $\tau$ candidates. The non-closure is taken as an uncertainty as a function of the Optimal Observable.  Additional uncertainties are added taking into account the statistical uncertainty in the computation of $\mathcal{F}$ and the relative composition in the SR,  and the uncertainty in the background subtraction in the \antiTauCR and in the $\mathcal{F}$ computation.

Other sources of backgrounds from misidentified objects are estimated directly from simulations.

In the \LL channel, processes including a top quark have a sizeable contribution. This background is estimated directly from the MC simulations, but it is normalised to the data in a dedicated CR (\TopCR) obtained with the same selection as the SR but inverting the veto on the number of $b-$jet in the events. In the \LH and \HH channels the contributions from top-quark processes are estimated directly from MC.

In all three channels, other smaller prompt contributions are estimated directly from MC.


\section{Systematic uncertainties}
\label{sec:systematics}

Systematic uncertainties affecting both the shape of the Optimal Observable and the event yields in the signal and control regions are considered. These can be grouped in three main categories: experimental uncertainties, uncertainties in data-driven modelling, and theoretical uncertainties.

Experimental systematic uncertainties include those related to the trigger, reconstruction, and identification of objects in the final state. These sources include electron and muon energy or momentum scale and resolution~\cite{EGAM-2021-01,MUON-2018-03},  $\tauhadvis$ energy scale and resolution~\cite{ATL-PHYS-PUB-2019-033}, $b$-tagging efficiency~\cite{FTAG-2018-01,FTAG-2019-02}, $\etmiss$ reconstruction~\cite{JETM-2020-03-fix}, jet energy scale and resolution~\cite{JETM-2018-05}, pile-up modelling~\cite{PERF-2014-03}, and the luminosity measurement~\cite{DAPR-2021-01}.
Jet and $\etmiss$ reconstruction  and theory uncertainties constitute the dominant sources of systematic uncertainty, at $20\%$ and 15\%, respectively.

Uncertainties in the data-driven background modelling affect the three channels differently.
For \tll, the uncertainties in the matrix method estimates are related to the limited statistical precision in the computation of the real and fake-lepton efficiencies, the MC background subtraction, and the flavour composition difference between the signal region and the fake control region.
For \tlhad and \thadhad, statistical uncertainties in the individual fake factors, and uncertainties due to differences in composition between the SR and the regions where the fake factors are computed, are considered.
Uncertainties affecting the \Ztt embedded MC sample are evaluated by propagating all experimental uncertainties listed above. In addition the effects of trigger efficiencies, muon and $\tauhadvis$ reconstruction and isolation efficiencies, and $e/\mu$ energy scale and resolution are considered.

Theory uncertainties in the signal and background cross-sections are evaluated by varying the PDF sets and the renormalisation and factorisation scales, as well as the value of the strong coupling constant $\alpha_s$, used in MC generation.

For $\Zttjets$ production, 100 PDF variations are considered, taking the standard deviation as the uncertainty.
The PDF sets are evaluated for up- and down-variations of $\alpha_s$, and a combined uncertainty is derived by adding the uncertainties in PDF and $\alpha_s$ variations in quadrature.
Uncertainties in renormalisation and factorisation scales are determined based on the maximum absolute difference when independently varying their values up or down by a factor of two, resulting in a seven-point variation.
In addition, uncertainties in the jet-to-parton matching scheme (CKKW) and resummation scale are evaluated using parameterisations as function of the jet multiplicity and $Z$ transverse momentum.

For top processes, uncertainties related to the choice of matrix element and parton shower generators are assessed by comparing to alternative generators, while uncertainties in initial- and final-state radiation are evaluated by varying the scale of emission, as described in Ref.~\cite{ATL-PHYS-PUB-2017-007}.
The PDF uncertainty is evaluated by varying the PDF sets for both the production and the parton shower, and the two are treated as separate uncertainties.

Uncertainties in signal processes are similarly evaluated by PDF and $\alpha_s$ variations, and uncertainties in the matrix element and parton shower are assessed through alternative generators. The effect of choosing a different matrix element generator is assessed by comparing the nominal signal sample with an alternative sample using \MGNLO~\cite{Alwall:2014hca},  but keeping the same parton shower model. Instead, the parton shower uncertainty is obtained through the comparison with an alternative sample using \HERWIG[7]~\cite{Bellm:2017jjp} instead of  \PYTHIA[8].

For VBF, VH and \ttH production, QCD scale uncertainties from missing higher-order corrections are evaluated through the seven-point variation of renormalisation and factorisation scales.
For ggF Higgs production, additional uncertainties related to the ggF kinematics are considered, following the approach described in Ref.~\cite{ATL-PHYS-PUB-2023-031}. These include effects from the Higgs \pt shape, VBF topology selection, and top-quark-mass dependence.

Table~\ref{tab:SystGroupedComb} shows the impact as a percentage of the main systematic uncertainties in the final \dtil 68\% CL interval obtained after the fit described in Section~\ref{sec:fitmodel}. The impact of a given systematic uncertainty in the fitted parameter \dtil is estimated from the covariance matrix as described in Ref.~\cite{Pinto:2023yob}. The main limitation on the sensitivity comes from the statistical uncertainties in the actual observed data, while all the other sources only account for a small fraction.

\begin{table}[!tb]
\centering
\caption{Summary of the different sources of systematic uncertainty and their impact on the final fit results,  expressed as a percentage of the fitted \dtil 68\% CL interval. Experimental systematic uncertainties in the reconstruction of the objects include the reconstruction efficiency, energy scale and resolution, and trigger efficiencies. "Sample size" includes the per-bin statistical uncertainties in the simulated MC samples, and the data-driven backgrounds of misidentified $\tau$-leptons. }
\begin{large}
\begin{tabular}{lc }
\toprule
Systematic source&  Uncertainty $[\%]$\\
\midrule

Jet/\MET reconstruction & $\pm$ 20 \\
Signal theory & $\pm$ 15  \\
Background theory & $\pm$ 11\\
Normalisation factors & $^{+6.0}_{-5.5}$\\
Misidentified $\tau$-leptons & $\pm$ 4.8  \\
$\tau$-leptons reconstruction & $\pm$ 4.0 \\
Sample size & $\pm$ 3.0\\
Leptons reconstruction & $\pm$ 2.4\\
Luminosity & $\pm$ 0.4 \\
Flavour tagging & $\pm$ 0.3\\
Embedding & $\pm$ 0.2\\
\midrule
Total systematic uncertainty			& 				$\pm$ 30\\
Total statistical uncertainty 				& $\pm$ 95\\
\bottomrule
\end{tabular}
\end{large}
\label{tab:SystGroupedComb}
\end{table}

%


\section{Fitting procedure}
\label{sec:fitmodel}

The statistical analysis is performed using a binned maximum-likelihood fit, where the likelihood function is defined as the product of Poisson distributions for each bin of each fit region (SRs and CRs), and Gaussian
terms for the systematic uncertainties.  The probability distributions $\mathcal{L}(\boldsymbol{x};\mu,\boldsymbol{\theta}) $ depends on the actual data $\boldsymbol{x}$,  the parameter of interest $\mu$, and further nuisance parameters $\boldsymbol{\theta}$, corresponding to the systematic uncertainties. The sensitivity to the CP-odd coupling (either in the HISZ or the Warsaw basis) is estimated by doing a negative log-likelihood scan (NLL). The NLL is computed for different hypotheses of the parameter of interest, using signal template samples  constructed by reweighting the SM VBF Higgs signal as described in Section~\ref{sec:samples}.  The best-fit value is the minimum of the NLL curve with respect to the SM NLL ($\Delta$NLL),  with a confidence interval of $[\mu-\sigma_{\mu},\mu+\sigma_{\mu}]$. The 68\% and 95\% CL intervals are obtained by reading where the $\Delta$NLL curves intersect the points at 0.5 and 1.92, respectively.

Each of the SRs defined in Table~\ref{tab:NNValues}, \SRHighNN and \SRLowNN, are binned in \OO. The binning of the signal regions in the three channels is optimised to have an equal number of expected SM signal events in each bin.  All three channels have a total of 10 bins. The bin boundaries are then fixed to be symmetric around zero.

For each SR, one CR (\ZttCRHighNN and \ZttCRLowNN) is defined to constrain the \Ztt background.  These regions have the same kinematic cuts as the nominal corresponding SR, including the NN score, but using embedded results, as described in Section~\ref{sec:bkg}. These CRs each consist of only one bin. A single common normalisation factor is applied to scale the \Zll MC in the \Ztt CR as well as the \Ztt MC in the corresponding SRs. One common normalisation factor is used for each channel.
Table~\ref{tab:EmbNF} shows the embedding normalisation factors (NF) obtained for the three channels.
In the \LL channel, the \TopCR is used to constrain the top backgrounds (\ttbar and single top) with one nuisance parameter, with a value of $NF^{\textrm{top}} =0.95^{+0.08}_{-0.07}$.  In total six SRs and seven CRs are used in the fit.

\begin{table}[b!]
\centering
\caption{Normalisation factors for the $Z\rightarrow\tau\tau$  MC background obtained with the embedding procedure.}
\begin{tabular}{ccc}
\toprule
Channel &  $Z\rightarrow\tau\tau$  NF\\
\midrule
\LL  & $1.03^{+0.10}_{-0.09}$ \\[5pt]
\LH  &  $0.95^{+0.18}_{-0.15}$ \\[5pt]
\HH  &  $0.93^{+0.07}_{-0.07}$ \\

\bottomrule
\end{tabular}
\label{tab:EmbNF}
\end{table}

The normalisation of the signal sample ( $NF^{\textrm{VBF} H} $) is added as a common free floating normalisation factor. In this way,  the  analysis is not sensitive to overall changes in the total cross-section that might be due to additional CP-even terms, but only considers differences in shapes of \OO, and is therefore only sensitive to the interference term of the EFT model.  All the other Higgs production modes are assumed as background and their normalisation is set to the SM prediction. The observed value for the signal normalisation factor is found to be $NF^{\textrm{VBF} H} =0.87^{+0.14}_{-0.13}$, which is consistent with the SM prediction.

Two fit setups are considered, one where the three channels (\thadhad,  \tlhad, \tll) are separately fitted, and one where they are fitted at the same time.  In both fits, the experimental and theory systematic uncertainties are considered 100\% correlated between channels, while the systematic uncertainties in the modelling of misidentified $\tau$ candidates are uncorrelated.  All results are shown using the combined fit to the three channels, while the separate channels fit is used to provide the single channel $\Delta$NLL curves.

%
%


\section{Results}
\label{sec:results}

The combined fit procedure described in Section~\ref{sec:fitmodel} is performed, and the post-fit Optimal Observable distributions in the SRs are shown in Figure~\ref{fig:SRHigh}.  The value of the parameter of interest, systematic uncertainties, and background and signal normalisations are adjusted within their allowed constraints to minimize the $\Delta$NLL. Tables~\ref{tab:SRYields} and~\ref{tab:CRYields} show the signal and background predicted yields and the observed data events in the SRs and CRs.  Overall, good agreement is observed between SM predictions and data within the systematic uncertainties.  In the \SRHighNN, the \HH channels shows a small excess in the highest bin of the \OO, and an overprediction for negative \OO, while in the \LH channel, the lowest bins show a slight underprediction.  In the \SRLowNN the agreement is also good within systematic uncertainties, with just one bin in the \LH channel where the prediction is lower than the observed data, a trend in OO opposite to that seen in the  \SRHighNN. In the \LL channel,  the SM predictions are in good agreement with the observations in both SRs.

\begin{figure}[!tb]
\centering
\subfloat[]{    \includegraphics[width=.3\textwidth]{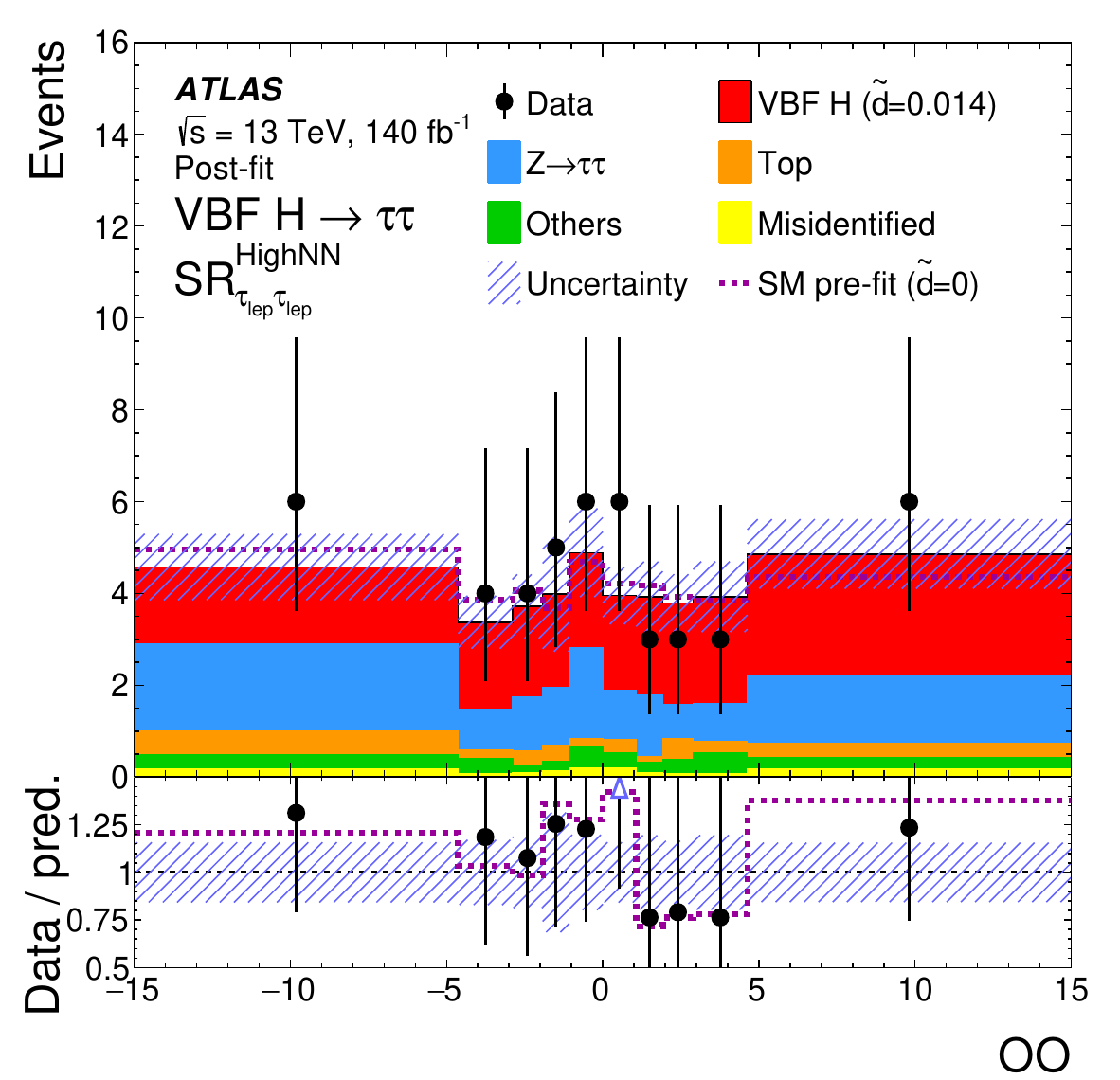}}
\subfloat[]{    \includegraphics[width=.3\textwidth]{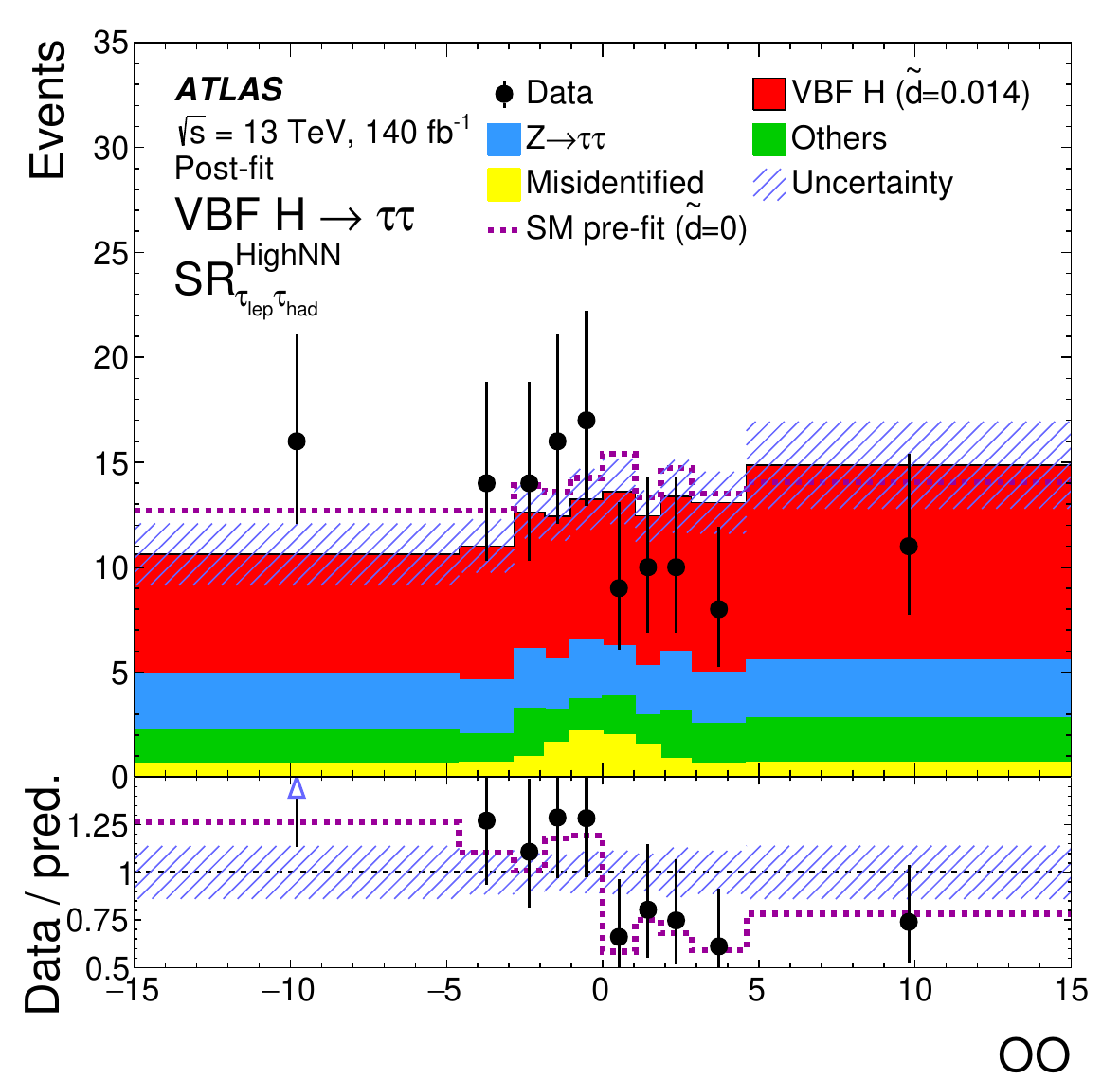}}
\subfloat[]{    \includegraphics[width=.3\textwidth]{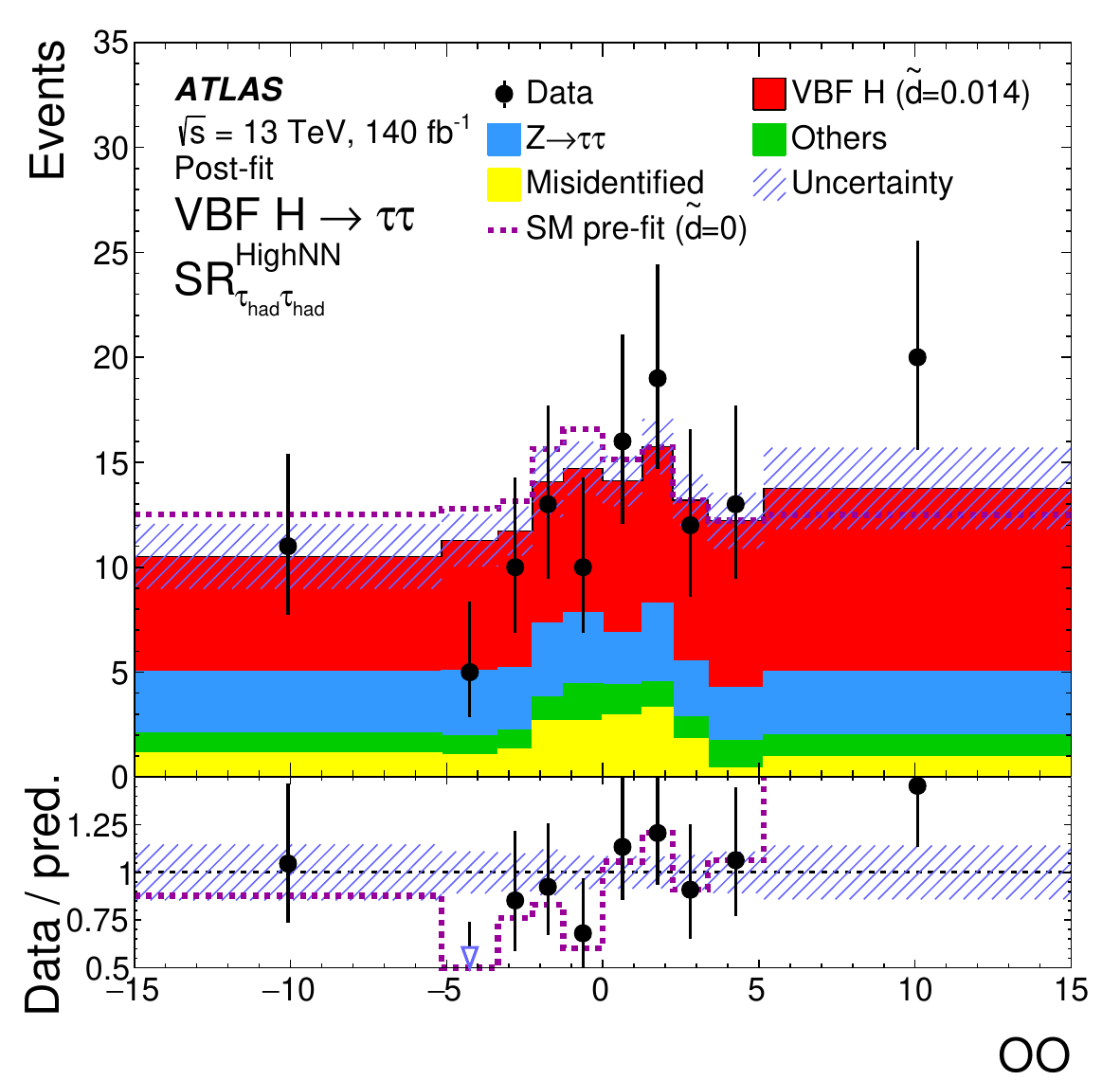}} \\
\subfloat[]{    \includegraphics[width=.3\textwidth]{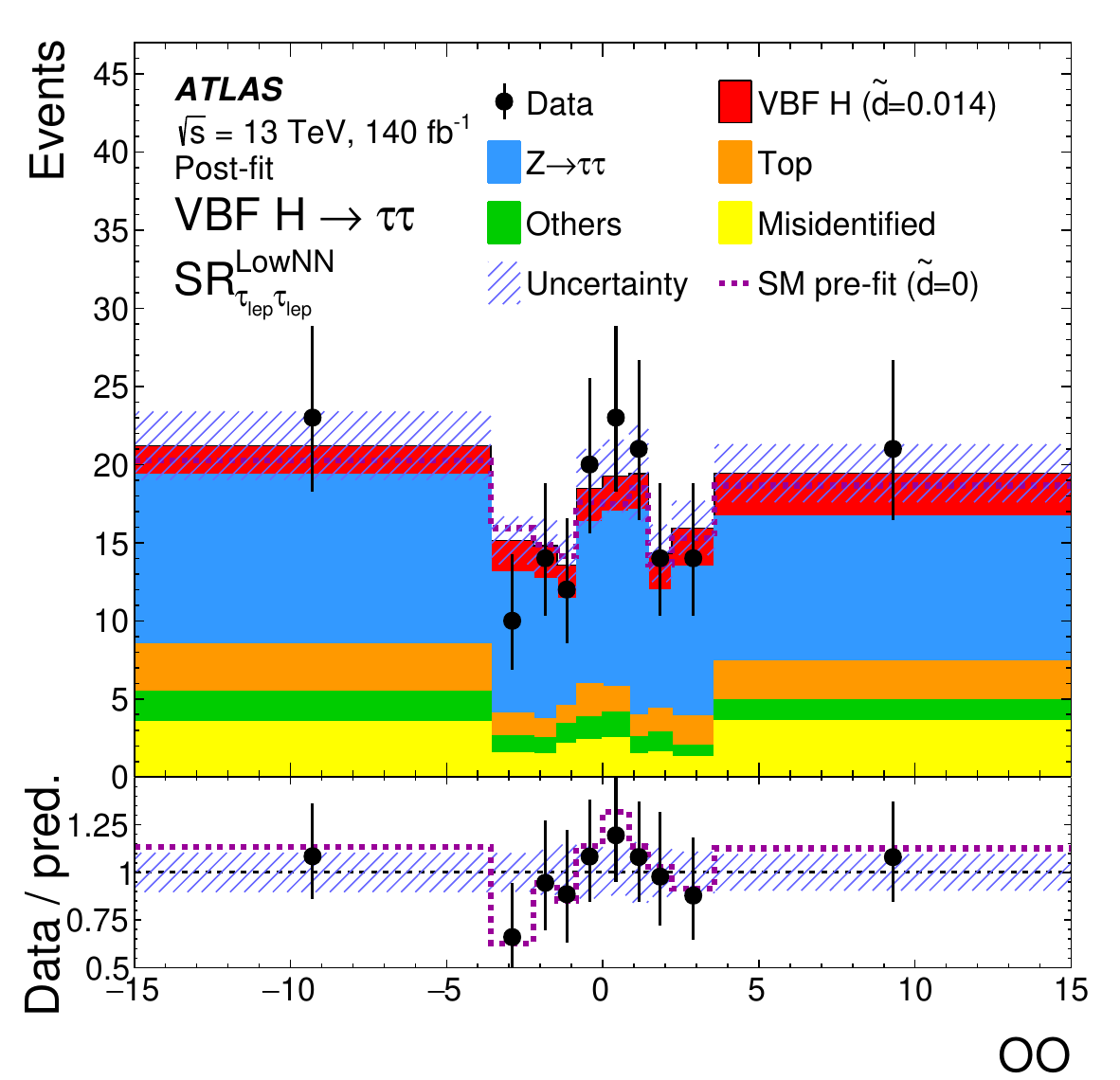}}
\subfloat[]{    \includegraphics[width=.3\textwidth]{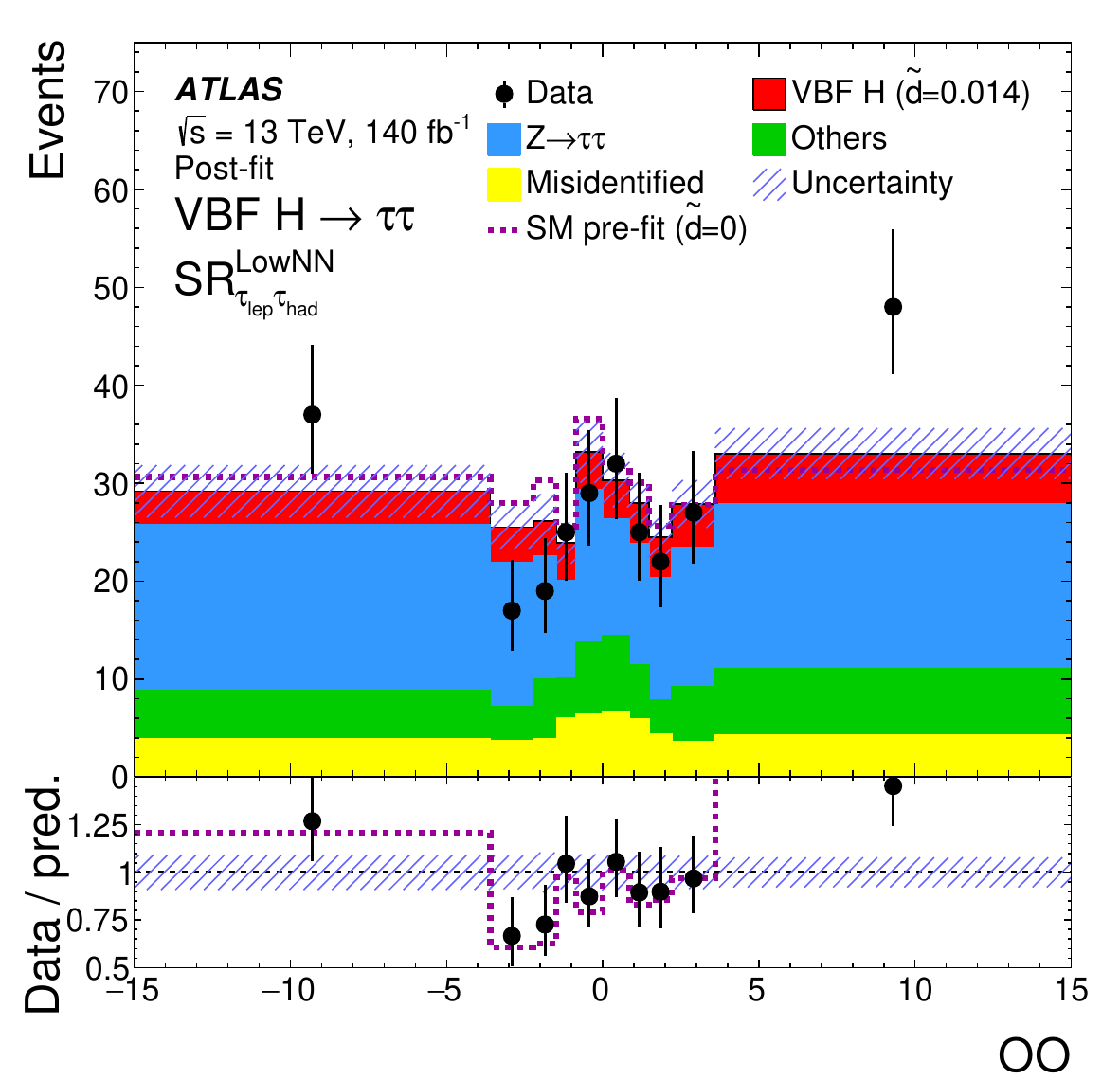}}
\subfloat[]{    \includegraphics[width=.3\textwidth]{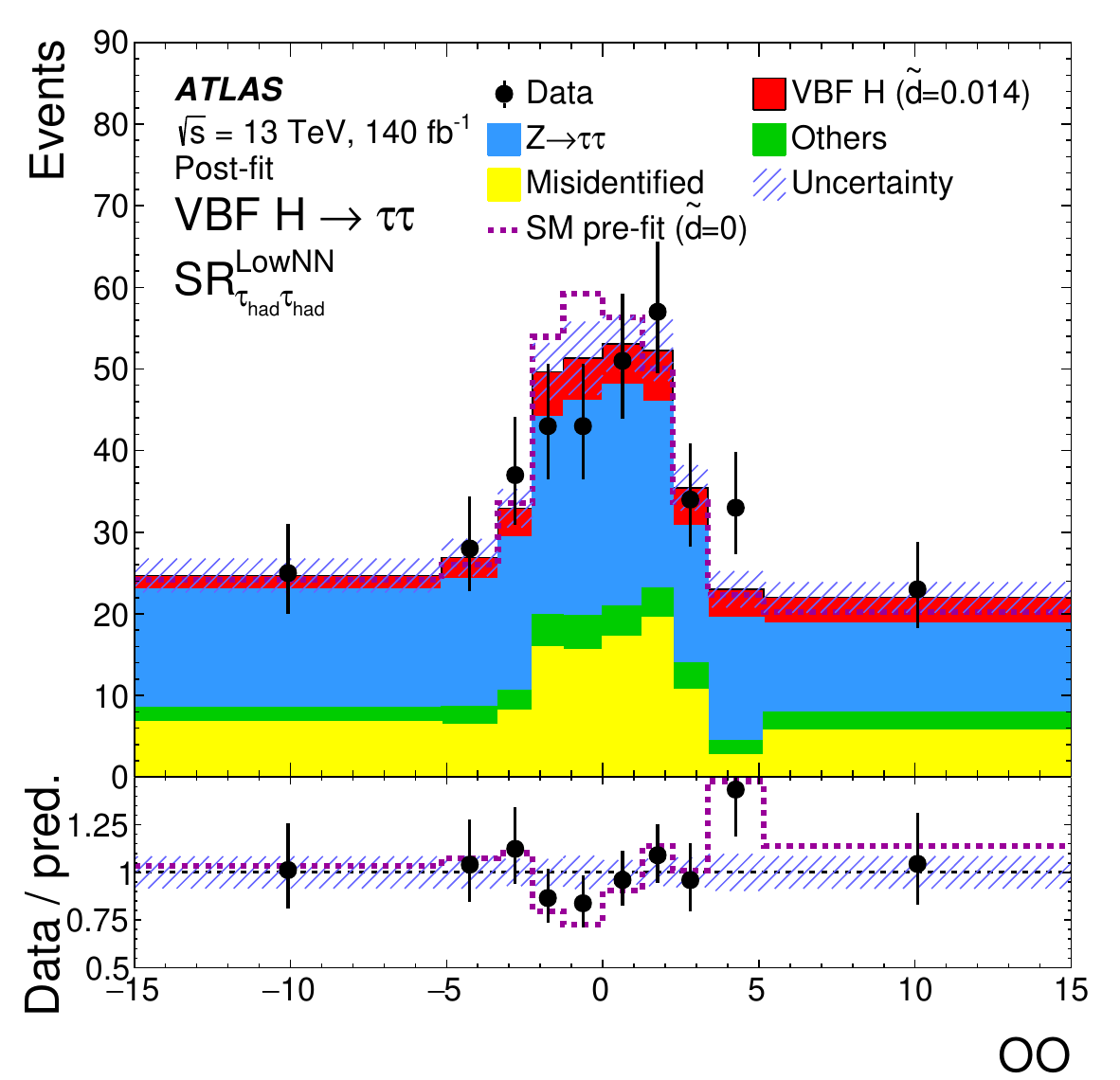}}

\caption{
Post-fit distributions of the \OO in the (a,b,c) \SRHighNN , (d,e,f) \SRLowNN  and  for the three channels: (a,d) \LL, (b,e) \LH, (c,f) \HH. The value of the parameter of interest, the signal strength, the normalisation of backgrounds and the systematic uncertainties are set to the value from the best fit result.  The dashed line in the bottom panel shows the total pre-fit SM predictions (with $\dtil=0$).  The dashed area represent the total post-fit systematic uncertainty in the final SM prediction.}
\label{fig:SRHigh}
\end{figure}

\begin{table}[!t]
\centering
\begin{small}
\caption{Post-fit event yields in the SRs and the Top CR for the three channels in the analysis. The ``Other bkg'' includes diboson, $Z\rightarrow\ell\ell$, and $W\rightarrow\tau\nu+$jets, and other non-VBF-Higgs processes.   The uncertainties in the MC predictions include theory, experimental and statistical uncertainties.}
\label{tab:SRYields}
\begin{tabular}{lr @{${}\pm{}$} lr @{${}\pm{}$} lr @{${}\pm{}$} lr @{${}\pm{}$} lr @{${}\pm{}$} lr @{${}\pm{}$} lr @{${}\pm{}$} l}
\toprule
& \multicolumn{5}{c}{\LL}
& \multicolumn{4}{c}{\LH}
& \multicolumn{4}{c}{\HH}
\\[3pt]
\cmidrule(lr){2-7}\cmidrule(lr){8-11}\cmidrule(lr){12-15}
\multirow{2}{*}{ }
& \multicolumn{2}{c}{SR}
& \multicolumn{2}{c}{SR}
& \multicolumn{2}{c}{\multirow{2}{*}{CR Top}}
& \multicolumn{2}{c}{SR}
& \multicolumn{2}{c}{SR}
& \multicolumn{2}{c}{SR}
& \multicolumn{2}{c}{SR}
\\
& \multicolumn{2}{c}{High NN}
& \multicolumn{2}{c}{Low NN}
& \multicolumn{2}{c}{}
& \multicolumn{2}{c}{High NN}
& \multicolumn{2}{c}{Low NN}
& \multicolumn{2}{c}{High NN}
& \multicolumn{2}{c}{Low NN}
\\
\midrule
$Z\rightarrow\tau\tau$  &  12.7 &  3.0 &   97 &  11 &   246 &  29 &   26 &   5 & 138 & 18 & 30 &  5 & 192 & 24 \\
Other bkg       		&   3.1 &  0.7 & 12.8 & 2.1 &  28.5 & 1.4 &   18 &   5 &  55 & 12 & 12 &  4 &  29 &  7 \\
Misid.\ $\tau$-leptons  &  1.27 & 0.35 &   22 &   6 & 155.7 & 1.6 & 11.9 & 2.5 &  49 & 10 & 18 &  5 & 109 & 18 \\
Top                     &   2.9 &  1.1 &   18 &   4 &  2130 &  60 &    0 &   0 &   0 &  0 &  0 &  0 &   0 &  0 \\
VBF $H$                    &  21.1 &  3.1 & 22.3 & 3.5 &   7.8 & 0.5 &   71 &  10 &  40 &  7 & 71 & 10 &  41 &  7 \\
\midrule
Total                   & 41 & 4 & 172 & 10 & 2570 & 50 & 127 & 9 & 282 & 15 &  131 & 9 & 371 & 18\\
\midrule
Data
& \multicolumn{2}{c}{46}
& \multicolumn{2}{c}{172}
& \multicolumn{2}{c}{2571}
& \multicolumn{2}{c}{125}
& \multicolumn{2}{c}{281}
& \multicolumn{2}{c}{129}
& \multicolumn{2}{c}{374}
\\
\bottomrule
\end{tabular}
\end{small}
\end{table}

\begin{table}[!b]
\centering
\caption{Post-fit event yields in the embedding CRs for the three channels in the analysis. The uncertainties in the MC predictions include theory, experimental and statistical uncertainties.}
\label{tab:CRYields}
\begin{footnotesize}
\begin{tabular}{lr @{${}\pm{}$} lr @{${}\pm{}$} lr @{${}\pm{}$} lr @{${}\pm{}$} lr @{${}\pm{}$} lr @{${}\pm{}$} l}
\toprule
& \multicolumn{4}{c}{\LL}
& \multicolumn{4}{c}{\LH}
& \multicolumn{4}{c}{\HH}
\\[3pt]
\cmidrule(lr){2-5}\cmidrule(lr){6-9}\cmidrule(lr){10-13}
& \multicolumn{2}{c}{\ZttCRHighNNLL}
& \multicolumn{2}{c}{\ZttCRLowNNLL}
& \multicolumn{2}{c}{\ZttCRHighNNLH}
& \multicolumn{2}{c}{\ZttCRLowNNLH}
& \multicolumn{2}{c}{\ZttCRHighNNHH}
& \multicolumn{2}{c}{\ZttCRLowNNHH}
\\
\midrule
$Z\rightarrow \ell\ell$ (emb)	& 15.4 &  2.4 &   86 &    8  &   44 &    6 &  158 &   12 & 34.6 &  3.2 &  193 &   13\\
Emb.\ bkg						& 0.16 & 0.01 & 1.13 & 0.03  & 1.04 & 0.06 & 2.67 & 0.09 & 1.00 & 0.05 & 5.07 & 0.13\\
\midrule
Total    						& 15.6 &  2.4 &   87 &    8  &   45 &    6 &  161 &   12 & 35.6 &  3.2 &  198 &   13\\
\midrule
Data
& \multicolumn{2}{c}{16}
& \multicolumn{2}{c}{84}
& \multicolumn{2}{c}{44}
& \multicolumn{2}{c}{162}
& \multicolumn{2}{c}{40}
& \multicolumn{2}{c}{192}
\\
\bottomrule
\end{tabular}
\end{footnotesize}

\end{table}

The best estimator of the CP-violating strength parameters in the different EFT bases are extracted from the fit procedure, together with the 68\% and 95\% CL intervals.
Table~\ref{tab:com_fit_results} shows the summary of the best fit results of the strength parameters, and their corresponding 68\% and 95\% CL intervals for \dtil and \chwtil. The other Warsaw basis parameters are not considered, since their sensitivity is between 10 and 20 times worse than \chwtil. Figure~\ref{fig:NLL} shows a comparison of the expected and observed $\Delta$NLL curves for each interpretation.  The observed $\Delta$NLL curves for the single-channel fits are also shown with dashed lines.  The agreement between the combined fit and the single-channel fits is found to be at 2.8$\sigma$, due to the statistical fluctuations in the first and last bins as seen in Figure~\ref{fig:SRHigh}.

\begin{table}[!t]
\centering
\caption{Expected and observed 68\% and 95\% CL intervals on the CP-violating parameters obtained from the combined fit of the Optimal Observable. For the \chwtil results  a new-physics scale at $\Lambda=1~\TeV$ is assumed.  }
\begin{footnotesize}
\label{tab:com_fit_results}
\begin{tabular}{l c r @{{}, {}} lr @{{}, {}} lr @{{}, {}} lr @{{}, {}} lr @{{}, {}} l }
\toprule
\multirow{2}{*}{Parameter}
& Observed
& \multicolumn{2}{c}{\multirow{2}{*}{68\% CL (Exp.)}}
& \multicolumn{2}{c}{\multirow{2}{*}{95\% CL (Exp.)}}
& \multicolumn{2}{c}{\multirow{2}{*}{68\% CL (Obs.)}}
& \multicolumn{2}{c}{\multirow{2}{*}{95\% CL (Obs.)}} \\
& best value  & \multicolumn{2}{c}{} & \multicolumn{2}{c}{} & \multicolumn{2}{c}{} & \multicolumn{2}{c}{} \\
\midrule
$\dtil$ (lin.+quad.)	& $0.014$ & $[-0.011$ & $0.011]$  & $[-0.022$ & $0.022]$  & $[0.001$ & $0.028]$ & $[-0.012$ & $0.044]$ \\
$\dtil$ (lin.\ only)     & $0.011$ & $[-0.011$ & $0.011]$  & $[-0.021$ & $0.022]$  & $[0.000$ & $0.023]$  & $[-0.012$ & $0.034]$ \\
\midrule
$\chwtil$ (lin.+quad.)	& $0.26$  & $[-0.22$ & $0.22]$  & $[-0.43$ & $0.44]$  & $[0.01$ & $0.53]$  & $[-0.24$ & $0.83]$ \\
$\chwtil$ (lin.\ only)   & $0.21$  & $[-0.21$ & $0.22]$  & $[-0.41$ & $0.44]$  & $[-0.02$ & $0.45]$  & $[-0.23$ & $0.70]$ \\
\bottomrule
\end{tabular}
\end{footnotesize}

\end{table}

\begin{figure}[!b]
\centering

\subfloat[]{  \includegraphics[width=.45\textwidth]{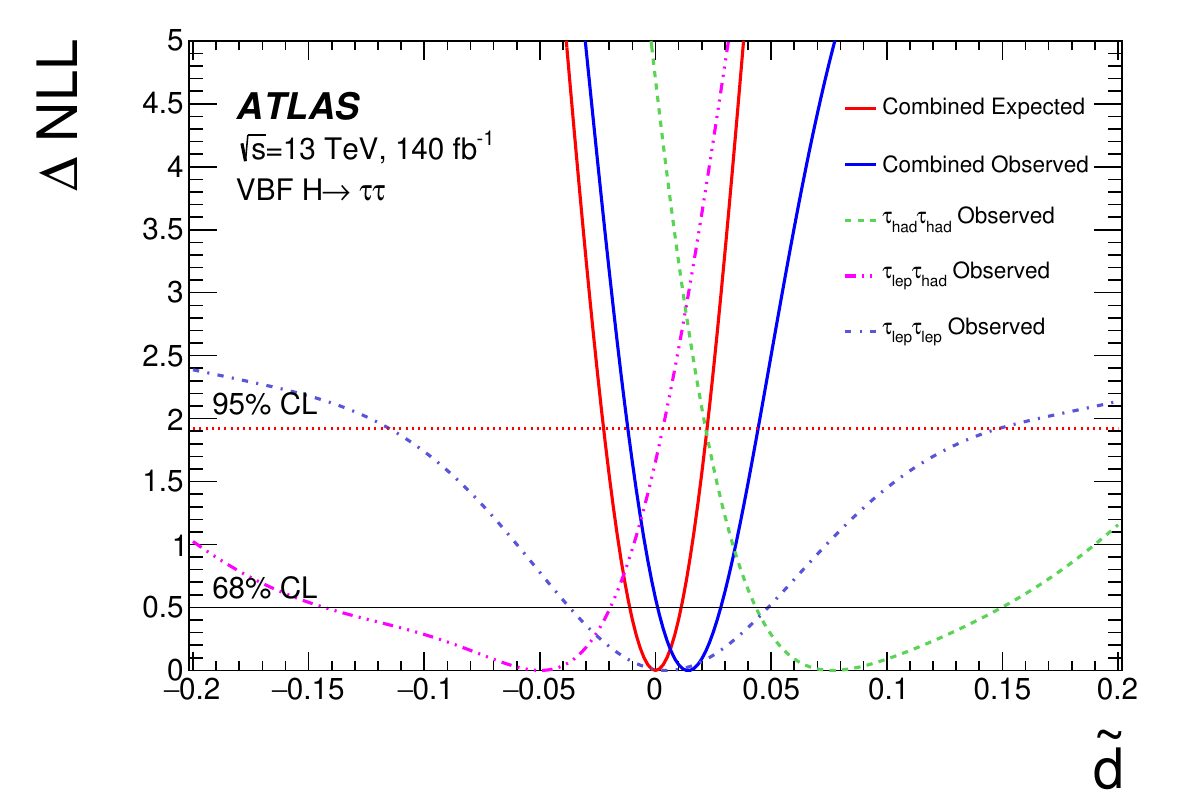}}
\subfloat[]{  \includegraphics[width=.45\textwidth]{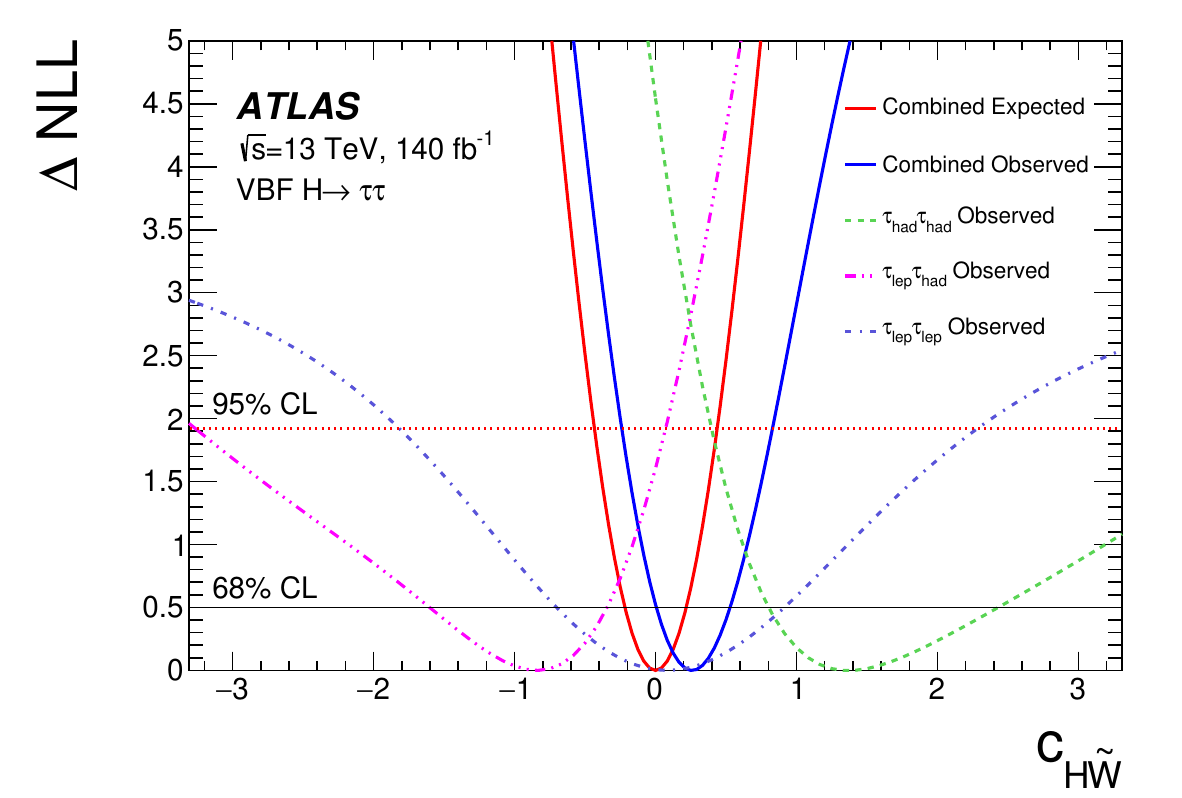}}\\

\caption{Expected and observed $\Delta$NLL distributions for the combined fit as a function of the  CP-violating strength parameters (a) \dtil, and (b) \chwtil.  The dashed lines show the observed $\Delta$NLL curves for the fit to the single channels.  The expected curve is obtained assuming the SM predictions. The horizontal lines correspond to the $\Delta$NLL values used to determine the 68\% and 95\% confidence intervals.  }
\label{fig:NLL}
\end{figure}

The fit is repeated on a simulated dataset but using the \OOReg and $\Delta\phi_{jj}^{\textrm{sign}}$ observables and expected 95\% CL limits are obtained. The expected limits obtained with the Optimal Observable are 5\% stronger than the expected limits obtained using the \OOReg, and 30\% stronger than those obtained with $\Delta\phi_{jj}^{\textrm{sign}}$.

Figure~\ref{fig:LinQuad} shows a comparison of the expected and observed best fit values, using in the reweighting processes both the linear and quadratic terms or the linear term only. For all the interpretations, the difference of adding the quadratic term is small, relaxing the limits by about 10\%.  In all cases, the CP-violating strength parameters are compatible with zero. The expected confidence interval on \dtil represents a large improvement relative to what was set in Ref~\cite{HIGG-2018-14}, and improves also the limits relative to the analysis based on $\Delta\phi_{jj}^{\textrm{sign}}$~\cite{HIGG-2022-07}.

\begin{figure}[!tb]
\centering

\includegraphics[width=.65\textwidth]{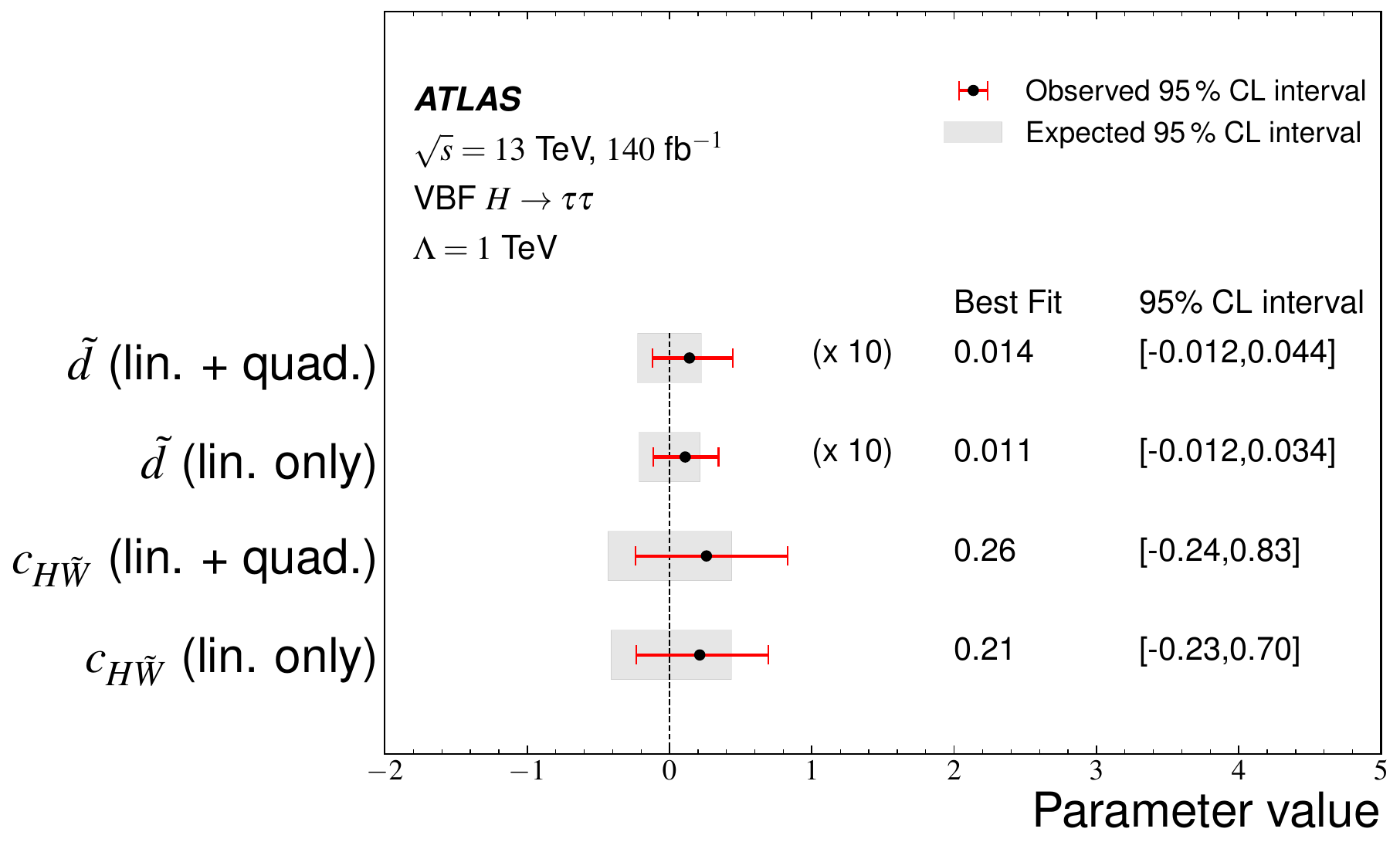}%

\caption{Expected and observed best fit values and 95\% CL intervals for all the CP-violating strength parameters considered in the analysis.  The \dtil values are scaled by a factor ten for illustrative purposes. For the \chwtil results  a new-physics scale at $\Lambda=1~\TeV$ is assumed.  }
\label{fig:LinQuad}
\end{figure}

\begin{figure}[!hb]
\centering

\includegraphics[width=.6\textwidth]{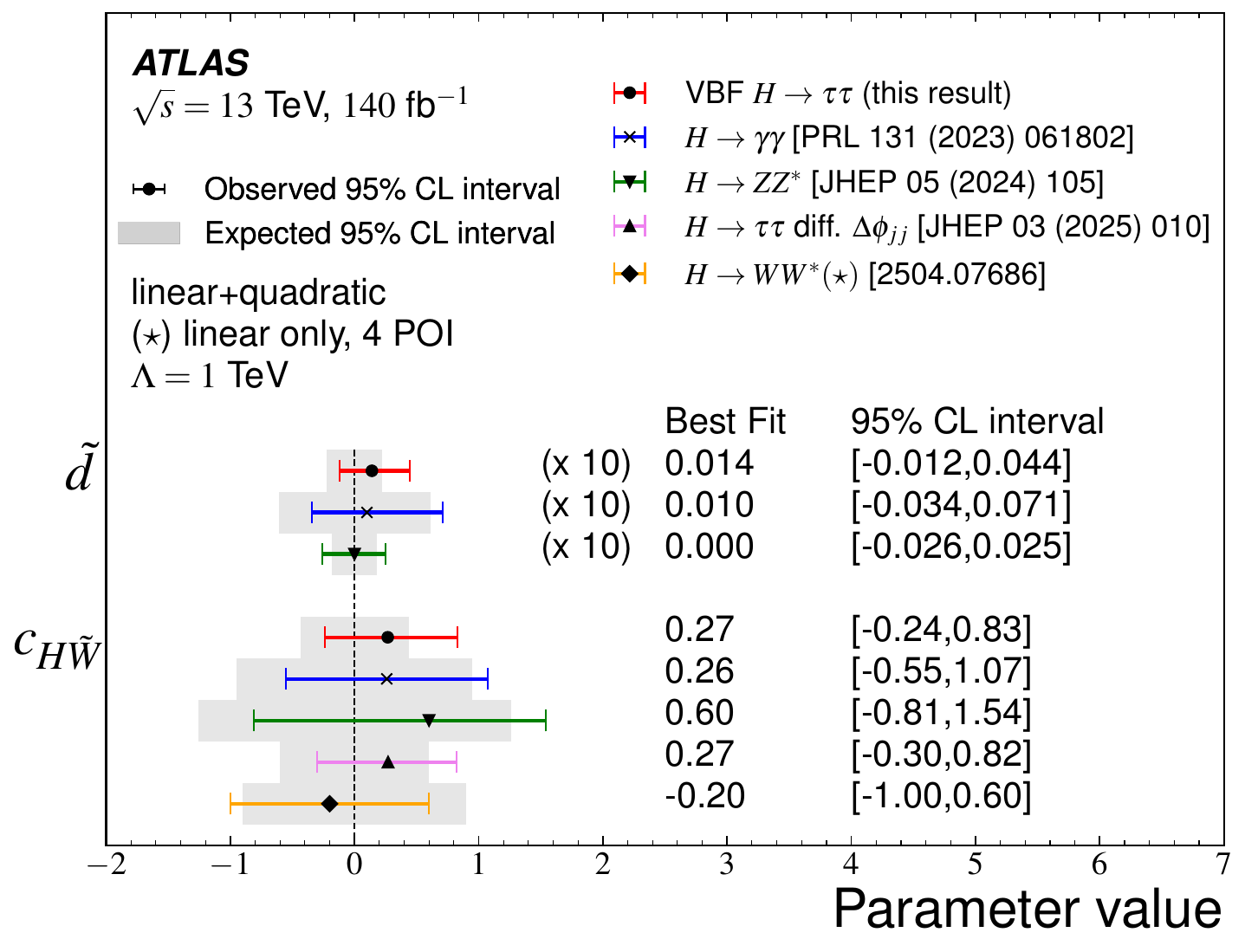}%
\caption{
Comparison of results of the analysis presented with the ATLAS $H\rightarrow ZZ^*\rightarrow 4\ell $ analysis~\cite{HIGG-2018-30-fix} and the ATLAS $H\rightarrow \gamma\gamma $ VBF measurements~\cite{HIGG-2020-08} for \dtil and \chwtil. The results for \chwtil are also compared with the ATLAS $H\rightarrow \tau\tau$ differential cross-section measurement based on $\Delta\phi_{jj}^{\textrm{sign}}$~\cite{HIGG-2022-07} and the ATLAS $H\rightarrow WW^*\rightarrow \ell\nu\ell\nu $ analysis~\cite{HIGP-2024-07}, based on pure shape information.%
The data points show the observed results, and the uncertainty bands correspond to the 95 \% CL interval including statistical and systematic uncertainties.  The \dtil values are scaled by a factor ten for illustrative purposes. For the \chwtil results a new-physics scale at $\Lambda=1~\TeV$ is assumed. The EFT parameterisation includes both the linear and quadratic terms, and \chwtil is fitted while the other CP-odd operators are assumed to be zero, except for the ATLAS $H\rightarrow WW^*\rightarrow \ell\nu\ell\nu $ analysis, where only the linear term is considered and all the Warsaw basis CP-odd and CP-even coefficients are fitted simultaneously (and therefore labelled as 4 POI), with a correlation between the coupling measurements of less than 5\%.
}
\label{fig:CompareAnalysis}
\end{figure}

Figure~\ref{fig:CompareAnalysis} shows a comparison of the expected and observed best fit values between different ATLAS analyses. The results are compared with the $H\rightarrow ZZ\rightarrow 4\ell $ analysis~\cite{HIGG-2018-30-fix} and the $H\rightarrow \gamma\gamma $ VBF measurements~\cite{HIGG-2020-08} for \dtil and \chwtil. The results for \chwtil are also compared with the $H\rightarrow \tau\tau$ differential cross-section measurement based on $\Delta\phi_{jj}^{\textrm{sign}}$~\cite{HIGG-2022-07} and the $H\rightarrow WW^*\rightarrow \ell\nu\ell\nu $ analysis~\cite{HIGP-2024-07}.
This analysis, the $H\rightarrow ZZ\rightarrow 4\ell $ analysis, and the $H\rightarrow \gamma\gamma $ VBF measurements use datasets with no overlap. However, the $H\rightarrow \tau\tau$ $\Delta\phi_{jj}$ differential measurement uses similar final states in the same dataset as this analysis.
In the \chwtil results for all the analyses a new-physics scale at $\Lambda=1~\TeV$ is assumed, and \chwtil is fitted while the other CP-odd operators are assumed to be zero, except in the $H\rightarrow WW^*\rightarrow \ell\nu\ell\nu $ analysis where all the Warsaw basis CP-odd and CP-even coefficients are fitted simultaneously, with a correlation between the coupling measurements of less than 5\%. For both \chwtil and \dtil the EFT parameterisation includes both the linear and quadratic terms, except for the $H\rightarrow WW^*\rightarrow \ell\nu\ell\nu $ analysis, where only the linear term is considered.


\FloatBarrier

\section{Conclusion}

A test of the CP invariance of the Higgs boson in the vertex with vector bosons was presented, in the final state with the Higgs boson decaying into pairs of $\tau$-leptons, using the full LHC Run 2  proton--proton collision dataset of 140~\ifb collected by the ATLAS experiment at $\sqrt{s}=13~\TeV$. The Optimal Observable is used to probe the CP properties of the Higgs boson.

No statistically significant deviation relative to the Standard Model prediction of a CP-even coupling between the Higgs boson and vector bosons is observed.  The results are interpreted using two different EFT bases, in particular the HISZ and the Warsaw bases, with sensitivity mainly on \dtil and \chwtil. The HISZ parameter \dtil is constrained to a 95\% CL interval of $\left[-0.012,0.044 \right]$,  while the Warsaw basis parameter \chwtil is constrained to $\left[-0.24,0.83 \right]$ for $\Lambda=1~\TeV $, in both cases considering the linear and quadratic terms in the reweighting procedure.  The expected 95\% CL interval  based on the Standard Model predictions for the two operators are $\dtil\in \left[-0.022, 0.022 \right]$ and  $\chwtil \in \left[-0.43, 0.44 \right]$, also considering both the linear and quadratic terms in the parameterisation.  The observed limits on these operators are among the most stringent so far, and greatly improve on the previous ATLAS analysis that was based on a partial Run 2 dataset using 36~\ifb by being able to also set constraints at the 95\% CL on the observed parameters.  Additionally, new interpretations are added, which can be used for future combination with other analyses.


\section*{Acknowledgements}

%
%

%
%

We thank CERN for the very successful operation of the LHC and its injectors, as well as the support staff at
CERN and at our institutions worldwide without whom ATLAS could not be operated efficiently.

The crucial computing support from all WLCG partners is acknowledged gratefully, in particular from CERN, the ATLAS Tier-1 facilities at TRIUMF/SFU (Canada), NDGF (Denmark, Norway, Sweden), CC-IN2P3 (France), KIT/GridKA (Germany), INFN-CNAF (Italy), NL-T1 (Netherlands), PIC (Spain), RAL (UK) and BNL (USA), the Tier-2 facilities worldwide and large non-WLCG resource providers. Major contributors of computing resources are listed in Ref.~\cite{ATL-SOFT-PUB-2025-001}.

We gratefully acknowledge the support of ANPCyT, Argentina; YerPhI, Armenia; ARC, Australia; BMWFW and FWF, Austria; ANAS, Azerbaijan; CNPq and FAPESP, Brazil; NSERC, NRC and CFI, Canada; CERN; ANID, Chile; CAS, MOST and NSFC, China; Minciencias, Colombia; MEYS CR, Czech Republic; DNRF and DNSRC, Denmark; IN2P3-CNRS and CEA-DRF/IRFU, France; SRNSFG, Georgia; BMFTR, HGF and MPG, Germany; GSRI, Greece; RGC and Hong Kong SAR, China; ICHEP and Academy of Sciences and Humanities, Israel; INFN, Italy; MEXT and JSPS, Japan; CNRST, Morocco; NWO, Netherlands; RCN, Norway; MNiSW, Poland; FCT, Portugal; MNE/IFA, Romania; MSTDI, Serbia; MSSR, Slovakia; ARIS and MVZI, Slovenia; DSI/NRF, South Africa; MICIU/AEI, Spain; SRC and Wallenberg Foundation, Sweden; SERI, SNSF and Cantons of Bern and Geneva, Switzerland; NSTC, Taipei; TENMAK, T\"urkiye; STFC/UKRI, United Kingdom; DOE and NSF, United States of America.

Individual groups and members have received support from BCKDF, CANARIE, CRC and DRAC, Canada; CERN-CZ, FORTE and PRIMUS, Czech Republic; COST, ERC, ERDF, Horizon 2020, ICSC-NextGenerationEU and Marie Sk{\l}odowska-Curie Actions, European Union; Investissements d'Avenir Labex, Investissements d'Avenir Idex and ANR, France; DFG and AvH Foundation, Germany; Herakleitos, Thales and Aristeia programmes co-financed by EU-ESF and the Greek NSRF, Greece; BSF-NSF and MINERVA, Israel; NCN and NAWA, Poland; La Caixa Banking Foundation, CERCA Programme Generalitat de Catalunya and PROMETEO and GenT Programmes Generalitat Valenciana, Spain; G\"{o}ran Gustafssons Stiftelse, Sweden; The Royal Society and Leverhulme Trust, United Kingdom.

In addition, individual members wish to acknowledge support from CERN: European Organization for Nuclear Research (CERN DOCT); Chile: Agencia Nacional de Investigaci\'on y Desarrollo (FONDECYT 1230812, FONDECYT 1240864, Fondecyt 3240661); China: Chinese Ministry of Science and Technology (MOST-2023YFA1605700, MOST-2023YFA1609300), National Natural Science Foundation of China (NSFC - 12175119, NSFC 12275265); Czech Republic: Czech Science Foundation (GACR - 24-11373S), Ministry of Education Youth and Sports (ERC-CZ-LL2327, FORTE CZ.02.01.01/00/22\_008/0004632), PRIMUS Research Programme (PRIMUS/21/SCI/017); EU: H2020 European Research Council (ERC - 101002463); European Union: European Research Council (BARD No. 101116429, ERC - 948254, ERC 101089007), European Regional Development Fund (SMASH COFUND 101081355, SLO ERDF), Horizon 2020 Framework Programme (MUCCA - CHIST-ERA-19-XAI-00), European Union, Future Artificial Intelligence Research (FAIR-NextGenerationEU PE00000013), Italian Center for High Performance Computing, Big Data and Quantum Computing (ICSC, NextGenerationEU); France: Agence Nationale de la Recherche (ANR-21-CE31-0022, ANR-22-EDIR-0002); Germany: Baden-Württemberg Stiftung (BW Stiftung-Postdoc Eliteprogramme), Deutsche Forschungsgemeinschaft (DFG - 469666862, DFG - CR 312/5-2); China: Research Grants Council (GRF); Italy: Istituto Nazionale di Fisica Nucleare (ICSC, NextGenerationEU), Ministero dell'Università e della Ricerca (NextGenEU 153D23001490006 M4C2.1.1, NextGenEU I53D23000820006 M4C2.1.1, NextGenEU I53D23001490006 M4C2.1.1, SOE2024\_0000023); Japan: Japan Society for the Promotion of Science (JSPS KAKENHI JP22H01227, JSPS KAKENHI JP22H04944, JSPS KAKENHI JP22KK0227, JSPS KAKENHI JP24K23939, JSPS KAKENHI JP25H00650, JSPS KAKENHI JP25H01291, JSPS KAKENHI JP25K01023); Norway: Research Council of Norway (RCN-314472); Poland: Ministry of Science and Higher Education (IDUB AGH, POB8, D4 no 9722), Polish National Science Centre (NCN 2021/42/E/ST2/00350, NCN OPUS 2023/51/B/ST2/02507, NCN OPUS nr 2022/47/B/ST2/03059, NCN UMO-2019/34/E/ST2/00393, UMO-2022/47/O/ST2/00148, UMO-2023/49/B/ST2/04085, UMO-2023/51/B/ST2/00920, UMO-2024/53/N/ST2/00869); Portugal: Foundation for Science and Technology (FCT); Spain: Ministry of Science and Innovation (MCIN \& NextGenEU PCI2022-135018-2, MICIN \& FEDER PID2021-125273NB, RYC2019-028510-I, RYC2020-030254-I, RYC2021-031273-I, RYC2022-038164-I); Sweden: Carl Trygger Foundation (Carl Trygger Foundation CTS 22:2312), Swedish Research Council (Swedish Research Council 2023-04654, VR 2021-03651, VR 2022-03845, VR 2022-04683, VR 2023-03403, VR 2024-05451), Knut and Alice Wallenberg Foundation (KAW 2018.0458, KAW 2022.0358, KAW 2023.0366); Switzerland: Swiss National Science Foundation (SNSF - PCEFP2\_194658); United Kingdom: Leverhulme Trust (Leverhulme Trust RPG-2020-004), Royal Society (NIF-R1-231091); United States of America: U.S. Department of Energy (ECA DE-AC02-76SF00515), Neubauer Family Foundation.

%
%


\clearpage
\appendix
\printbibliography

@article{Bothmann:2019yzt-fix,
      author = "{Sherpa collaboration}",
%      author = "Bothmann, Enrico and others",
      title = "{Event generation with Sherpa 2.2}",
      journal = "SciPost Phys.",
      volume = "7",
      year = "2019",
      number = "3",
      pages = "034",
      doi = "10.21468/SciPostPhys.7.3.034",
      reportNumber = "FERMILAB-PUB-19-218-T, SLAC-PUB-17433, IPPP/19/42, MCNET-19-11",
      eprint = "1905.09127",
      archivePrefix = "arXiv",
      primaryClass = "hep-ph",
}

@article{Pinto:2023yob,
    author = "Pinto, Andr\'es and Wu, Zhibo and Balli, Fabrice and Berger, Nicolas and Boonekamp, Maarten and Chapon, \'Emilien and Kawamoto, Tatsuo and Malaescu, Bogdan",
    title = "{Uncertainty components in profile likelihood fits}",
    eprint = "2307.04007",
    archivePrefix = "arXiv",
    primaryClass = "physics.data-an",
    doi = "10.1140/epjc/s10052-024-12877-5",
    journal = "Eur. Phys. J. C",
    volume = "84",
    number = "6",
    pages = "593",
    year = "2024"
}

@article{BUCHMULLER1986621,
title = {Effective lagrangian analysis of new interactions and flavour conservation},
journal = "{Nucl. Phys. B}",
volume = {268},
number = {3},
pages = {621-653},
year = {1986},
%issn = {0550-3213},
%doi = {https://doi.org/10.1016/0550-3213(86)90262-2},
doi            = "10.1016/0550-3213(86)90262-2",
%url = {https://www.sciencedirect.com/science/article/pii/0550321386902622},
author = {W. Buchmüller and D. Wyler},
abstract = {New interactions with a scale Λ larger than the Fermi scale GF−12 will manifest themselves at energies below Λ through small deviations from the standard model, which can be described by an effective lagrangian containing non-renormalizable SU(3) × SU(2) × U(1) invariant operators. We construct the first two terms of this lagrangian in an expansion in powers of 1/Λ and study systematically possible effects of new interactions such as anomalous magnetic moments, deviations from universality in weak interactions and rare processes. Among the flavour conserving processes the universality of the charged current weak interactions yields the strongest bound on the new interaction scale, Λ > 5 TeV, whereas flavour non-conserving processes imply the bound Λ > 3000 TeV. We derive conditions for natural flavour conservation. Although it cannot be excluded that all flavour changing operators are dynamically suppressed, this appears difficult to understand without a symmetry for which the standard electroweak theory seems to provide no hint. We emphasize the importance of searching for rare decays of D-mesons.}
}

@article{PhysRevD.48.2182,
  title = {Low energy effects of new interactions in the electroweak boson sector},
  author = {Hagiwara, K. and Ishihara, S. and Szalapski, R. and Zeppenfeld, D.},
  journal = {Phys. Rev. D},
  volume = {48},
  %issue = {5},
  pages = {2182--2203},
  numpages = {0},
  year = {1993},
  month = {Sep},
  publisher = {American Physical Society},
  doi = {10.1103/PhysRevD.48.2182},
 % url = {https://link.aps.org/doi/10.1103/PhysRevD.48.2182}
}

@article{Elagin:2010aw,
    author = "Elagin, A. and Murat, P. and Pranko, A. and Safonov, A.",
    title = "{A new mass reconstruction technique for resonances decaying to $\tau\tau$}",
    eprint = "1012.4686",
    archivePrefix = "arXiv",
    primaryClass = "hep-ex",
    reportNumber = "FERMILAB-PUB-11-918-PPD",
    doi = "10.1016/j.nima.2011.07.009",
    journal = "Nucl. Instrum. Meth. A",
    volume = "654",
    pages = "481--489",
    year = "2011"
}

@software{tensorflow,
  author       = "{TensorFlow Developers}",
  title        = {TensorFlow},
  month        = nov,
  year         = 2023,
  publisher    = {Zenodo},
  version      = {v2.15.0},
  doi          = {10.5281/zenodo.10126399},
  url          = {https://doi.org/10.5281/zenodo.10126399}
}

@misc{adam,
      title={Adam: A Method for Stochastic Optimization}, 
      author={Diederik P. Kingma and Jimmy Ba},
      year={2017},
      eprint={1412.6980},
      archivePrefix={arXiv},
      primaryClass={cs.LG}
}

@article{relu,
   title={A high-bias, low-variance introduction to Machine Learning for physicists},
   volume={810},
%   ISSN={0370-1573},
%   url={http://dx.doi.org/10.1016/j.physrep.2019.03.001},
   DOI={10.1016/j.physrep.2019.03.001},
   %journal="{Physics Reports}",
   journal="{Phys. Rept.}",
   publisher={Elsevier BV},
   author={Mehta, Pankaj and Bukov, Marin and Wang, Ching-Hao and Day, Alexandre G.R. and Richardson, Clint and Fisher, Charles K. and Schwab, David J.},
   year={2019},
   month=may, pages={1–124},
   eprint={1803.08823},
   archivePrefix={arXiv},

}

@misc{l2,
      title={L2 Regularization for Learning Kernels}, 
      author={Corinna Cortes and Mehryar Mohri and Afshin Rostamizadeh},
      year={2012},
      eprint={1205.2653},
      archivePrefix={arXiv},
      primaryClass={cs.LG}
}

@article{Ellis:1987xu,
    author = "Ellis, R. Keith and Hinchliffe, I. and Soldate, M. and van der Bij, J. J.",
    title = "{Higgs Decay to $\tau^+$$\tau^-$ A Possible Signature of Intermediate Mass Higgs Bosons at
high energy hadron colliders}",
    reportNumber = "FERMILAB-PUB-87-100-T, LBL-23504",
    doi = "10.1016/0550-3213(88)90019-3",
    journal = "Nucl. Phys. B",
    volume = "297",
    pages = "221--243",
    year = "1988"
}

@article{PhysRevD.74.095001,
  title = {Anomalous Higgs boson couplings in vector boson fusion at the CERN LHC},
  author = {Hankele, V. and Kl\"amke, G. and Zeppenfeld, D. and Figy, T.},
  journal = {Phys. Rev. D},
  volume = {74},
  %issue = {9},
  pages = {095001},
  numpages = {8},
  year = {2006},
  month = {Nov},
  publisher = {American Physical Society},
  doi = {10.1103/PhysRevD.74.095001},
  %url = {https://link.aps.org/doi/10.1103/PhysRevD.74.095001}
}

@Article{DELPHI:2008uqu,
    author = "Abdallah, J. and others",
    collaboration = "DELPHI",
    title = "{Study of W-boson polarisations and triple gauge boson couplings in the reaction e+e- ---\ensuremath{>} W+W- at LEP 2}",
    eprint = "0801.1235",
    archivePrefix = "arXiv",
    primaryClass = "hep-ex",
    reportNumber = "CERN-PH-EP-2007-036",
    doi = "10.1140/epjc/s10052-008-0528-3",
    journal = "Eur. Phys. J. C",
    volume = "54",
    pages = "345--364",
    year = "2008"
}

@Article{LEPDtil2001,
   title={Measurement of W boson polarisations and CP-violating triple gauge couplings from W$^{+}$W$^{-}$production at LEP},
   volume={19},
%   ISSN={1434-6052},
%   url={http://dx.doi.org/10.1007/s100520100602},
   DOI={10.1007/s100520100602},
   number={2},
   journal={Eur. Phys. J. C},
   publisher={Springer Science and Business Media LLC},
   author={{OPAL Collaboration}},
   year={2001},
   month={feb},
   pages={229–240},
   eprint = "hep-ex/0009021",
   archivePrefix = "arXiv",

}

@Article{20057,
      author = {{ALEPH Collaboration}},
      title = {Improved measurement of the triple gauge-boson couplings $\gamma WW$ and $ZWW$ in $e^+e^-$ collisions},
      journal = {Phys. Lett. B},
      volume = {614},
      number = {1},
      pages = {7-26},
      year = {2005},
      %issn = {0370-2693},
      doi = {https://doi.org/10.1016/j.physletb.2005.03.058},
      reportNumber = "CERN-PH-EP-2004-065",
      %url = {https://www.sciencedirect.com/science/article/pii/S0370269305004120},
}

@article{Sakharov:1967dj,
      author         = "Sakharov, A. D.",
      title          = "{Violation of CP invariance, C asymmetry, and baryon
                        asymmetry of the universe}",
      journal        = "Pisma Zh. Eksp. Teor. Fiz.",
      volume         = "5",
      year           = "1967",
      pages          = "32-35",
      doi            = "10.1070/PU1991v034n05ABEH002497",
%      note           = "[Usp. Fiz. Nauk161,61(1991)]",
      SLACcitation   = "%%CITATION = ZFPRA,5,32;%%"
}

@article{Huet:1994jb,
      author         = "Huet, P.  and Sather, E.",
      title          = "{Electroweak baryogenesis and standard model CP
                        violation}",
      journal        = "Phys. Rev. D",
      volume         = "51",
      year           = "1995",
      pages          = "379-394",
      doi            = "10.1103/PhysRevD.51.379",
      eprint         = "hep-ph/9404302",
      archivePrefix  = "arXiv",
      %primaryClass   = "hep-ph",
      reportNumber   = "SLAC-PUB-6479",
      SLACcitation   = "%%CITATION = HEP-PH/9404302;%%"
}

@article{Gavela:1993ts,
      author         = "Gavela, M. B. and Hern\'{a}ndez, P. and Orloff, J. and P\`{e}ne,
                        O.",
      title          = "{Standard Model CP-violation and Baryon asymmetry}",
      journal        = "Mod. Phys. Lett. A",
      volume         = "9",
      year           = "1994",
      pages          = "795-810",
      doi            = "10.1142/S0217732394000629",
      eprint         = "hep-ph/9312215",
      archivePrefix  = "arXiv",
      %primaryClass   = "hep-ph",
      reportNumber   = "CERN-TH-7081-93, LPTHE-ORSAY-93-48, HUTP-93-A036,
                        HD-THEP-93-45",
      SLACcitation   = "%%CITATION = HEP-PH/9312215;%%"
}

@article{Cohen:1993nk,
      author         = "Cohen, A. G. and Kaplan, D. B. and Nelson, A. E.",
      title          = "{Progress in Electroweak Baryogenesis}",
      journal        = "Ann. Rev. Nucl. Part. Sci.",
      volume         = "43",
      year           = "1993",
      pages          = "27-70",
      doi            = "10.1146/annurev.ns.43.120193.000331",
      eprint         = "hep-ph/9302210",
      archivePrefix  = "arXiv",
      %primaryClass   = "hep-ph",
      reportNumber   = "UCSD-PTH-93-02, BUHEP-93-4",
      SLACcitation   = "%%CITATION = HEP-PH/9302210;%%"
}

@article{Riotto:1999yt,
      author         = "Riotto, A. and Trodden, M.",
      title          = "{Recent Progress in Baryogenesis}",
      journal        = "Ann. Rev. Nucl. Part. Sci.",
      volume         = "49",
      year           = "1999",
      pages          = "35-75",
      doi            = "10.1146/annurev.nucl.49.1.35",
      eprint         = "hep-ph/9901362",
      archivePrefix  = "arXiv",
      %primaryClass   = "hep-ph",
      reportNumber   = "CERN-TH-99-04, CWRU-P6-99",
      SLACcitation   = "%%CITATION = HEP-PH/9901362;%%"
}

@article{Hou:2008xd,
      author = "Wei-Shu Hou",
      title ="{Source of CP Violation for Baryon Asymmetry of the Universe}",
      journal ="Chin. J. Phys.",
      volume ="47",
      year ="2009",
      pages ="134-141",
      eprint         = "0803.1234",
      archivePrefix  = "arXiv",
      doi = "https://doi.org/10.48550/arXiv.0803.1234",
}

@article{Grzadkowski:2010es,
    author = "Grzadkowski, Bohdan and Iskrzy{\'n}ski, M and Misiak, Mikolaj and Rosiek, Janusyz",
    title = "{Dimension-Six Terms in the Standard Model Lagrangian}",
    eprint = "1008.4884",
    archivePrefix = "arXiv",
    primaryClass = "hep-ph",
    reportNumber = "IFT-9-2010, TTP10-35",
    doi = "10.1007/JHEP10(2010)085",
    journal = "JHEP",
    volume = "10",
    pages = "085",
    year = "2010"
}

@article{Brivio:2020onw,
    author = "Brivio, Ilaria",
    title = "{SMEFTsim 3.0 - a practical guide}",
    eprint = "2012.11343",
    archivePrefix = "arXiv",
    primaryClass = "hep-ph",
    doi = "10.1007/JHEP04(2021)073",
    journal = "JHEP",
    volume = "04",
    pages = "073",
    year = "2021"
}

@article{Hankele:2006ma,
      author         = "Hankele, V. and Kl{\"a}mke, G. and Zeppenfeld, D. and Figy,
                        T.",
      title          = "{Anomalous Higgs boson couplings in vector boson fusion
                        at the CERN LHC}",
      journal        = "Phys. Rev. D",
      volume         = "74",
      year           = "2006",
      pages          = "095001",
      doi            = "10.1103/PhysRevD.74.095001",
      eprint         = "hep-ph/0609075",
      archivePrefix  = "arXiv",
      %primaryClass   = "hep-ph",
      reportNumber   = "KA-TP-09-2006",
      SLACcitation   = "%%CITATION = HEP-PH/0609075;%%"
}

@article{Atwood:1991ka,
      author         = "Atwood, D. and Soni, A.",
      title          = "{Analysis for magnetic moment and electric dipole moment
                        form factors of the top quark via $\ensuremath{e^{+}e^{-} \to t \bar{t}}$}",
      journal        = "Phys. Rev. D",
      volume         = "45",
      year           = "1992",
      pages          = "2405-2413",
      doi            = "10.1103/PhysRevD.45.2405",
      reportNumber   = "BNL-46892",
      SLACcitation   = "%%CITATION = PHRVA,D45,2405;%%"
}

@article{Davier:1992nw,
      author         = "Davier, M. and Duflot, L. and Le Diberder, F. and Rougé, A.",
      title          = "{The optimal method for the measurement of tau
                        polarization}",
      journal        = "Phys. Lett. B",
      volume         = "306",
      year           = "1993",
      pages          = "411-417",
      doi            = "10.1016/0370-2693(93)90101-M",
      reportNumber   = "LAL-92-73, X-LPNHE-92-22",
      SLACcitation   = "%%CITATION = PHLTA,B306,411;%%"
}

@article{Diehl:1993br,
      author         = "Diehl, M. and Nachtmann, O.",
      title          = "{Optimal observables for the measurement of three gauge
                        boson couplings in $\ensuremath{e^{+}e^{-} \to W^{+}W^{-}}$}",
      journal        = "Z. Phys. C",
      volume         = "62",
      year           = "1994",
      pages          = "397-412",
      doi            = "10.1007/BF01555899",
      reportNumber   = "HD-THEP-93-37",
      SLACcitation   = "%%CITATION = ZEPYA,C62,397;%%"
}

@article{Brehmer:2017lrt,
      author         = "Brehmer, J. and Kling, F. and Plehn, T. and
                        Tait, T. M. P.",
      title          = "{Better Higgs-CP tests through information geometry}",
      journal        = "Phys. Rev. D",
      volume         = "97",
      year           = "2018",
      number         = "9",
      pages          = "095017",
      doi            = "10.1103/PhysRevD.97.095017",
      eprint         = "1712.02350",
      archivePrefix  = "arXiv",
      primaryClass   = "hep-ph",
      reportNumber   = "UCI-HEP-TR-2017-14",
      SLACcitation   = "%%CITATION = ARXIV:1712.02350;%%"
}

@article{Ball:2012cx-fix,
      author         = "{NNPDF Collaboration}",
      title          = "{Parton distributions with LHC data}",
      journal        = "Nucl. Phys. B",
      volume         = "867",
      year           = "2013",
      pages          = "244",
      doi            = "10.1016/j.nuclphysb.2012.10.003",
      eprint         = "1207.1303",
      archivePrefix  = "arXiv",
      primaryClass   = "hep-ph",
      reportNumber   = "EDINBURGH-2012-08, IFUM-FT-997, FR-PHENO-2012-014,
                        RWTH-TTK-12-25, CERN-PH-TH-2012-037, SFB-CPP-12-47\,
                        --CERN-PH-TH-2012-037",
      SLACcitation   = "%%CITATION = ARXIV:1207.1303;%%"
}

@article{Zyla:2020zbs-fix,
    author = "{Particle Data Group}",
    title = "{Review of Particle Physics}",
    journal = "PTEP",
    volume = "2020",
    number = "8",
    pages = "083C01",
    year = "2020",
    doi = "10.1093/ptep/ptaa104",
}

@article{Bredenstein:2006nk-fix,
      author         = "Bredenstein, A. and Denner, Ansgar and Dittmaier, S. and Weber, M. M.",
      title          = "{Precision calculations for the Higgs decays \(H \to ZZ/WW \to 4\) leptons}",
      booktitle      = "{Proceedings, 8th DESY Workshop on Elementary Particle
                        Theory: Loops and Legs in Quantum Field Theory}",
      venue          = "Eisenach, Germany",
      eventdate      = "2006-04-23/2006-04-28",
      journal        = "Nucl. Phys. B Proc. Suppl.",
      volume         = "160",
      year           = "2006",
      pages          = "131-135",
      doi            = "10.1016/j.nuclphysbps.2006.09.104",
      eprint         = "hep-ph/0607060",
      archivePrefix  = "arXiv",
      primaryClass   = "hep-ph",
      reportNumber   = "PSI-PR-06-09",
      SLACcitation   = "%%CITATION = HEP-PH/0607060;%%"
}

@Article{HIGG-2018-30-fix,
    author         = "{ATLAS Collaboration}",
    title          = "{Test of CP-invariance of the Higgs boson in vector-boson fusion production and in its decay into four leptons}",
    journal        = "JHEP",
    volume         = "05",
    year           = "2024",
    pages          = "105",
    doi            = "10.1007/JHEP05(2024)105",
    reportNumber   = "CERN-EP-2023-030",
    eprint         = "2304.09612",
    archivePrefix  = "arXiv",
    primaryClass   = "hep-ex",
}

@Article{JETM-2020-03-fix,
    author         = "{ATLAS Collaboration}",
    title          = "{The performance of missing transverse momentum reconstruction and its significance with the ATLAS detector using \(140\,\text{fb}^{-1}\) of \(\sqrt{s}=13\,\text{TeV}\) \(pp\) collisions}",
    year           = "2025",
    reportNumber   = "CERN-EP-2024-023",
    eprint         = "2402.05858",
    archivePrefix  = "arXiv",
    primaryClass   = "hep-ex",
    doi            = "10.1140/epjc/s10052-025-14062-8",
    journal        = "Eur. Phys. J. C",
    volume         = "85",
    number         = "6",
    pages          = "606",
}

@article{Djouadi:2006bz-fix,
      author         = "Djouadi, A. and M{\"u}hlleitner, M. M. and Spira, M.",
      title          = "{Decays of Supersymmetric Particles: the Program SUSY-HIT
                        (SUspect-SdecaY-Hdecay-InTerface)}",
      booktitle      = "{Proceedings, Physics at LHC, 3rd Conference}",
      venue          = "Cracow, Poland",
      eventdate      = "2006-07-03/2006-07-08",
      journal        = "Acta Phys. Polon. B",
      volume         = "38",
      year           = "2006",
      pages          = "635-644",
      eprint         = "hep-ph/0609292",
      archivePrefix  = "arXiv",
      XprimaryClass   = "hep-ph",
      reportNumber   = "CERN-PH-TH-2006-200",
      SLACcitation   = "%%CITATION = HEP-PH/0609292;%%"
}

@Article{Evans:2008zzb,
      author         = "Evans, Lyndon and Bryant, Philip",
      title          = "{LHC Machine}",
      journal        = "JINST",
      volume         = "3",
      pages          = "S08001",
      doi            = "10.1088/1748-0221/3/08/S08001",
      year           = "2008",
      SLACcitation   = "%%CITATION = JINST,3,S08001;%%",
}

@Article{Butterworth:2015oua,
      author         = "Butterworth, Jon and others",
      title          = "{PDF4LHC recommendations for LHC Run II}",
      journal        = "J. Phys. G",
      volume         = "43",
      year           = "2016",
      pages          = "023001",
      doi            = "10.1088/0954-3899/43/2/023001",
      eprint         = "1510.03865",
      archivePrefix  = "arXiv",
      primaryClass   = "hep-ph",
      reportNumber   = "OUTP-15-17P, SMU-HEP-15-12, TIF-UNIMI-2015-14,
                        LCTS-2015-27, CERN-PH-TH-2015-249",
      SLACcitation   = "%%CITATION = ARXIV:1510.03865;%%"
}

@Article{Pumplin:2002vw,
     author    = "Pumplin, J. and others",
     title     = "{New Generation of Parton Distributions with Uncertainties from Global QCD Analysis}",
     journal   = "JHEP",
     volume    = "07",
     year      = "2002",
     pages     = "012",
     doi       = "10.1088/1126-6708/2002/07/012",
     eprint    = "hep-ph/0201195",
     archivePrefix = "arXiv",
     SLACcitation  = "%%CITATION = HEP-PH/0201195;%%"
}

@article{Czakon:2012pz,
      author         = "Czakon, Michal and Mitov, Alexander",
      title          = "{NNLO corrections to top pair production at hadron
                        colliders: the quark-gluon reaction}",
      journal        = "JHEP",
      volume         = "01",
      pages          = "080",
      doi            = "10.1007/JHEP01(2013)080",
      year           = "2013",
      eprint         = "1210.6832",
      archivePrefix  = "arXiv",
      primaryClass   = "hep-ph",
      SLACcitation   = "%%CITATION = ARXIV:1210.6832;%%",
}

@Article{Sjostrand:2007gs,
	author    = "Sj{\"o}strand, T. and Mrenna, S. and Skands, P.",
	title     = "{A brief introduction to PYTHIA 8.1}",
	journal   = "Comput. Phys. Commun.",
	volume    = "178",
	year      = "2008",
	pages     = "852-867",
	eprint    = "0710.3820",
	archivePrefix = "arXiv",
	primaryClass  =  "hep-ph",
	doi       = "10.1016/j.cpc.2008.01.036",
	SLACcitation  = "%%CITATION = 0710.3820;%%"
}

@article{Sjostrand:2014zea,
      author         = "Sj{\"o}strand, Torbj{\"o}rn and Ask, Stefan and Christiansen,
                        Jesper R. and Corke, Richard and Desai, Nishita and Ilten,
                        Philip and Mrenna, Stephen and Prestel, Stefan and
                        Rasmussen, Christine O. and Skands, Peter Z.",
      title          = "{An introduction to PYTHIA 8.2}",
      journal        = "Comput. Phys. Commun.",
      volume         = "191",
      year           = "2015",
      pages          = "159",
      doi            = "10.1016/j.cpc.2015.01.024",
      eprint         = "1410.3012",
      archivePrefix  = "arXiv",
      primaryClass   = "hep-ph",
      reportNumber   = "LU-TP-14-36, MCNET-14-22, CERN-PH-TH-2014-190,
                        FERMILAB-PUB-14-316-CD, DESY-14-178, SLAC-PUB-16122,
                        --FERMILAB-PUB-14-316-CD",
      SLACcitation   = "%%CITATION = ARXIV:1410.3012;%%"
}

@Article{Lange:2001uf,
      author         = "Lange, D. J.",
      title          = "{The EvtGen particle decay simulation package}",
      booktitle      = "{Proceedings, 7th International Conference on B physics
                        at hadron machines (BEAUTY 2000)}",
      journal        = "Nucl. Instrum. Meth. A",
      volume         = "462",
      year           = "2001",
      pages          = "152",
      doi            = "10.1016/S0168-9002(01)00089-4",
      SLACcitation   = "%%CITATION = NUIMA,A462,152;%%"
}

@Article{Frixione:2008yi,
     author    = "Frixione, Stefano and Laenen, Eric and Motylinski, Patrick and White, Chris and Webber, Bryan R.",
     title     = "{Single-top hadroproduction in association with a \(W\) boson}",
     journal   = "JHEP",
     volume    = "07",
     year      = "2008",
     pages     = "029",
     eprint    = "0805.3067",
     archivePrefix = "arXiv",
     primaryClass  =  "hep-ph",
     doi       = "10.1088/1126-6708/2008/07/029"
}

@Article{Alwall:2014hca,
      author         = "Alwall, J. and Frederix, R. and Frixione, S. and Hirschi,
                        V. and Maltoni, F. and Mattelaer, O. and Shao, H. -S. and
                        Stelzer, T. and Torrielli, P. and Zaro, M.",
      title          = "{The automated computation of tree-level and
                        next-to-leading order differential cross sections, and
                        their matching to parton shower simulations}",
      journal        = "JHEP",
      volume         = "07",
      year           = "2014",
      pages          = "079",
      doi            = "10.1007/JHEP07(2014)079",
      eprint         = "1405.0301",
      archivePrefix  = "arXiv",
      primaryClass   = "hep-ph",
      reportNumber   = "CERN-PH-TH-2014-064, CP3-14-18, LPN14-066, MCNET-14-09,
                        ZU-TH-14-14",
      SLACcitation   = "%%CITATION = ARXIV:1405.0301;%%"
}

@article{Frederix:2012ps,
      author         = "Frederix, Rikkert and Frixione, Stefano",
      title          = "{Merging meets matching in MC@NLO}",
      journal        = "JHEP",
      volume         = "12",
      year           = "2012",
      pages          = "061",
      doi            = "10.1007/JHEP12(2012)061",
      eprint         = "1209.6215",
      archivePrefix  = "arXiv",
      primaryClass   = "hep-ph",
      reportNumber   = "CERN-PH-TH-2012-247, ZU-TH-21-12",
      SLACcitation   = "%%CITATION = ARXIV:1209.6215;%%"
}

@article{Bellm:2017jjp,
      author         = "Bellm, Johannes and others",
      title          = "{Herwig 7.1 Release Note}",
      eprint         = "1705.06919",
      year           = "2017",
      archivePrefix  = "arXiv",
      primaryClass   = "hep-ph",
      reportNumber   = "CERN-PH-TH-2017-109, MAN/HEP/2017/08, UWTHPH-2017-10, 
                        IFJPAN-IV-2017-7, NIKHEF 2017-026, HERWIG-2017-02, 
                        KA-TP-19-2017, MCnet-17-08, IPPP/17/40",
      SLACcitation   = "%%CITATION = ARXIV:1705.06919;%%"
}

@Article{Nason:2004rx,
      author         = "Nason, Paolo",
      title          = "{A new method for combining NLO QCD with shower Monte Carlo algorithms}",
      journal        = "JHEP",
      volume         = "11",
      pages          = "040",
      doi            = "10.1088/1126-6708/2004/11/040",
      year           = "2004",
      eprint         = "hep-ph/0409146",
      archivePrefix  = "arXiv",
}

@Article{Frixione:2007vw,
      author         = "Frixione, Stefano and Nason, Paolo and Oleari, Carlo",
      title          = "{Matching NLO QCD computations with parton shower
                        simulations: the POWHEG method}",
      journal        = "JHEP",
      volume         = "11",
      pages          = "070",
      doi            = "10.1088/1126-6708/2007/11/070",
      year           = "2007",
      eprint         = "0709.2092",
      archivePrefix  = "arXiv",
      primaryClass   = "hep-ph",
}

@Article{Alioli:2010xd,
      author         = "Alioli, Simone and Nason, Paolo and Oleari, Carlo and Re,
                        Emanuele",
      title          = "{A general framework for implementing NLO calculations in
                        shower Monte Carlo programs: the POWHEG BOX}",
      journal        = "JHEP",
      volume         = "06",
      pages          = "043",
      doi            = "10.1007/JHEP06(2010)043",
      year           = "2010",
      eprint         = "1002.2581",
      archivePrefix  = "arXiv",
      primaryClass   = "hep-ph",
}

@article{Hamilton:2012rf,
      author         = "Hamilton, Keith and Nason, Paolo and Oleari, Carlo and
                        Zanderighi, Giulia",
      title          = "{Merging H/W/Z + 0 and 1 jet at NLO with no merging
                        scale: a path to parton shower + NNLO matching}",
      journal        = "JHEP",
      volume         = "05",
      year           = "2013",
      pages          = "082",
      doi            = "10.1007/JHEP05(2013)082",
      eprint         = "1212.4504",
      archivePrefix  = "arXiv",
      primaryClass   = "hep-ph",
      reportNumber   = "CERN-PH-TH-2012-356",
      SLACcitation   = "%%CITATION = ARXIV:1212.4504;%%"
}

@article{Hamilton:2012np,
      author         = "Hamilton, Keith and Nason, Paolo and Zanderighi, Giulia",
      title          = "{MINLO: multi-scale improved NLO}",
      journal        = "JHEP",
      volume         = "10",
      year           = "2012",
      pages          = "155",
      doi            = "10.1007/JHEP10(2012)155",
      eprint         = "1206.3572",
      archivePrefix  = "arXiv",
      primaryClass   = "hep-ph",
      reportNumber   = "CERN-PH-TH-2012-166, OUTP-12-11P, MCNET-12-07",
      SLACcitation   = "%%CITATION = ARXIV:1206.3572;%%"
}

@article{Kidonakis:2010ux,
      author         = "Kidonakis, Nikolaos",
      title          = "{Two-loop soft anomalous dimensions for single top quark
                        associated production with a \(W^{-}\) or \(H^{-}\)}",
      journal        = "Phys. Rev. D",
      volume         = "82",
      year           = "2010",
      pages          = "054018",
      doi            = "10.1103/PhysRevD.82.054018",
      eprint         = "1005.4451",
      archivePrefix  = "arXiv",
      primaryClass   = "hep-ph",
      SLACcitation   = "%%CITATION = ARXIV:1005.4451;%%"
}

@inproceedings{Kidonakis:2013zqa,
      author         = "Kidonakis, Nikolaos",
      title          = "{Top Quark Production}",
      booktitle      = "{Proceedings, Helmholtz International Summer School on Physics of Heavy Quarks and Hadrons (HQ2013)}",
      eventdate      = "2013-07-15/2013-07-28",
      venue          = "JINR, Dubna, Russia",
      pages          = "139-168",
      doi            = "10.3204/DESY-PROC-2013-03/Kidonakis",
      eprint         = "1311.0283",
      archivePrefix  = "arXiv",
      primaryClass   = "hep-ph",
      reportNumber   = "KSU-HEP-110113",
      SLACcitation   = "%%CITATION = ARXIV:1311.0283;%%"
}

@article{Beneke:2011mq,
      author         = "Beneke, M. and Falgari, P. and Klein, S. and Schwinn, C.",
      title          = "{Hadronic top-quark pair production with NNLL threshold
                        resummation}",
      journal        = "Nucl. Phys. B",
      volume         = "855",
      year           = "2012",
      pages          = "695-741",
      doi            = "10.1016/j.nuclphysb.2011.10.021",
      eprint         = "1109.1536",
      archivePrefix  = "arXiv",
      primaryClass   = "hep-ph",
      reportNumber   = "TTK-11-38, ITP-UU-11-26, SPIN-11-19, FR-PHENO-2011-015,
                        SFB-CPP-11-49",
      SLACcitation   = "%%CITATION = ARXIV:1109.1536;%%"
}

@article{Anastasiou:2003ds,
      author         = "Anastasiou, Charalampos and Dixon, Lance and Melnikov,
                        Kirill and Petriello, Frank",
      title          = "{High-precision QCD at hadron colliders: Electroweak
                        gauge boson rapidity distributions at next-to-next-to leading order}",
      journal        = "Phys. Rev. D",
      volume         = "69",
      year           = "2004",
      pages          = "094008",
      doi            = "10.1103/PhysRevD.69.094008",
      eprint         = "hep-ph/0312266",
      archivePrefix  = "arXiv",
      reportNumber   = "SLAC-PUB-10288, UH-511-1042-03",
      SLACcitation   = "%%CITATION = HEP-PH/0312266;%%"
}

@article{Cacciari:2011hy,
      author         = "Cacciari, Matteo and Czakon, Michal and Mangano,
                        Michelangelo and Mitov, Alexander and Nason, Paolo",
      title          = "{Top-pair production at hadron colliders with
                        next-to-next-to-leading logarithmic soft-gluon
                        resummation}",
      journal        = "Phys. Lett. B",
      volume         = "710",
      pages          = "612-622",
      doi            = "10.1016/j.physletb.2012.03.013",
      year           = "2012",
      eprint         = "1111.5869",
      primaryClass   = "hep-ph",
      archivePrefix  = "arXiv",
      reportNumber   = "CERN-PH-TH-2011-277, TTK-11-54",
      SLACcitation   = "%%CITATION = ARXIV:1111.5869;%%",
}

@article{Czakon:2012zr,
      author         = "Czakon, Michal and Mitov, Alexander",
      title          = "{NNLO corrections to top-pair production at hadron
                        colliders: the all-fermionic scattering channels}",
      journal        = "JHEP",
      volume         = "12",
      pages          = "054",
      doi            = "10.1007/JHEP12(2012)054",
      year           = "2012",
      eprint         = "1207.0236",
      archivePrefix  = "arXiv",
      primaryClass   = "hep-ph",
      SLACcitation   = "%%CITATION = ARXIV:1207.0236;%%",
}

@article{Czakon:2013goa,
      author         = "Czakon, Michal and Fiedler, Paul and Mitov, Alexander",
      title          = "{Total Top-Quark Pair-Production Cross Section at
                        Hadron Colliders Through \(\mathcal{O}(\alpha_S^4)\)}",
      journal        = "Phys. Rev. Lett.",
      volume         = "110",
      pages          = "252004",
      doi            = "10.1103/PhysRevLett.110.252004",
      year           = "2013",
      eprint         = "1303.6254",
      archivePrefix  = "arXiv",
      primaryClass   = "hep-ph",
      reportNumber   = "CERN-PH-TH-2013-056, TTK-13-08",
      SLACcitation   = "%%CITATION = ARXIV:1303.6254;%%",
}

@article{Baernreuther:2012ws,
      author         = "B{\"a}rnreuther, Peter and Czakon, Michal and Mitov,
                        Alexander",
      title          = "{Percent-Level-Precision Physics at the
                  Tevatron: Next-to-Next-to-Leading Order QCD
                  Corrections to \(q \bar{q} \to t \bar{t} + X\)}",
      journal        = "Phys. Rev. Lett.",
      volume         = "109",
      pages          = "132001",
      doi            = "10.1103/PhysRevLett.109.132001",
      year           = "2012",
      eprint         = "1204.5201",
      archivePrefix  = "arXiv",
      primaryClass   = "hep-ph",
      SLACcitation   = "%%CITATION = ARXIV:1204.5201;%%",
}

@article{deFlorian:2016spz,
      author         = "de Florian, D. and others",
      title          = "{Handbook of LHC Higgs Cross Sections: 4. Deciphering the Nature of the Higgs Sector}",
      collaboration  = "LHC Higgs Cross Section Working Group",
      doi            = "10.23731/CYRM-2017-002",
      year           = "2017",
      eprint         = "1610.07922",
      archivePrefix  = "arXiv",
      primaryClass   = "hep-ph",
      reportNumber   = "FERMILAB-FN-1025-T, CERN-2017-002-M",
      SLACcitation   = "%%CITATION = ARXIV:1610.07922;%%"
}

@Article{Hoeche:2009rj,
      author         = "H{\"o}che, Stefan and Krauss, Frank and Schumann, Steffen and Siegert, Frank",
      title          = "{QCD matrix elements and truncated showers}",
      journal        = "JHEP",
      volume         = "05",
      pages          = "053",
      doi            = "10.1088/1126-6708/2009/05/053",
      year           = "2009",
      eprint         = "0903.1219",
      archivePrefix  = "arXiv",
      primaryClass   = "hep-ph",
}

@Article{Gleisberg:2008fv,
      author         = "Gleisberg, Tanju and H{\"o}che, Stefan",
      title          = "{Comix, a new matrix element generator}",
      journal        = "JHEP",
      volume         = "12",
      pages          = "039",
      doi            = "10.1088/1126-6708/2008/12/039",
      year           = "2008",
      eprint         = "0808.3674",
      archivePrefix  = "arXiv",
      primaryClass   = "hep-ph",
}

@Article{Schumann:2007mg,
      author         = "Schumann, Steffen and Krauss, Frank",
      title          = "{A parton shower algorithm based on Catani--Seymour dipole factorisation}",
      journal        = "JHEP",
      volume         = "03",
      pages          = "038",
      doi            = "10.1088/1126-6708/2008/03/038",
      year           = "2008",
      eprint         = "0709.1027",
      archivePrefix  = "arXiv",
      primaryClass   = "hep-ph",
}

@article{Hoeche:2012yf,
      author         = "H{\"o}che, Stefan and Krauss, Frank and Sch{\"o}nherr, Marek and
                        Siegert, Frank",
      title          = "{QCD matrix elements + parton showers. The NLO case}",
      journal        = "JHEP",
      volume         = "04",
      year           = "2013",
      pages          = "027",
      doi            = "10.1007/JHEP04(2013)027",
      eprint         = "1207.5030",
      archivePrefix  = "arXiv",
      primaryClass   = "hep-ph",
      reportNumber   = "SLAC-PUB-15191, IPPP-12-52, DCPT-12-104, LPN12-081,
                        FR-PHENO-2012-017, MCNET-12-09, --FR-PHENO-2012-017",
      SLACcitation   = "%%CITATION = ARXIV:1207.5030;%%"
}

@article{Hoeche:2011fd,
      author         = "H{\"o}che, Stefan and Krauss, Frank and Sch{\"o}nherr, Marek and
                        Siegert, Frank",
      title          = "{A critical appraisal of NLO+PS matching methods}",
      journal        = "JHEP",
      volume         = "09",
      year           = "2012",
      pages          = "049",
      doi            = "10.1007/JHEP09(2012)049",
      eprint         = "1111.1220",
      archivePrefix  = "arXiv",
      primaryClass   = "hep-ph",
      reportNumber   = "SLAC-PUB-14661, IPPP-11-67, DCPT-11-134, LPN11-58,
                        FR-PHENO-2011-019, MCNET-11-24",
      SLACcitation   = "%%CITATION = ARXIV:1111.1220;%%"
}

@article{Catani:2001cc,
      author         = "Catani, S. and Krauss, F. and Webber, B. R. and Kuhn, R.",
      title          = "{QCD Matrix Elements + Parton Showers}",
      journal        = "JHEP",
      volume         = "11",
      year           = "2001",
      pages          = "063",
      doi            = "10.1088/1126-6708/2001/11/063",
      eprint         = "hep-ph/0109231",
      archivePrefix  = "arXiv",
      reportNumber   = "CERN-TH-2000-367, CAVENDISH-HEP-00-03",
      SLACcitation   = "%%CITATION = HEP-PH/0109231;%%"
}

@article{Cascioli:2011va,
      author         = "Cascioli, Fabio and Maierh{\"o}fer, Philipp and Pozzorini, Stefano",
      title          = "{Scattering Amplitudes with Open Loops}",
      journal        = "Phys. Rev. Lett.",
      volume         = "108",
      year           = "2012",
      pages          = "111601",
      doi            = "10.1103/PhysRevLett.108.111601",
      eprint         = "1111.5206",
      archivePrefix  = "arXiv",
      primaryClass   = "hep-ph",
      reportNumber   = "ZU-TH-23-11, LPN11-66",
      SLACcitation   = "%%CITATION = ARXIV:1111.5206;%%"
}

@article{Buccioni:2019sur,
    author = "Buccioni, Federico and Lang, Jean-Nicolas and Lindert, Jonas M. and Maierh{\"o}fer, Philipp and Pozzorini, Stefano and Zhang, Hantian and Zoller, Max F.",
    title = "{OpenLoops 2}",
    eprint = "1907.13071",
    archivePrefix = "arXiv",
    primaryClass = "hep-ph",
    reportNumber = "IPPP/19/62, FR-PHENO-2019-12, PSI-PR-19-15, ZU-TH 37/19",
    doi = "10.1140/epjc/s10052-019-7306-2",
    journal = "Eur. Phys. J. C",
    volume = "79",
    number = "10",
    pages = "866",
    year = "2019"
}

@article{Denner:2016kdg,
      author         = "Denner, Ansgar and Dittmaier, Stefan and Hofer, Lars",
      title          = "{\textsc{Collier}: A fortran-based complex one-loop library in
                        extended regularizations}",
      journal        = "Comput. Phys. Commun.",
      volume         = "212",
      year           = "2017",
      pages          = "220-238",
      doi            = "10.1016/j.cpc.2016.10.013",
      eprint         = "1604.06792",
      archivePrefix  = "arXiv",
      primaryClass   = "hep-ph",
      reportNumber   = "FR-PHENO-2016-003, ICCUB-16-016",
      SLACcitation   = "%%CITATION = ARXIV:1604.06792;%%"
}

@article{Czakon:2011xx,
  author         = "Czakon, Michal and Mitov, Alexander",
  title          = "{Top++: A program for the calculation of the top-pair cross-section at hadron colliders}",
  journal        = "Comput. Phys. Commun.",
  volume         = "185",
  year           = "2014",
  pages          = "2930",
  doi            = "10.1016/j.cpc.2014.06.021",
  eprint         = "1112.5675",
  archivePrefix  = "arXiv",
  primaryClass   = "hep-ph",
  reportNumber   = "CERN-PH-TH-2011-303, TTK-11-58",
}

@article{Campbell:2012am,
      author         = "Campbell, John M. and Ellis, R. Keith and Frederix,
                        Rikkert and Nason, Paolo and Oleari, Carlo and Williams,
                        Ciaran",
      title          = "{NLO Higgs boson production plus one and two jets using
                        the POWHEG BOX, MadGraph4 and MCFM}",
      journal        = "JHEP",
      volume         = "07",
      year           = "2012",
      pages          = "092",
      doi            = "10.1007/JHEP07(2012)092",
      eprint         = "1202.5475",
      archivePrefix  = "arXiv",
      primaryClass   = "hep-ph",
      reportNumber   = "FERMILAB-PUB-12-040-T, CERN-PH-TH-2012-048",
      SLACcitation   = "%%CITATION = ARXIV:1202.5475;%%"
}

@article{Catani:2007vq,
      author         = "Catani, Stefano and Grazzini, Massimiliano",
      title          = "{Next-to-Next-to-Leading-Order Subtraction Formalism in Hadron Collisions and
                        its Application to Higgs-Boson Production at the Large Hadron Collider}",
      journal        = "Phys. Rev. Lett.",
      volume         = "98",
      year           = "2007",
      pages          = "222002",
      doi            = "10.1103/PhysRevLett.98.222002",
      eprint         = "hep-ph/0703012",
      archivePrefix  = "arXiv",
      SLACcitation   = "%%CITATION = HEP-PH/0703012;%%"
}

@article{Nason:2009ai,
      author         = "Nason, Paolo and Oleari, Carlo",
      title          = "{NLO Higgs boson production via vector-boson fusion
                        matched with shower in POWHEG}",
      journal        = "JHEP",
      volume         = "02",
      year           = "2010",
      pages          = "037",
      doi            = "10.1007/JHEP02(2010)037",
      eprint         = "0911.5299",
      archivePrefix  = "arXiv",
      primaryClass   = "hep-ph",
      SLACcitation   = "%%CITATION = ARXIV:0911.5299;%%"
}

@article{Anastasiou:2016cez,
      author         = "Anastasiou, Charalampos and Duhr, Claude and Dulat, Falko
                        and Furlan, Elisabetta and Gehrmann, Thomas and Herzog,
                        Franz and Lazopoulos, Achilleas and Mistlberger, Bernhard",
      title          = "{High precision determination of the gluon fusion Higgs
                        boson cross-section at the LHC}",
      journal        = "JHEP",
      volume         = "05",
      year           = "2016",
      pages          = "058",
      doi            = "10.1007/JHEP05(2016)058",
      eprint         = "1602.00695",
      archivePrefix  = "arXiv",
      primaryClass   = "hep-ph",
      reportNumber   = "CP3-16-01, ZU-TH-27-15, NIKHEF-2016-004,
                        CERN-TH-2016-006",
      SLACcitation   = "%%CITATION = ARXIV:1602.00695;%%"
}

@article{Anastasiou:2015ema,
      author         = "Anastasiou, Charalampos and Duhr, Claude and Dulat, Falko
                        and Herzog, Franz and Mistlberger, Bernhard",
      title          = "{Higgs Boson Gluon-Fusion Production in QCD at Three
                        Loops}",
      journal        = "Phys. Rev. Lett.",
      volume         = "114",
      year           = "2015",
      pages          = "212001",
      doi            = "10.1103/PhysRevLett.114.212001",
      eprint         = "1503.06056",
      archivePrefix  = "arXiv",
      primaryClass   = "hep-ph",
      reportNumber   = "CERN-PH-TH-2015-055, CP3-15-07",
      SLACcitation   = "%%CITATION = ARXIV:1503.06056;%%"
}

@article{Dulat:2018rbf,
      author         = "Dulat, Falko and Lazopoulos, Achilleas and Mistlberger,
                        Bernhard",
      title          = "{iHixs 2 -- Inclusive Higgs cross sections}",
      journal        = "Comput. Phys. Commun.",
      volume         = "233",
      year           = "2018",
      pages          = "243-260",
      doi            = "10.1016/j.cpc.2018.06.025",
      eprint         = "1802.00827",
      archivePrefix  = "arXiv",
      primaryClass   = "hep-ph",
      reportNumber   = "CERN-TH-2018-019, SLAC-PUB-17222",
      SLACcitation   = "%%CITATION = ARXIV:1802.00827;%%"
}

@article{Harlander:2009mq,
      author         = "Harlander, Robert V. and Ozeren, Kemal J.",
      title          = "{Finite top mass effects for hadronic Higgs production at
                        next-to-next-to-leading order}",
      journal        = "JHEP",
      volume         = "11",
      year           = "2009",
      pages          = "088",
      doi            = "10.1088/1126-6708/2009/11/088",
      eprint         = "0909.3420",
      archivePrefix  = "arXiv",
      primaryClass   = "hep-ph",
      reportNumber   = "WUB-09-10",
      SLACcitation   = "%%CITATION = ARXIV:0909.3420;%%"
}

@article{Harlander:2009bw,
      author         = "Harlander, Robert V. and Ozeren, Kemal J.",
      title          = "{Top mass effects in Higgs production at
                        next-to-next-to-leading order QCD: Virtual corrections}",
      journal        = "Phys. Lett. B",
      volume         = "679",
      year           = "2009",
      pages          = "467-472",
      doi            = "10.1016/j.physletb.2009.08.012",
      eprint         = "0907.2997",
      archivePrefix  = "arXiv",
      primaryClass   = "hep-ph",
      reportNumber   = "WUB-09-08",
      SLACcitation   = "%%CITATION = ARXIV:0907.2997;%%"
}

@article{Harlander:2009my,
      author         = "Harlander, Robert V. and Mantler, Hendrik and Marzani,
                        Simone and Ozeren, Kemal J.",
      title          = "{Higgs production in gluon fusion at
                        next-to-next-to-leading order QCD for finite top mass}",
      journal        = "Eur. Phys. J. C",
      volume         = "66",
      year           = "2010",
      pages          = "359-372",
      doi            = "10.1140/epjc/s10052-010-1258-x",
      eprint         = "0912.2104",
      archivePrefix  = "arXiv",
      primaryClass   = "hep-ph",
      reportNumber   = "WUB-09-18, MAN-HEP-2009-44",
      SLACcitation   = "%%CITATION = ARXIV:0912.2104;%%"
}

@article{Pak:2009dg,
      author         = "Pak, Alexey and Rogal, Mikhail and Steinhauser, Matthias",
      title          = "{Finite top quark mass effects in NNLO Higgs boson
                        production at LHC}",
      journal        = "JHEP",
      volume         = "02",
      year           = "2010",
      pages          = "025",
      doi            = "10.1007/JHEP02(2010)025",
      eprint         = "0911.4662",
      archivePrefix  = "arXiv",
      primaryClass   = "hep-ph",
      reportNumber   = "SFB-CPP-09-116, TTP09-43",
      SLACcitation   = "%%CITATION = ARXIV:0911.4662;%%"
}

@article{Actis:2008ug,
      author         = "Actis, Stefano and Passarino, Giampiero and Sturm,
                        Christian and Uccirati, Sandro",
      title          = "{NLO electroweak corrections to Higgs boson production at
                        hadron colliders}",
      journal        = "Phys. Lett. B",
      volume         = "670",
      year           = "2008",
      pages          = "12-17",
      doi            = "10.1016/j.physletb.2008.10.018",
      eprint         = "0809.1301",
      archivePrefix  = "arXiv",
      primaryClass   = "hep-ph",
      reportNumber   = "PITHA-08-20, SFB-CPP-08-62, TTP08-38",
      SLACcitation   = "%%CITATION = ARXIV:0809.1301;%%"
}

@article{Actis:2008ts,
      author         = "Actis, Stefano and Passarino, Giampiero and Sturm,
                        Christian and Uccirati, Sandro",
      title          = "{NNLO computational techniques: The cases \(H \to \gamma\gamma\) and \(H \to gg\)}",
      journal        = "Nucl. Phys. B",
      volume         = "811",
      year           = "2009",
      pages          = "182-273",
      doi            = "10.1016/j.nuclphysb.2008.11.024",
      eprint         = "0809.3667",
      archivePrefix  = "arXiv",
      primaryClass   = "hep-ph",
      reportNumber   = "PITHA-08-24, SFB-CPP-08-73, TTP08-42",
      SLACcitation   = "%%CITATION = ARXIV:0809.3667;%%"
}

@article{Bonetti:2018ukf,
      author         = "Bonetti, Marco and Melnikov, Kirill and Tancredi,
                        Lorenzo",
      title          = "{Higher order corrections to mixed QCD-EW contributions
                        to Higgs boson production in gluon fusion}",
      journal        = "Phys. Rev. D",
      volume         = "97",
      year           = "2018",
      number         = "5",
      pages          = "056017",
      doi            = "10.1103/PhysRevD.97.056017",
      eprint         = "1801.10403",
      archivePrefix  = "arXiv",
      primaryClass   = "hep-ph",
      reportNumber   = "CERN-TH-2018-011, TTP18-004",
      SLACcitation   = "%%CITATION = ARXIV:1801.10403;%%",
      related = "Bonetti:2018ukf-err",
      relatedstring  = "Erratum:",
}

@article{Ciccolini:2007jr,
     author         = "Ciccolini, M. and Denner, Ansgar and Dittmaier, S.",
     title          = "{Strong and Electroweak Corrections to the Production of a Higgs Boson + 2 Jets via Weak Interactions at the Large Hadron Collider}",
     journal        = "Phys. Rev. Lett.",
     volume         = "99",
     pages          = "161803",
     doi            = "10.1103/PhysRevLett.99.161803",
     year           = "2007",
     eprint         = "0707.0381",
     archivePrefix  = "arXiv",
     primaryClass   = "hep-ph",
}

@Article{Ciccolini:2007ec,
    author    = "Ciccolini, Mariano and Denner, Ansgar and Dittmaier, Stefan",
    title     = "{Electroweak and QCD corrections to Higgs production via
                 vector-boson fusion at the CERN LHC}",
    journal   = "Phys. Rev. D",
    volume    = "77",
    year      = "2008",
    pages     = "013002",
    eprint    = "0710.4749",
    archivePrefix = "arXiv",
    primaryClass  =  "hep-ph",
    doi       = "10.1103/PhysRevD.77.013002",
    SLACcitation  = "%%CITATION = 0710.4749;%%"
}

@Article{Bolzoni:2010xr,
    author    = "Bolzoni, Paolo and Maltoni, Fabio and Moch, Sven-Olaf and
                 Zaro, Marco",
    title     = "{Higgs Boson Production via Vector-Boson Fusion at Next-to-Next-to-Leading Order in QCD}",
    journal   = "Phys. Rev. Lett.",
    volume    = "105",
    year      = "2010",
    pages     = "011801",
    eprint    = "1003.4451",
    archivePrefix = "arXiv",
    primaryClass  =  "hep-ph",
    doi       = "10.1103/PhysRevLett.105.011801",
    SLACcitation  = "%%CITATION = 1003.4451;%%"
}

@article{Djouadi:1997yw,
    author    = "Djouadi, A. and Kalinowski, J. and Spira, M.",
    title     = "{HDECAY: a program for Higgs boson decays in the Standard
                 Model and its  supersymmetric extension}",
    journal   = "Comput. Phys. Commun.",
    volume    = "108",
    year      = "1998",
    pages     = "56",
    eprint    = "hep-ph/9704448",
    archivePrefix = "arXiv",
    doi       = "10.1016/S0010-4655(97)00123-9",
    reportNumber   = "DESY-97-079, IFT-96-29, PM-97-04",
    SLACcitation  = "%%CITATION = HEP-PH/9704448;%%"
}

@article{Spira:1997dg,
      author         = "Spira, Michael",
      title          = "{QCD Effects in Higgs Physics}",
      journal        = "Fortsch. Phys.",
      volume         = "46",
      year           = "1998",
      pages          = "203-284",
      doi            = "10.1002/(SICI)1521-3978(199804)46:3<203::AID-PROP203>3.0.CO;2-4",
      eprint         = "hep-ph/9705337",
      archivePrefix  = "arXiv",
      reportNumber   = "CERN-TH-97-068, CERN-TH-97-68",
      SLACcitation   = "%%CITATION = HEP-PH/9705337;%%"
}

@article{Bredenstein:2006ha,
      author         = "Bredenstein, A. and Denner, Ansgar and Dittmaier, S. and Weber, M. M.",
      title          = "{Radiative corrections to the semileptonic and hadronic
                        Higgs-boson decays \(H \to WW / ZZ \to 4\) fermions}",
      journal        = "JHEP",
      volume         = "02",
      year           = "2007",
      pages          = "080",
      doi            = "10.1088/1126-6708/2007/02/080",
      eprint         = "hep-ph/0611234",
      archivePrefix  = "arXiv",
      xprimaryClass   = "hep-ph",
      reportNumber   = "KEK-CP-187, KEK-PREPRINT-2006-47, MPP-2006-138,
                        PSI-PR-06-11, WUB-06-08",
      SLACcitation   = "%%CITATION = HEP-PH/0611234;%%"
}

@article{Bredenstein:2006rh,
      author         = "Bredenstein, A. and Denner, Ansgar and Dittmaier, S. and Weber, M. M.",
      title          = "{Precise predictions for the Higgs-boson decay $H \to WW/ZZ \to 4$ leptons}",
      journal        = "Phys. Rev. D",
      volume         = "74",
      year           = "2006",
      pages          = "013004",
      doi            = "10.1103/PhysRevD.74.013004",
      eprint         = "hep-ph/0604011",
      archivePrefix  = "arXiv",
      primaryClass   = "hep-ph",
      reportNumber   = "MPP-2005-24, PSI-PR-06-05",
      SLACcitation   = "%%CITATION = HEP-PH/0604011;%%"
}

@article{Ciccolini:2003jy,
      author         = "Ciccolini, M. L. and Dittmaier, S. and Kr{\"a}mer, M.",
      title          = "{Electroweak radiative corrections to associated \(WH\) and
                        \(ZH\) production at hadron colliders}",
      journal        = "Phys. Rev. D",
      volume         = "68",
      year           = "2003",
      pages          = "073003",
      doi            = "10.1103/PhysRevD.68.073003",
      eprint         = "hep-ph/0306234",
      archivePrefix  = "arXiv",
      reportNumber   = "EDINBURGH-2003-05, MPI-PHT-2003-24",
      SLACcitation   = "%%CITATION = HEP-PH/0306234;%%"
}

@article{Brein:2003wg,
      author         = "Brein, Oliver and Djouadi, Abdelhak and Harlander, Robert",
      title          = "{NNLO QCD corrections to the Higgs-strahlung processes at
                        hadron colliders}",
      journal        = "Phys. Lett. B",
      volume         = "579",
      year           = "2004",
      pages          = "149-156",
      doi            = "10.1016/j.physletb.2003.10.112",
      eprint         = "hep-ph/0307206",
      archivePrefix  = "arXiv",
      XprimaryClass   = "hep-ph",
      reportNumber   = "MPP-2003-35, CERN-TH-2003-161, PM-03-16",
      SLACcitation   = "%%CITATION = HEP-PH/0307206;%%"
}

@article{Brein:2011vx,
      author         = "Brein, Oliver and Harlander, Robert V. and Wiesemann, Marius and Zirke, Tom",
      title          = "{Top-quark mediated effects in hadronic Higgs-Strahlung}",
      journal        = "Eur. Phys. J. C",
      volume         = "72",
      year           = "2012",
      pages          = "1868",
      doi            = "10.1140/epjc/s10052-012-1868-6",
      eprint         = "1111.0761",
      archivePrefix  = "arXiv",
      primaryClass   = "hep-ph",
      reportNumber   = "CERN-PH-TH-2011-268, FR-PHENO-2011-016, WUB-11-15",
      SLACcitation   = "%%CITATION = ARXIV:1111.0761;%%"
}

@article{Altenkamp:2012sx,
      author         = "Altenkamp, Lukas and Dittmaier, Stefan and Harlander,
                        Robert V. and Rzehak, Heidi and Zirke, Tom J. E.",
      title          = "{Gluon-induced Higgs-strahlung at next-to-leading order
                        QCD}",
      journal        = "JHEP",
      volume         = "02",
      year           = "2013",
      pages          = "078",
      doi            = "10.1007/JHEP02(2013)078",
      eprint         = "1211.5015",
      archivePrefix  = "arXiv",
      primaryClass   = "hep-ph",
      reportNumber   = "CERN-PH-TH-2012-312, FR-PHENO-2012-023, WUB-12-21",
      SLACcitation   = "%%CITATION = ARXIV:1211.5015;%%"
}

@article{Denner:2014cla,
      author         = "Denner, Ansgar and Dittmaier, Stefan and Kallweit, Stefan
                        and M{\"u}ck, Alexander",
      title          = "{HAWK 2.0: A Monte Carlo program for Higgs production
                        in vector-boson fusion and Higgs strahlung at hadron
                        colliders}",
      journal        = "Comput. Phys. Commun.",
      volume         = "195",
      year           = "2015",
      pages          = "161-171",
      doi            = "10.1016/j.cpc.2015.04.021",
      eprint         = "1412.5390",
      archivePrefix  = "arXiv",
      primaryClass   = "hep-ph",
      reportNumber   = "FR-PHENO-2014-013, MITP-14-101, TTK-14-36",
      SLACcitation   = "%%CITATION = ARXIV:1412.5390;%%"
}

@article{Brein:2012ne,
      author         = "Brein, Oliver and Harlander, Robert V. and Zirke, Tom J.
                        E.",
      title          = "{vh@nnlo -- Higgs Strahlung at hadron colliders}",
      journal        = "Comput. Phys. Commun.",
      volume         = "184",
      year           = "2013",
      pages          = "998-1003",
      doi            = "10.1016/j.cpc.2012.11.002",
      eprint         = "1210.5347",
      archivePrefix  = "arXiv",
      primaryClass   = "hep-ph",
      SLACcitation   = "%%CITATION = ARXIV:1210.5347;%%"
}

@article{Harlander:2014wda,
      author         = "Harlander, Robert V. and Kulesza, Anna and Theeuwes,
                        Vincent and Zirke, Tom",
      title          = "{Soft gluon resummation for gluon-induced Higgs
                        Strahlung}",
      journal        = "JHEP",
      volume         = "11",
      year           = "2014",
      pages          = "082",
      doi            = "10.1007/JHEP11(2014)082",
      eprint         = "1410.0217",
      archivePrefix  = "arXiv",
      primaryClass   = "hep-ph",
      reportNumber   = "MS-TP-14-17, WUB-14-10, LPN14-112",
      SLACcitation   = "%%CITATION = ARXIV:1410.0217;%%"
}

@article{Agostinelli:2002hh,
      author         = "Agostinelli, S. and others",
      title          = "{\textsc{Geant4} -- a simulation toolkit}",
      journal        = "Nucl. Instrum. Meth. A",
      volume         = "506",
      year           = "2003",
      pages          = "250",
      doi            = "10.1016/S0168-9002(03)01368-8",
      reportNumber   = "SLAC-PUB-9350, FERMILAB-PUB-03-339",
      SLACcitation   = "%%CITATION = NUIMA,A506,250;%%"
}

@online{chollet2015keras,
      author         = {Chollet, Fran\c{c}ois and others},
      title          = {Keras},
      year           = {2015},
      url            = {https://keras.io},
}

@Article{PERF-2007-01,
    author         = "{ATLAS Collaboration}",
    title          = "{The ATLAS Experiment at the CERN Large Hadron Collider}",
    journal        = "JINST",
    volume         = "3",
    year           = "2008",
    pages          = "S08003",
    doi            = "10.1088/1748-0221/3/08/S08003",
    primaryClass   = "hep-ex",
}

@Article{SOFT-2010-01,
    author         = "{ATLAS Collaboration}",
    title          = "{The ATLAS Simulation Infrastructure}",
    journal        = "Eur. Phys. J. C",
    volume         = "70",
    year           = "2010",
    pages          = "823",
    doi            = "10.1140/epjc/s10052-010-1429-9",
    eprint         = "1005.4568",
    archivePrefix  = "arXiv",
    primaryClass   = "physics.ins-det",
}

@Article{HIGG-2012-27,
    author         = "{ATLAS Collaboration}",
    title          = "{Observation of a new particle in the search for the Standard Model Higgs boson with the ATLAS detector at the LHC}",
    journal        = "Phys. Lett. B",
    volume         = "716",
    year           = "2012",
    pages          = "1",
    doi            = "10.1016/j.physletb.2012.08.020",
    reportNumber   = "CERN-PH-EP-2012-218",
    eprint         = "1207.7214",
    archivePrefix  = "arXiv",
    primaryClass   = "hep-ex",
}

@Article{STDM-2012-23,
    author         = "{ATLAS Collaboration}",
    title          = "{Measurement of the \(Z/\gamma^*\) boson transverse momentum distribution in \(pp\) collisions at \(\sqrt{s} = 7\,\text{TeV}\) with the ATLAS detector}",
    journal        = "JHEP",
    volume         = "09",
    year           = "2014",
    pages          = "145",
    doi            = "10.1007/JHEP09(2014)145",
    reportNumber   = "CERN-PH-EP-2014-075",
    eprint         = "1406.3660",
    archivePrefix  = "arXiv",
    primaryClass   = "hep-ex",
}

@Article{HIGG-2014-09,
    author         = "{ATLAS Collaboration}",
    title          = "{Modelling \(Z\rightarrow\tau\tau\) processes in ATLAS with \(\tau\)-embedded \(Z\rightarrow\mu\mu\) data}",
    journal        = "JINST",
    volume         = "10",
    year           = "2015",
    pages          = "P09018",
    doi            = "10.1088/1748-0221/10/09/P09018",
    reportNumber   = "CERN-PH-EP-2015-130",
    eprint         = "1506.05623",
    archivePrefix  = "arXiv",
    primaryClass   = "hep-ex",
}

@Article{PERF-2014-03,
    author         = "{ATLAS Collaboration}",
    title          = "{Performance of pile-up mitigation techniques for jets in \(pp\) collisions at \(\sqrt{s} = 8\,\text{TeV}\) using the ATLAS detector}",
    journal        = "Eur. Phys. J. C",
    volume         = "76",
    year           = "2016",
    pages          = "581",
    doi            = "10.1140/epjc/s10052-016-4395-z",
    reportNumber   = "CERN-PH-EP-2015-206",
    eprint         = "1510.03823",
    archivePrefix  = "arXiv",
    primaryClass   = "hep-ex",
}

@Article{HIGG-2015-06,
    author         = "{ATLAS Collaboration}",
    title          = "{Test of \(CP\) invariance in vector-boson fusion production of the Higgs boson using the \textit{Optimal Observable} method in the ditau decay channel with the ATLAS detector}",
    journal        = "Eur. Phys. J. C",
    volume         = "76",
    year           = "2016",
    pages          = "658",
    doi            = "10.1140/epjc/s10052-016-4499-5",
    reportNumber   = "CERN-EP-2016-002",
    eprint         = "1602.04516",
    archivePrefix  = "arXiv",
    primaryClass   = "hep-ex",
}

@Article{TRIG-2016-01,
    author         = "{ATLAS Collaboration}",
    title          = "{Performance of the ATLAS trigger system in 2015}",
    journal        = "Eur. Phys. J. C",
    volume         = "77",
    year           = "2017",
    pages          = "317",
    doi            = "10.1140/epjc/s10052-017-4852-3",
    reportNumber   = "CERN-EP-2016-241",
    eprint         = "1611.09661",
    archivePrefix  = "arXiv",
    primaryClass   = "hep-ex",
}

@Article{DAPR-2018-01,
    author         = "{ATLAS Collaboration}",
    title          = "{ATLAS data quality operations and performance for 2015--2018 data-taking}",
    journal        = "JINST",
    volume         = "15",
    year           = "2020",
    pages          = "P04003",
    doi            = "10.1088/1748-0221/15/04/P04003",
    reportNumber   = "CERN-EP-2019-207",
    eprint         = "1911.04632",
    archivePrefix  = "arXiv",
    primaryClass   = "physics.ins-det",
}

@Article{FTAG-2018-01,
    author         = "{ATLAS Collaboration}",
    title          = "{ATLAS \(b\)-jet identification performance and efficiency measurement with \(t\bar{t}\) events in \(pp\) collisions at \(\sqrt{s} = 13\,\text{TeV}\)}",
    journal        = "Eur. Phys. J. C",
    volume         = "79",
    year           = "2019",
    pages          = "970",
    doi            = "10.1140/epjc/s10052-019-7450-8",
    reportNumber   = "CERN-EP-2019-132",
    eprint         = "1907.05120",
    archivePrefix  = "arXiv",
    primaryClass   = "hep-ex",
}

@Article{HIGG-2018-14,
    author         = "{ATLAS Collaboration}",
    title          = "{Test of CP invariance in vector-boson fusion production of the Higgs boson in the \(H\to\tau\tau\) channel in proton--proton collisions at \(\sqrt{s} = 13\,\text{TeV}\) with the ATLAS detector}",
    journal        = "Phys. Lett. B",
    volume         = "805",
    year           = "2020",
    pages          = "135426",
    doi            = "10.1016/j.physletb.2020.135426",
    reportNumber   = "CERN-EP-2020-009",
    eprint         = "2002.05315",
    archivePrefix  = "arXiv",
    primaryClass   = "hep-ex",
}

@Article{JETM-2018-05,
    author         = "{ATLAS Collaboration}",
    title          = "{Jet energy scale and resolution measured in proton--proton collisions at \(\sqrt{s} = 13\,\text{TeV}\) with the ATLAS detector}",
    journal        = "Eur. Phys. J. C",
    volume         = "81",
    year           = "2021",
    pages          = "689",
    doi            = "10.1140/epjc/s10052-021-09402-3",
    reportNumber   = "CERN-EP-2020-083",
    eprint         = "2007.02645",
    archivePrefix  = "arXiv",
    primaryClass   = "hep-ex",
}

@Article{MUON-2018-03,
    author         = "{ATLAS Collaboration}",
    title          = "{Muon reconstruction and identification efficiency in ATLAS using the full Run~2 \(pp\) collision data set at \(\sqrt{s} = 13\,\text{TeV}\)}",
    journal        = "Eur. Phys. J. C",
    volume         = "81",
    year           = "2021",
    pages          = "578",
    doi            = "10.1140/epjc/s10052-021-09233-2",
    reportNumber   = "CERN-EP-2020-199",
    eprint         = "2012.00578",
    archivePrefix  = "arXiv",
    primaryClass   = "hep-ex",
}

@Article{PIX-2018-001,
    author         = "Abbott, B. and others",
    title          = "{Production and integration of the ATLAS Insertable B-Layer}",
    journal        = "JINST",
    volume         = "13",
    year           = "2018",
    pages          = "T05008",
    doi            = "10.1088/1748-0221/13/05/T05008",
    eprint         = "1803.00844",
    archivePrefix  = "arXiv",
    primaryClass   = "physics.ins-det",
}

@Article{TRIG-2018-01,
    author         = "{ATLAS Collaboration}",
    title          = "{Performance of the ATLAS muon triggers in Run~2}",
    journal        = "JINST",
    volume         = "15",
    year           = "2020",
    pages          = "P09015",
    doi            = "10.1088/1748-0221/15/09/p09015",
    reportNumber   = "CERN-EP-2020-031",
    eprint         = "2004.13447",
    archivePrefix  = "arXiv",
    primaryClass   = "physics.ins-det",
}

@Article{TRIG-2018-05,
    author         = "{ATLAS Collaboration}",
    title          = "{Performance of electron and photon triggers in ATLAS during LHC Run~2}",
    journal        = "Eur. Phys. J. C",
    volume         = "80",
    year           = "2020",
    pages          = "47",
    doi            = "10.1140/epjc/s10052-019-7500-2",
    reportNumber   = "CERN-EP-2019-169",
    eprint         = "1909.00761",
    archivePrefix  = "arXiv",
    primaryClass   = "hep-ex",
}

@Article{EGAM-2019-01,
    author         = "{ATLAS Collaboration}",
    title          = "{Tools for estimating fake/non-prompt lepton backgrounds with the ATLAS detector at the LHC}",
    journal        = "JINST",
    volume         = "18",
    year           = "2023",
    pages          = "T11004",
    doi            = "10.1088/1748-0221/18/11/T11004",
    reportNumber   = "CERN-EP-2022-214",
    eprint         = "2211.16178",
    archivePrefix  = "arXiv",
    primaryClass   = "hep-ex",
}

@Article{FTAG-2019-02,
    author         = "{ATLAS Collaboration}",
    title          = "{Calibration of the light-flavour jet mistagging efficiency of the \(b\)-tagging algorithms with \(Z\)+jets events using \(139\,\text{fb}^{-1}\) of ATLAS proton--proton collision data at \(\sqrt{s} = 13\,\text{TeV}\)}",
    journal        = "Eur. Phys. J. C",
    volume         = "83",
    year           = "2023",
    pages          = "728",
    doi            = "10.1140/epjc/s10052-023-11736-z",
    reportNumber   = "CERN-EP-2022-211",
    eprint         = "2301.06319",
    archivePrefix  = "arXiv",
    primaryClass   = "hep-ex",
}

@Article{FTAG-2019-07,
    author         = "{ATLAS Collaboration}",
    title          = "{ATLAS flavour-tagging algorithms for the LHC Run~2 \(pp\) collision dataset}",
    journal        = "Eur. Phys. J. C",
    volume         = "83",
    year           = "2023",
    pages          = "681",
    doi            = "10.1140/epjc/s10052-023-11699-1",
    reportNumber   = "CERN-EP-2022-226",
    eprint         = "2211.16345",
    archivePrefix  = "arXiv",
    primaryClass   = "physics.data-an",
}

@Article{TRIG-2019-02,
    author         = "{ATLAS Collaboration}",
    title          = "{Performance of the ATLAS Level-1 topological trigger in Run~2}",
    journal        = "Eur. Phys. J. C",
    volume         = "82",
    year           = "2022",
    pages          = "7",
    doi            = "10.1140/epjc/s10052-021-09807-0",
    reportNumber   = "CERN-EP-2021-040",
    eprint         = "2105.01416",
    archivePrefix  = "arXiv",
    primaryClass   = "hep-ex",
}

@Article{TRIG-2019-03,
    author         = "{ATLAS Collaboration}",
    title          = "{The ATLAS inner detector trigger performance in \(pp\) collisions at \(13\,\text{TeV}\) during LHC Run~2}",
    journal        = "Eur. Phys. J. C",
    volume         = "82",
    year           = "2022",
    pages          = "206",
    doi            = "10.1140/epjc/s10052-021-09920-0",
    reportNumber   = "CERN-EP-2021-076",
    eprint         = "2107.02485",
    archivePrefix  = "arXiv",
    primaryClass   = "hep-ex",
}

@Article{HIGG-2020-08,
    author         = "{ATLAS Collaboration}",
    title          = "{Test of \(CP\) Invariance in Higgs Boson Vector-Boson-Fusion Production Using the \(H \to \gamma\gamma\) Channel with the ATLAS Detector}",
    journal        = "Phys. Rev. Lett.",
    volume         = "131",
    year           = "2023",
    pages          = "061802",
    doi            = "10.1103/PhysRevLett.131.061802",
    reportNumber   = "CERN-EP-2022-134",
    eprint         = "2208.02338",
    archivePrefix  = "arXiv",
    primaryClass   = "hep-ex",
}

@Article{HIGG-2020-16,
    author         = "{ATLAS Collaboration}",
    title          = "{Measurement of the properties of Higgs boson production at \(\sqrt{s} = 13\,\text{TeV}\) in the \(H \to \gamma\gamma\) channel using \(139\,\text{fb}^{-1}\) of \(pp\) collision data with the ATLAS experiment}",
    journal        = "JHEP",
    volume         = "07",
    year           = "2023",
    pages          = "088",
    doi            = "10.1007/JHEP07(2023)088",
    reportNumber   = "CERN-EP-2022-094",
    eprint         = "2207.00348",
    archivePrefix  = "arXiv",
    primaryClass   = "hep-ex",
}

@Article{DAPR-2021-01,
    author         = "{ATLAS Collaboration}",
    title          = "{Luminosity determination in \(pp\) collisions at \(\sqrt{s} = 13\,\text{TeV}\) using the ATLAS detector at the LHC}",
    journal        = "Eur. Phys. J. C",
    volume         = "83",
    year           = "2023",
    pages          = "982",
    doi            = "10.1140/epjc/s10052-023-11747-w",
    reportNumber   = "CERN-EP-2022-281",
    eprint         = "2212.09379",
    archivePrefix  = "arXiv",
    primaryClass   = "hep-ex",
}

@Article{EGAM-2021-01,
    author         = "{ATLAS Collaboration}",
    title          = "{Electron and photon efficiencies in LHC Run~2 with the ATLAS experiment}",
    journal        = "JHEP",
    volume         = "05",
    year           = "2024",
    pages          = "162",
    doi            = "10.1007/JHEP05(2024)162",
    reportNumber   = "CERN-EP-2023-182",
    eprint         = "2308.13362",
    archivePrefix  = "arXiv",
    primaryClass   = "hep-ex",
}

@Article{HIGG-2021-20,
    author         = "{ATLAS Collaboration}",
    title          = "{Measurements of Higgs boson production by gluon--gluon fusion and vector-boson fusion using \(H \to W W^{*} \to e\nu \mu\nu\) decays in \(pp\) collisions at \(\sqrt{s} = 13\,\text{TeV}\) with the ATLAS detector}",
    journal        = "Phys. Rev. D",
    volume         = "108",
    year           = "2023",
    pages          = "032005",
    doi            = "10.1103/PhysRevD.108.032005",
    reportNumber   = "CERN-EP-2022-078",
    eprint         = "2207.00338",
    archivePrefix  = "arXiv",
    primaryClass   = "hep-ex",
}

@Article{HIGG-2022-07,
    author         = "{ATLAS Collaboration}",
    title          = "{Differential cross-section measurements of Higgs boson production in the \(H \rightarrow \tau^+\tau^-\) decay channel in \(pp\) collisions at \(\sqrt{s} = 13\,\text{TeV}\) with the ATLAS detector}",
    journal        = "JHEP",
    volume         = "03",
    year           = "2025",
    pages          = "010",
    doi            = "10.1007/JHEP03(2025)010",
    reportNumber   = "CERN-EP-2024-198",
    eprint         = "2407.16320",
    archivePrefix  = "arXiv",
    primaryClass   = "hep-ex",
}

@Article{SOFT-2022-02,
    author         = "{ATLAS Collaboration}",
    title          = "{Software and computing for Run~3 of the ATLAS experiment at the LHC}",
    journal        = "Eur. Phys. J. C",
    volume         = "85",
    year           = "2025",
    pages          = "234",
    doi            = "10.1140/epjc/s10052-024-13701-w",
    reportNumber   = "CERN-EP-2024-100",
    eprint         = "2404.06335",
    archivePrefix  = "arXiv",
    primaryClass   = "hep-ex",
}

@Article{HIGP-2024-07,
    author         = "{ATLAS Collaboration}",
    title          = "{Measurements of Higgs boson production via gluon--gluon fusion and vector-boson fusion using \(H \rightarrow WW^{*} \rightarrow \ell\nu\ell\nu\) decays in \(pp\) collisions with the ATLAS detector and their effective field theory interpretations}",
    year           = "2025",
    reportNumber   = "CERN-EP-2025-054",
    eprint         = "2504.07686",
    archivePrefix  = "arXiv",
    primaryClass   = "hep-ex",
}

@Article{HIGP-2024-09,
    author         = "{ATLAS Collaboration}",
    title          = "{Probing the Higgs boson CP properties in vector-boson fusion production in the \(H \rightarrow \tau^{+}\tau^{-}\) channel with the ATLAS detector}",
    year           = "2025",
    reportNumber   = "CERN-EP-2025-127",
    eprint         = "2506.19395",
    archivePrefix  = "arXiv",
    primaryClass   = "hep-ex",
}

@Booklet{ATL-SOFT-PUB-2025-001,
    author         = "{ATLAS Collaboration}",
    title          = "{ATLAS Computing Acknowledgements}",
    howpublished   = "{ATL-SOFT-PUB-2025-001}",
    url            = "https://cds.cern.ch/record/2922210",
    year           = "2025",
}

@Report{ATLAS-TDR-19,
    author         = "{ATLAS Collaboration}",
    title          = "{ATLAS Insertable B-Layer: Technical Design Report}",
    type           = "ATLAS-TDR-19; CERN-LHCC-2010-013",
    year           = "2010",
    url            = "https://cds.cern.ch/record/1291633",
    related        = "ATLAS-TDR-19-addm",
    relatedstring  = "Addendum:",
}

@Article{LUCID2,
    author         = "Avoni, G. and others",
    title          = "{The new LUCID-2 detector for luminosity measurement and monitoring in ATLAS}",
    journal        = "JINST",
    volume         = "13",
    year           = "2018",
    pages          = "P07017",
    doi            = "10.1088/1748-0221/13/07/P07017",
}

@Article{CMS-HIG-12-028,
    author         = "{CMS Collaboration}",
    title          = "{Observation of a new boson at a mass of 125 GeV with the CMS experiment at the LHC}",
    journal        = "Phys. Lett. B",
    volume         = "716",
    year           = "2012",
    pages          = "30",
    doi            = "10.1016/j.physletb.2012.08.021",
    reportNumber   = "CERN-PH-EP-2012-220",
    eprint         = "1207.7235",
    archivePrefix  = "arXiv",
    primaryClass   = "hep-ex",
}

@Article{CMS-HIG-14-018,
    author         = "{CMS Collaboration}",
    title          = "{Constraints on the spin-parity and anomalous HVV couplings of the Higgs boson in proton collisions at \(7\) and \(8\,\text{TeV}\)}",
    journal        = "Phys. Rev. D",
    volume         = "92",
    year           = "2015",
    pages          = "012004",
    doi            = "10.1103/PhysRevD.92.012004",
    reportNumber   = "CERN-PH-EP-2014-265",
    eprint         = "1411.3441",
    archivePrefix  = "arXiv",
    primaryClass   = "hep-ex",
}

@Article{CMS-HIG-19-009,
    author         = "{CMS Collaboration}",
    title          = "{Constraints on anomalous Higgs boson couplings to vector bosons and fermions in its production and decay using the four-lepton final state}",
    journal        = "Phys. Rev. D",
    volume         = "104",
    year           = "2021",
    pages          = "052004",
    doi            = "10.1103/PhysRevD.104.052004",
    reportNumber   = "CERN-EP-2021-054",
    eprint         = "2104.12152",
    archivePrefix  = "arXiv",
    primaryClass   = "hep-ex",
}

@Article{CMS-HIG-20-007,
    author         = "{CMS Collaboration}",
    title          = "{Constraints on anomalous Higgs boson couplings to vector bosons and fermions from the production of Higgs bosons using the \(\tau\tau\) final state}",
    journal        = "Phys. Rev. D",
    volume         = "108",
    year           = "2023",
    pages          = "032013",
    doi            = "10.1103/PhysRevD.108.032013",
    reportNumber   = "CERN-EP-2022-021",
    eprint         = "2205.05120",
    archivePrefix  = "arXiv",
    primaryClass   = "hep-ex",
}

@Article{CMS-HIG-23-016,
    author         = "{CMS Collaboration}",
    title          = "{Constraints on standard model effective field theory for a Higgs boson produced in association with \(W\) or \(Z\) bosons in the \(H \rightarrow b\bar{b}\) decay channel in proton--proton collisions at \(\sqrt{s} = 13\,\text{TeV}\)}",
    journal        = "JHEP",
    volume         = "03",
    year           = "2025",
    pages          = "114",
    doi            = "10.1007/JHEP03(2025)114",
    reportNumber   = "CERN-EP-2024-294",
    eprint         = "2411.16907",
    archivePrefix  = "arXiv",
    primaryClass   = "hep-ex",
}

@Booklet{ATL-PHYS-PUB-2014-021,
    author         = "{ATLAS Collaboration}",
    title          = "{ATLAS Pythia~8 tunes to \(7~\text{TeV}\) data}",
    howpublished   = "{ATL-PHYS-PUB-2014-021}",
    url            = "https://cds.cern.ch/record/1966419",
    year           = "2014",
}

@Booklet{ATL-PHYS-PUB-2016-017,
    author         = "{ATLAS Collaboration}",
    title          = "{The Pythia~8 A3 tune description of ATLAS minimum bias and inelastic measurements incorporating the Donnachie--Landshoff diffractive model}",
    howpublished   = "{ATL-PHYS-PUB-2016-017}",
    url            = "https://cds.cern.ch/record/2206965",
    year           = "2016",
}

@Booklet{ATL-PHYS-PUB-2016-020,
    author         = "{ATLAS Collaboration}",
    title          = "{Studies on top-quark Monte Carlo modelling for Top2016}",
    howpublished   = "{ATL-PHYS-PUB-2016-020}",
    url            = "https://cds.cern.ch/record/2216168",
    year           = "2016",
}

@Booklet{ATL-PHYS-PUB-2017-007,
    author         = "{ATLAS Collaboration}",
    title          = "{Studies on top-quark Monte Carlo modelling with Sherpa and MG5\_aMC@NLO}",
    howpublished   = "{ATL-PHYS-PUB-2017-007}",
    url            = "https://cds.cern.ch/record/2261938",
    year           = "2017",
}

@Booklet{ATL-PHYS-PUB-2019-033,
    author         = "{ATLAS Collaboration}",
    title          = "{Identification of hadronic tau lepton decays using neural networks in the ATLAS experiment}",
    howpublished   = "{ATL-PHYS-PUB-2019-033}",
    url            = "https://cds.cern.ch/record/2688062",
    year           = "2019",
}

@Booklet{ATL-PHYS-PUB-2023-031,
    author         = "{ATLAS Collaboration}",
    title          = "{Evaluation of QCD uncertainties for Higgs boson production through gluon fusion and in association with two top quarks for simplified template cross-section measurements}",
    howpublished   = "{ATL-PHYS-PUB-2023-031}",
    url            = "https://cds.cern.ch/record/2878797",
    year           = "2023",
}

\clearpage
 
\begin{flushleft}
\hypersetup{urlcolor=black}
{\Large The ATLAS Collaboration}

\bigskip

\AtlasOrcid[0000-0002-6665-4934]{G.~Aad}$^\textrm{\scriptsize 104}$,
\AtlasOrcid[0000-0001-7616-1554]{E.~Aakvaag}$^\textrm{\scriptsize 17}$,
\AtlasOrcid[0000-0002-5888-2734]{B.~Abbott}$^\textrm{\scriptsize 123}$,
\AtlasOrcid[0000-0002-0287-5869]{S.~Abdelhameed}$^\textrm{\scriptsize 119a}$,
\AtlasOrcid[0000-0002-1002-1652]{K.~Abeling}$^\textrm{\scriptsize 55}$,
\AtlasOrcid[0000-0001-5763-2760]{N.J.~Abicht}$^\textrm{\scriptsize 49}$,
\AtlasOrcid[0000-0002-8496-9294]{S.H.~Abidi}$^\textrm{\scriptsize 30}$,
\AtlasOrcid[0009-0003-6578-220X]{M.~Aboelela}$^\textrm{\scriptsize 45}$,
\AtlasOrcid[0000-0002-9987-2292]{A.~Aboulhorma}$^\textrm{\scriptsize 36e}$,
\AtlasOrcid[0000-0001-5329-6640]{H.~Abramowicz}$^\textrm{\scriptsize 157}$
% ...
\bigskip
\\

% ...
$^{17}$Department for Physics and Technology, University of Bergen, Bergen; Norway.\\
% ...
$^{30}$Physics Department, Brookhaven National Laboratory, Upton NY; United States of America.\\
% ...
$^{36}$$^{(a)}$Facult\'e des Sciences Ain Chock, Universit\'e Hassan II de Casablanca;$^{(b)}$Facult\'{e} des Sciences, Universit\'{e} Ibn-Tofail, K\'{e}nitra;$^{(c)}$Facult\'e des Sciences Semlalia, Universit\'e Cadi Ayyad, LPHEA-Marrakech;$^{(d)}$LPMR, Facult\'e des Sciences, Universit\'e Mohamed Premier, Oujda;$^{(e)}$Facult\'e des sciences, Universit\'e Mohammed V, Rabat;$^{(f)}$Institute of Applied Physics, Mohammed VI Polytechnic University, Ben Guerir; Morocco.\\
% ...
$^{45}$Physics Department, Southern Methodist University, Dallas TX; United States of America.\\
% ...
$^{49}$Fakult\"{a}t Physik , Technische Universit{\"a}t Dortmund, Dortmund; Germany.\\
% ...
$^{55}$II. Physikalisches Institut, Georg-August-Universit\"{a}t G\"ottingen, G\"ottingen; Germany.\\
% ...
$^{104}$CPPM, Aix-Marseille Universit\'e, CNRS/IN2P3, Marseille; France.\\
% ...
$^{119}$$^{(a)}$New York University Abu Dhabi, Abu Dhabi;$^{(b)}$United Arab Emirates University, Al Ain; United Arab Emirates.\\
% ...
$^{123}$Homer L. Dodge Department of Physics and Astronomy, University of Oklahoma, Norman OK; United States of America.\\
% ...
$^{157}$Raymond and Beverly Sackler School of Physics and Astronomy, Tel Aviv University, Tel Aviv; Israel.\\
% ...
$^{a}$ Also at Affiliated with an institute formerly covered by a cooperation agreement with CERN.\\
$^{b}$ Also at An-Najah National University, Nablus; Palestine.\\
$^{c}$ Also at Borough of Manhattan Community College, City University of New York, New York NY; United States of America.\\
$^{d}$ Also at Center for Interdisciplinary Research and Innovation (CIRI-AUTH), Thessaloniki; Greece.\\
$^{e}$ Also at Centre of Physics of the Universities of Minho and Porto (CF-UM-UP); Portugal.\\
$^{f}$ Also at CERN, Geneva; Switzerland.\\
% ...

\end{flushleft}


\end{document}